\documentclass[runningheads]{llncs}
\usepackage{subfigure}
\usepackage{amscd}
\usepackage{latexsym}
\usepackage{amsmath}
\usepackage{amsfonts}
\usepackage{amssymb}
\DeclareMathSymbol\RRightarrow{\mathrel}{AMSa}{"56}
\DeclareMathSymbol\LLeftarrow{\mathrel}{AMSa}{"57}
\usepackage{enumerate}
\usepackage{graphicx}
\usepackage{verbatim}
\usepackage{color}
\usepackage[normalem]{ulem} 
\usepackage{wrapfig}
\newcommand\Supp[1] {\lceil#1\rceil}
\newcommand\Defs {\mathord{:=}\,}

\newcommand{\Et}{\mathsf{bv}}

\newcommand{\Ft}{\mathsf{ft}}
\newcommand{\Ht}{\mathsf{H}}
\newcommand{\Bt}{\mathsf{B}}
\newcommand\Gt[1]{\mathsf{W}_{#1}}

\newcommand\CCL {\mathsf{ccl}} 

\newcommand\Vh {\mathsf{h}}
\newcommand\Vv {\mathsf{v}}
\newcommand\Va {\mathsf{a}}
\newcommand\Vb {\mathsf{b}}
\newcommand\Vc {\mathsf{c}}
\newcommand\Vd {\mathsf{d}}

\newcommand\Vx {\mathsf{x}}
\newcommand\Vy {\mathsf{y}}
\newcommand\Vm {\mathsf{m}}
\newcommand\Vn {\mathsf{n}}

\newcommand\Point[1] {\PSet{}{}{#1}}

\newcommand\Sec[1] {Sec.~\ref{#1}}
\newcommand\App[1] {App.~\ref{#1}}
\renewcommand\Sec[1] {\S\ref{#1}}
\renewcommand\App[1] {\S\ref{#1}}

\newcommand\Wlog {\textit{wlog}}
\newcommand\Eqn[1] {(\ref{#1})}

\newcommand\Spot {\raisebox{-.1em}{\Large\boldmath$\cdot$}}

\newcommand\VV {{\cal V}}
\newcommand\NLg {\textrm{l}\overline{\textrm{g}}}
\renewcommand\NLg {\overline{\lg}}
\newcommand\CC {{\cal C}}
\newcommand\DD {{\cal D}}
\newcommand\SeqCC[1] {(-\,;\,#1)}
\newcommand\HH {{\cal H}}
\newcommand\Atomic[1] {\AtomicOpen#1\AtomicClose}
\newcommand\AtomicOpen {\langle\!\langle}
\newcommand\AtomicClose {\rangle\!\rangle}
\newcommand\Fun {\mathbin{\rightarrow}}
\newcommand\Rel {\mathbin{\leftrightarrow}}

\newcommand\Pow {{\mathbb P}}

\newcommand\Dist {{\mathbb D}}
\newcommand\TDist {\mathsf{D}}
\newcommand\Uniform[1] {\PSet{}{}{#1}}
\newcommand\Support[1] {{\lceil #1 \rceil}}
\newcommand\Norm[1]{[#1]}
\newcommand{\Transpose}[1]{{#1}^{\mathbf{T}}} 
\newcommand\Above {\mathbin{\raisebox{-.15em}{+}\kern-.39em\raisebox{.15em}{\makebox[0pt]{$+$}}\kern.45em}}
\newcommand\AboveRel {~\mathrel{\Above}~}
\newcommand\Beside {{+}\kern-.5em{+}}

\newcommand\IdM[1] {{\mathsf 1}_{#1}}
\newcommand\DiagM[1] {\backslash\kern-.35em\backslash\kern-.05em\mbox{\small$#1$}\kern-.05em\backslash\kern-.35em\backslash}

\newcommand\HMM {\textit{HMM}}

\newcommand\MAX {\sqcup}
\newcommand\MIN {\sqcap}
\newcommand\UMax[1] {{\def\Cond{#1}\ifx\Cond\empty {\sqcup} \else {\sqcup}^{#1} \fi}}
\newcommand\UMin[1] {{\def\Cond{#1}\ifx\Cond\empty {\sqcup} \else {\sqcap}^{#1} \fi}}

\newcommand\In {{:}\,}
\newcommand\List[1] {\langle#1\rangle}
\newcommand\Ref {\mathrel{\sqsubseteq}}
\newcommand\Finer {\mathrel{\raisebox{0.2ex}{$\sqsubset$}\kern-.8em\raisebox{-1.1ex}{$\sim$}}}
\newcommand\StrictFiner {\mathrel{\raisebox{0.2ex}{$\sqsubset$}\kern-.8em\raisebox{-1.1ex}{${\sim}\kern-.55em\raisebox{.15em}{\tiny$/$}$}\kern.4em}}
\newcommand\Coarser {\mathrel{\raisebox{0.2ex}{$\sqsupset$}\kern-.8em\raisebox{-1.1ex}{$\sim$}}}

\newcommand\Similar {\mathrel{\approx}}
\newcommand\NotSimilar {\mathrel{\not\approx}}
\newcommand\StrictRef {\mathrel{\sqsubset}}

\newcommand\NotRef {\mathrel{\not\sqsubseteq}}
\newcommand\ORef[1] {\mathrel{\preceq_{#1}}}
\newcommand\StrictORef[1] {\mathrel{\prec_{#1}}}

\newcommand\NotORef[1] {\mathrel{\not\preceq_{#1}}}
\newcommand\ERef {\mathrel{\preceq}}

\newcommand\NotERef {\mathrel{\not\preceq}}
\newcommand\BRef {\mathrel{\preceq_\Bt}}
\newcommand\NotBRef {\mathrel{\not\preceq_\Bt}}
\newcommand\StrictBRef {\mathrel{\prec_\Bt}}
\newcommand\HRef {\mathrel{\preceq_\Ht}}
\newcommand\NotHRef {\mathrel{\not\preceq_\Ht}}
\newcommand\StrictHRef {\mathrel{\prec_\Ht}}
\newcommand\GRef[1] {\mathrel{\preceq_{\Gt{#1}}}}
\newcommand\NotGRef[1] {\mathrel{\not\preceq_{\Gt{#1}}}}

\newcommand{\RefMatrices}{{\cal R}}
\newcommand{\FinerMatrices}{{\cal M}} 
\newcommand{\StratMatrices}{{\cal G}}
\newcommand{\UnitMatrix}{{\bf 1}}
\newcommand{\ZeroMatrix}{{\bf 0}}

\newcommand{\Fracs}{\textsf{fracs}}    
\newcommand\From[1][*] {{\def\z{#1}\if*\z {:}{\in}\,\else {:}{\in}_{\kern-.1em{#1}}\,\fi}}
\newcommand\Gets {{:}{=}\,}
\newcommand\NGets {{:}{\in}\,}
\newcommand\PGets {{\NGets}}
\newcommand\Implies {\mathop{\Rightarrow}}

\newcommand\RHS{\textit{rhs}}
\newcommand\If {\textbf{if}}
\newcommand\Then {\textbf{then}}
\newcommand\Else {\textbf{else}}

\newcommand\Fi {\textbf{fi}}
\newcommand\Reveal {\textbf{reveal}}
\newcommand\Sem[1] {[\![#1]\!]}
\newcommand\Vpk {v'\kern-.25em}
\newcommand\MatSem[2][*] {\def\z{#1}\if*\z (\!(#2)\!) \else \left(\kern-.3em\left(\rule{0pt}{#1}#2\right)\kern-.3em\right) \fi}
\newcommand\MatSemC[1] {(\kern-.21em\raisebox{.025em}{\small$\mid$}#1\raisebox{.025em}{\small$\mid$}\kern-.22em)}
\newcommand\SemC[1] {[\![#1]\!]_C}
\newcommand\Or {\mathbin{\sqcap}}
\newcommand\PC[1] {\mathbin{_{#1}\oplus}} 
\newcommand{\ITE}[3]{#1\,\IF\,#2\,\ELSE\,#3}
\newcommand\True {\textsf{true}}
\newcommand\False {\textsf{false}}

\newcommand\Xor {\mathbin{\nabla}}

\newcommand\Vis {\textbf{vis}}
\newcommand\Hid {\textbf{hid}}

\newcommand{\Begin}{\mbox{\boldmath\textbf{$\parallel\kern-.365em[$}}}
\newcommand{\End}{\mbox{\boldmath\textbf{$]\kern-.365em\parallel$}}}

\newcommand{\LeftPS}{ \{\!\!\{ }
\newcommand{\RightPS}{ \}\!\!\} }

\newcommand{\Hide}{\mathsf{embed}} 

\newcommand\Skip {\textbf{skip}}
\newcommand\Abort {\textbf{abort}}
\newcommand\HangRight[1] {\makebox[0pt][l]{#1}}
\newcommand\HangLeft[1] {\makebox[0pt][r]{#1}}
\newcommand\InQuadL[1] {\makebox[0pt][l]{$#1$}\quad}

\newcommand\IF {\textbf{if}}
\newcommand\Mod {\mathbin{\bf mod}}
\newcommand\Rnd {\mathbin{\bf rnd}}

\newcommand\Tr {\textsf{tr}}

\newcommand\ELSE {\textbf{else}}
\newcommand\Wide[1] {~~~#1~~~}
\newcommand\WideRm[1] {~~~\textrm{#1}~~~}
\newcommand\WIDE[1] {~~~~~~#1~~~~~~}
\newcommand\WIDERM[1] {~~~~~~\textrm{#1}~~~~~~}

\newcommand\At[1] {^{@#1}}
\newcommand\Att[2] {^{@\frac{#1}{#2}}}
\newcommand\IIFrac[2] {\mbox{$\frac{#1}{#2}$}} 
\newcommand\IFrac[2] {\mbox{\small$#1/#2$}} 

\newenvironment{NumMini}[1]{\begin{equation}\label{#1}\begin{minipage}{.8\linewidth}}{\end{minipage}\end{equation}}
\newenvironment{Reason}{\vspace{-.0em}\begin{tabbing}\hspace{2em}\= \hspace{1cm} \= \kill}{\end{tabbing}\vspace{-1em}}
\newcommand\Step[2] {#1 \> $\begin{array}[t]{@{}llll}#2\end{array}$ \\}
\newcommand\StepR[3] {#1 \> $\begin{array}[t]{@{}llll}#3\end{array}$ \` {\RF \makebox[0pt][r]{\begin{tabular}[t]{r}``#2''\end{tabular}}} \\}
\newcommand\WideStepR[3] {#1 \> $\begin{array}[t]{@{}ll}~\\#3\end{array}$ \` {\RF \makebox[0pt][r]{\begin{tabular}[t]{r}``#2''\end{tabular}}} \\} 
\newcommand\Space {~ \\}
\newcommand\RF {\small}
\newenvironment{EQN}[1]{\def\z{#1}\medskip\par\noindent\hfill}{\hfill\makebox[0pt][r]{(\z)}\medskip\par\noindent}
\newcommand\ImageInText[4]%
{\makebox[0pt][l]{%
{\def\Up{#1}\def\Right{#2}
\hfill%
\raisebox{\Up}[0pt][0pt]{
\makebox[0pt][l]{\hspace{\Right}
\includegraphics[scale=#3]{#4}%
}}}}}
\newcommand\Bag[3]{
 \bgroup\def\Bound{#1}\{\ifx\Bound\empty\egroup #3 \else\egroup #1
  \bgroup\def\Cond{#2}\ifx\Cond\empty\egroup\else\egroup\mathrel{\mid}#2\fi
  \bgroup\def\Elem{#3}\ifx\Elem\empty\egroup\else\egroup\mathrel{\Spot}#3\fi
 \fi\}
}
\newcommand\Set[3]{
 \bgroup\def\Bound{#1}\{\ifx\Bound\empty\egroup #3 \else\egroup #1
  \bgroup\def\Cond{#2}\ifx\Cond\empty\egroup\else\egroup\mathrel{\mid}#2\fi
  \bgroup\def\Elem{#3}\ifx\Elem\empty\egroup\else\egroup\mathrel{\Spot}#3\fi
 \fi\}
}
\newcommand\SetBig[3]{
  \bgroup\def\Cond{#2}\ifx\Cond\empty\egroup
   \begin{array}[t]{l}
	\{#1 \mathrel{\Spot}\\
    \hspace{3em}#3\}
   \end{array}
  \else\egroup
   \bgroup\def\Elem{#3}\ifx\Elem\empty\egroup
    \begin{array}[t]{l}
     \{#1 \mathrel{\mid} \\
     \hspace{3em}#2\}
   \end{array}
   \else\egroup
    \begin{array}[t]{l}
     \{#1 \\
     \hspace{3em} \mathrel{\mid}#2 \\
     \hspace{3em} \mathrel{\Spot}#3\}
    \end{array}
   \fi
  \fi
}
\newcommand\PSet[3]{
 \bgroup\def\Bound{#1}\LeftPS \ifx\Bound\empty\egroup #3 \else\egroup #1
  \bgroup\def\Cond{#2}\ifx\Cond\empty\egroup\else\egroup\mathrel{\mid}#2\fi
  \bgroup\def\Elem{#3}\ifx\Elem\empty\egroup\else\egroup\mathrel{\Spot}#3\fi
 \fi \RightPS
}
\newcommand\PSetBig[3]{
  \bgroup\def\Cond{#2}\ifx\Cond\empty\egroup
   \begin{array}[t]{l}
	\LeftPS#1 \mathrel{\Spot}\\
    \hspace{3em}#3\RightPS
   \end{array}
  \else\egroup
   \bgroup\def\Elem{#3}\ifx\Elem\empty\egroup
    \begin{array}[t]{l}
     \LeftPS#1 \mathrel{\mid} \\
     \hspace{3em}#2\RightPS
   \end{array}
   \else\egroup
    \begin{array}[t]{l}
     \LeftPS#1 \\
     \hspace{3em} \mathrel{\mid}#2 \\
     \hspace{3em} \mathrel{\Spot}#3\RightPS
    \end{array}
   \fi
  \fi
}
\newcommand\Exists[3]{
 (\begin{array}[c]{l}
  \exists#1
  \bgroup\def\Cond{#2}\ifx\Cond\empty\egroup\else\egroup\mathrel{\mid}#2\fi
  \bgroup\def\Elem{#3}\ifx\Elem\empty\egroup\else\egroup\mathrel{\Spot}#3\fi
 \end{array})
}
\newcommand\ForAll[3]{
 (\begin{array}[c]{l}
  \forall#1
  \bgroup\def\Cond{#2}\ifx\Cond\empty\egroup\else\egroup\mathrel{\mid}#2\fi
  \bgroup\def\Elem{#3}\ifx\Elem\empty\egroup\else\egroup\mathrel{\Spot}#3\fi
 \end{array})
}
\newcommand\Union[3]{
(\begin{array}[c]{l}
  \bigcup#1
  \bgroup\def\Cond{#2}\ifx\Cond\empty\egroup\else\egroup\mathrel{\mid}#2\fi
  \bgroup\def\Elem{#3}\ifx\Elem\empty\egroup\else\egroup\mathrel{\Spot}#3\fi
 \end{array})
}
\newcommand\Sum[3]{
 \begin{array}[c]{l}
  (\sum#1
  \bgroup\def\Cond{#2}\ifx\Cond\empty\egroup\else\egroup\mathrel{\mid}#2\fi
  \bgroup\def\Elem{#3}\ifx\Elem\empty\egroup\else\egroup\mathrel{\Spot}#3\fi )
 \end{array}
}

\newcommand{\EXP}{\mathop{\sum\kern-.45cm\sum}}
\renewcommand{\EXP}{+}
\renewcommand{\EXP}{{\cal E}}
\renewcommand{\EXP}{\otimes}
\renewcommand{\EXP}{\odot}
\newcommand\Exp[2]{(\kern-.25ex\EXP\,#1\bgroup\def\Elem{#2}\ifx\Elem\empty\egroup\else\egroup\mathrel{\Spot}#2\fi)}
\newcommand\ExpNorm[2]{[\kern-.1ex\EXP\,#1\bgroup\def\Elem{#2}\ifx\Elem\empty\egroup\else\egroup\mathrel{\Spot}#2\fi]}
\newcommand\CExp[3]{
 (\begin{array}[c]{l}
  \EXP#1
  \bgroup\def\Cond{#2}\ifx\Cond\empty\egroup\else\egroup\mathrel{\mid}#2\fi
  \bgroup\def\Elem{#3}\ifx\Elem\empty\egroup\else\egroup\mathrel{\Spot}#3\fi
 \end{array})
}
\newcommand\General[4]{
 \begin{array}[c]{l}
  (#1#2
  \bgroup\def\Cond{#3}\ifx\Cond\empty\egroup\else\egroup\mathrel{\mid}#3\fi
  \bgroup\def\Elem{#4}\ifx\Elem\empty\egroup\else\egroup\mathrel{\Spot}#4\fi)
 \end{array}
}
\newenvironment{Figure}[2][t]{\begin{figure}[#1]\def\Label{#2}\small}{\label{\Label}\end{figure}}
\newenvironment{Theorem}[2]{\begin{theorem}\label{#2}\textit{#1}\rm\quad}{\hfill$\Box$\end{theorem}}
\newenvironment{Lemma}[2]{\begin{lemma}\label{#2}\textit{#1}\rm\quad}{\hfill$\Box$\end{lemma}}
\newenvironment{Definition}[2]{\begin{definition}\label{#2}\textit{#1}\rm\quad}{\hfill$\Box$\end{definition}}

\newcommand\Itm[1] {(\ref{#1})}
\newcommand\Thm[1] {Thm.~\ref{#1}}
\newcommand\Lem[1] {Lem.~\ref{#1}}
\newcommand\Def[1] {Def.~\ref{#1}}
\newcommand\Fig[2][*] {{\def\z{#1}\if*\z Fig.~\ref{#2}\else Fig.~\ref{#2}(#1)\fi}}

\begin{document}
\title{Compositional closure \\
for Bayes Risk \\
in probabilistic noninterference}
\author{Annabelle McIver\inst{1} \and Larissa Meinicke\inst{1} \and Carroll Morgan\inst{2}%
\thanks{We acknowledge the support of the Australian Research Council Grant {DP0879529}.}}
\institute{Dept.~Computer Science, Macquarie University, NSW 2109 Australia
\and School of Comp.\ Sci.\ and Eng., Univ.~New South Wales, NSW 2052 Australia
}
\maketitle

\begin{abstract}
We give a sequential model for noninterference security including probability (but not demonic choice), thus supporting reasoning about the \emph{likelihood} that high-security values might be revealed by observations of low-security activity. Our novel methodological contribution is the definition of a \emph{refinement} order $(\Ref)$ and its use to \emph{compare} security measures between specifications and (their supposed) implementations. This contrasts with the more common practice of evaluating the security of individual programs in isolation.

\medskip
The appropriateness of our model and order is supported by our showing that $(\Ref)$ is the \emph{greatest} compositional relation --the compositional closure-- with respect to our semantics and an ``elementary'' order based on Bayes Risk --- a security measure already in widespread use. We also relate refinement to other measures such as Shannon Entropy.

\medskip
By applying the approach to a non-trivial example, the anonymous-majority \emph{Three-Judges} protocol, we demonstrate by example that correctness arguments can be simplified by the sort of layered developments --through levels of increasing detail-- that are allowed and encouraged by compositional semantics.
\end{abstract}

\tableofcontents

\newpage
\section{Introduction}
We apply notions of testing equivalence and refinement, based on \emph{Bayes Risk}, to the topic of \emph{noninterference security}~\cite{Goguen:84} \emph{with probability} but without demonic choice. Previously, we have studied noninterference for demonic systems without probabilistic choice \cite{Morgan:06,Morgan:07}, and we have studied probability and demonic choice without noninterfence \cite{Morgan:96d,McIver:05a}. Here thus we are completing a programme of treating these features ``pairwise.''

Our long-term aim --as we explain in the conclusion-- is to treat all three features together, based on the lessons we have learned by treating strict subsets of them. The benefit (should we succeed) would apply not only to security, but also to conventional program development where, in the presence of both probabilistic and demonic choice, the technique of data-transformation (aka.\ data \emph{refinement} or data \emph{reification}) becomes unexpectedly complex: variables inside local scopes must be treated analogously to ``high security'' variables in noninterference security.

We take the view, learned from others, that program/system development benefits from a comparison of specification programs with (putative) implementations of them, wherever this is possible, via a mathematically defined ``refinement'' relation whose formulation depends ultimately on a notion of testing that is agreed-to subjectively by all parties concerned \cite{Nicola:84}.\,%
\footnote{We say ``wherever this is possible'' since there are many aspects of system development that cannot be pinned down mathematically. But --we argue-- those that can be, should be.}
To explain our position unambiguously, we begin by recalling the well known effects of this approach for conventional, sequential programming.

\subsection{Elementary testing and refinement for conventional programs}\label{s1309}
Consider sequential programs operating over a state-space of named variables with fixed types, including a program \Abort\ that diverges (such as an infinite loop). We allow demonic nondeterminsm, statements such as $\Vx\Gets0\Or\Vx\Gets1$, in the now-conventional way in which they represent equally abstraction (we do not care whether $\Vx$ is assigned 0 or 1, as long as it is one or the other), on the one hand, or unpredictable and arbitrary run-time choice on the other.

Having determined a ``specification'' program $S$, we address the question of whether we are prepared to accept some program $I$ that purports to ``implement'' it. Although there is nowadays a widely accepted answer to this, we imagine that we are considering the question for the first time and that we are hoping to find an answer that everybody will accept. For that we search for a test on programs that is ``elementary'' in the sense that it is conceptually simple and that no ``reasonable'' person could ever argue that $S$ is implemented by $I$ if it is the case that $S$ always passes the test but $I$ might fail it.\,%
\footnote{There is a possibly dichotomy here between ``may testing'' and ``must testing,'' and we are taking the latter in this example: if $S$ must pass a certain test, then so must $I$ if it is to be considered an implementation.}

A common choice for such an \emph{elementary test} is ``can diverge,'' where divergence is considered to be a bad thing: using it, our criterion becomes ``if $I$ indeed implements $S$ and $I$ can diverge, then it must be possible for $S$ to diverge also.'' We note that the elementary test cannot be objectively justified: it is an ``axiom'' of the approach that will be built on it; and it is via the subjective axioms (in any approach) that we touch reality, where we avoid an infinite definitional regress.

The elementary test provides an ``only if'' answer to the implementation question, but not an ``if.'' That is, we do not say that $I$ implements $S$ \emph{if} either $I$ never fails the test or $S$ might fail it: this is not practical, because of context. For an example, let $S$ be $\If~\Vx{\neq}0~\Then~\Abort~\Fi$ and let $I$ be simply $\Abort$. Then indeed $S$ passes the test if $I$ does (because they both fail); but we cannot accept $I$ generally as a replacement for $S$ because context $\Vx\Gets0;S$ ``protects'' $S$, and passes the test as a whole; but the same context does not protect $I$, since $\Vx\Gets0;I$ (still) fails. This illustrates the inutility of the elementary view taken on its own, and it shows that we need a more sophisticated comparison in order to have a practical tool that respects contexts. (Thus it is clear above that we must add ``if executed from the same initial state.'') The story leads on from here to a definition, ultimately, of sequential-program refinement $(\Ref)$ as the unique relation such that\,%
\footnote{We say ``a'' rather thean ``the'' definition of refinement because this is just an example: other elementary tests, and other possible contexts, lead naturally to other definitions.}
\begin{enumerate}[(i)]
\item\label{i1253}\makebox[6em][l]{\textit{soundness}} If $S{\Ref}I$ then for all contexts $\CC$ we have that $\CC(I)$ passes the elementary test if $\CC(S)$ does, and
\item\label{i1254}\makebox[6em][l]{\textit{completeness}} If $S{\not\Ref}I$ then there is some context $\CC$ such that $\CC(I)$ fails the elementary test although $\CC(S)$ passes it.
\end{enumerate}
That relation turns out to have the \emph{direct} definition that $S{\Ref}I$ just when, for all initial states $s$,
if executing $I$ from $s$ can deliver some final state $s'$ then --from $s$ again-- either $S$ can deliver $s'$, as well, or $S$ can diverge. Crucially, it is the direct definition that allows $(\Ref)$ to be determined without examining all possible contexts.

\subsection{Elementary testing and refinement for probabilistic noninterference-secure programs}\label{s1028}
In attempting to follow the trajectory of \Sec{s1309} into the modern context of noninterference and probability, we immediately run into the problem that there are competing notions of elementary test. Here are just four of them:
\begin{description}
\item[Bayes Risk] \cite{Smith:07,Chatzikokolakis:07b,Braun:08,Braun:09} is based on the probability an attacker can reveal a high-security, ``hidden'' variable $\Vh$ using a single query ``Is $\Vh$ equal to $h$?''\ where $h$ is some value in $\Vh$'s type. Here (and below) the elementary testing of $S$ wrt.\ $I$ requires that the probability of revealing $\Vh$ in $I$ cannot be higher than it is in $S$.
\item[marginal guesswork] \cite{Pliam:00,Kopf:07} is measured in terms of how many queries of the form ``Is $\Vh$ equal to $h$?''\ are needed to determine $\Vh$'s value with a given probability.
\item[Shannon Entropy] \cite{Shannon:48} is related to the use of multiple queries of the form ``Is $\Vh$ in some set $H$?'' where $H$ is a subset of $\Vh$'s type.
\item[guessing entropy]~\cite{Massey:94,Kopf:07} is the average number of ``Is $\Vh$ equal to $h$?''\ guesses necessary to determine $\Vh$'s value.
\end{description}
Not only do these criteria compete for popularity, it turns out that on their own they are not even objectively comparable. For instance,  Pliam~\cite{Pliam:00} finds that there can be no general ordering between marginal guesswork and Shannon Entropy: that is, from a marginal-guesswork judgement of whether $S$ passes all tests that $I$ does, there is no way to determine whether the same would hold for Shannon-entropy judgements, nor vice versa.
Similarly, Smith has compared Bayes Risk and Shannon Entropy, and claims that these measures are inconsistent in the same sense~\cite{Smith:07}. The general view seems to be that none of these (four) methods can be said to be generally more- or less discriminating than any of the others.

In spite of the above, \emph{one of our contributions here} is to show that Bayes Risk is maximally discriminating among those four if context is taken into account.

\subsection{Features of our approach: a summary}

Our most significant deviation from traditional noninterference is that, rather than calculating security measures of programs in isolation, instead we focus on comparing security measures \emph{between} programs: typically one is supposed to be a specification, and another is supposed to be an implementation of it. What we are looking for is an implementation that is at least as secure as its specification. 

Since we never consider the security of programs in isolation, an advantage is that it is possible easily to arrange certain kinds of permissible information flow. For example whenever $s{\geq}i$ holds, a program $I$ that leaks only the $i$ low-order bits of a hidden integer $\Vh$ is secure with respect to a specification $S$ that leaks the $s$ low-order bits of $\Vh$ --- that is, for any implementation of $S$, the leaking of up to $s$ low-order bits of $\Vh$ is allowed but no more. This way we sometimes can avoid separate tools for declassification: to allow an implementation to release (partial) information, we simply arrange that its specification does so.

Typically it is both functional- and security properties (however we measure them) that are of interest. As such, we would like to define a relation $(\Ref)$ between these programs so that $S{\Ref}I$ just when \emph{implementation} $I$ has all the functional and the security properties that \emph{specification} $S$ does, where ``all'' is interpreted within our terms of reference. For incremental, compositional reasoning with such an order, it has been known from the very beginning~\cite{Wirth:71} that the \emph{refinement} relation $(\Ref)$ must satisfy two key technical properties:
\begin{description}
\item[Transitivity] If $S{\Ref}M{\Ref}I$ then also $S{\Ref}I$. Because of this a comparison between two large programs $S,I$ can be carried out via $S\Ref M_1\Ref\cdots\Ref M_N\Ref I$ through many small steps over a long time.
\item[Monotonicity of contexts] If $S\Ref I$ then also $\CC(S)\Ref\CC(I)$, where $\CC$ is any program context. Because of this, a large comparison can be carried out via many small steps \emph{independently} by a large programming team working in parallel. 
\end{description}
As argued above, since our comparisons rest ultimately on subjective criteria for failure, we reduce that dependency on what is essentially an arbitrary choice by making those criteria as elementary as possible: when can you be \emph{absolutely sure} that $S{\NotRef}I$, that refinement should fail? For this purpose we identify an elementary testing relation $(\ERef)$ based on Bayes Risk, such that if $S{\NotERef}I$ then $I$ ``certainly'' (but still subjectively) does not satisfy the specification $S$ in terms of ``reasonable'' functional- and probabilistically secure properties. 

Because our $(\ERef)$ is not respected by all contexts (there exist programs $S,I$ and context $\CC$ such that $S{\ERef}I$, yet $\CC(S){\NotERef}\CC(I)$ in spite of that) our relation $(\Ref)$ is chosen so that it is smaller --i.e. more restrictive-- than $(\ERef)$, so that it excludes just those ``apparent'' refinements that can be voided by context.

Our refinement relation is the \emph{compositional closure} of $(\ERef)$, the largest relation $(\Ref)$ such that $S {\Ref}I$ implies $\CC(S){\ERef}\CC(I)$ for all possible contexts $\CC$. Abusing terminology slightly, we will for simplicity say that $(\Ref)$ is \emph{compositional} just when it is respected by all possible contexts $\CC$ (whereas strictly speaking we should say that all such $\CC$'s are $(\Ref)$-monotonic). Further, we note that if we define equivalence $A{\sim}B$ to be ``bi-refinement'' $A{\Ref}B$ and $B{\Ref}A$ then monotonicity of $(\Ref)$ implies that $(\sim)$ is is a congruence for all contexts $\CC$.

There are two further, smaller idiosyncracies of our approach. The first is that we allow the high-security, ``hidden'' variables to be assigned-to by the program, so that it is the secrecy of the \emph{final} value $h'$ of $\Vh$ that is of concern to us, not the initial value $h$. This is because we could not otherwise meaningfully compare functional properties, nor would we be able to treat (sequential) compositional contexts. The other difference, more a position we take, is that we allow an attacker both \emph{perfect recall} and an awareness of \emph{implicit flow}: that the intermediate values of low-security ``visible'' program variables are observable, even if subsequently overwritten; and that the control-flow of non-atomic program statements is observable. 
As shown in our case study (\Sec{s1324}) it is this which allows us to model distributed applications: there, the values of intermediate variables can be observed (and recalled) if they are sent on an insecure channel, and the control flow of a program may be witnessed (for example) by observing which 
request an agent is instructed to fulfill.

\smallskip
In summary, our \underline{\textsc{technical contribution}} is that we (i) give a sequential semantics for probabilistic noninterference, (ii) define the above order $(\ERef)$ based on Bayes Risk, (iii) show it is \emph{not} compositional, (iv) identify a \textbf{compositional} subset of it, a \emph{refinement} order $(\Ref)$ such that $S{\Ref}I$ implies $\CC(S){\ERef}\CC(I)$ for all contexts $\CC$ and (v) show that $({\Ref})$ is in fact the compositional \textbf{closure} of $(\ERef)$, so that in fact we have $S{\NotRef}I$ \emph{only} when $\CC(S){\NotERef}\CC(I)$ for some $\CC$.

Finally, we note (vi) that $(\Ref)$ is sound for the other three, competing notions of elementary test and that therefore Bayes-Risk testing, with context, is maximally discriminating among them.

These technical contributions further our general goal of structuring secure protocols hierarchically and then designing/verifying them in separate pieces, a claim that we illustrate by showing how our model and our secure-program ordering may be used to give an incremental development of \emph{The Three Judges}, an ``anonymous majority'' protocol we constructed precisely to make this point.

\section{A probabilistic, noninterference sequential semantics}\label{s1121}
We identify \emph{visible} variables (low-security), typically $\Vv$ in  some finite type $\VV$, and \emph{hidden} variables (high-security), typically $\Vh$ in finite $\HH$. Variables are in \textsf{sans serif} to distinguish them from (decorated) values $v\In\VV,h\In\HH$ they might contain.\,%
\footnote{We say hidden and visible, rather than high- and low security, because of the connection with data refinement where the same technical issues occur but there are no security implications.}

As an example, let hidden $\Vh\In\Set{}{}{0,1,2}$ represent one of three boxes: Box 0 has two black balls; Box 1 has one black- and one white ball; and Box 2 has two white balls. Then let $\Vv\In\Set{}{}{w,b,\bot}$ represent a ball colour: \emph{white}, \emph{black} or \emph{unknown}. Our first experiment in this system is Program $S$, informally written $\Vh\Gets 0{\PC{}}1{\PC{}}2; \Vv\From \PSet{}{}{w\Att{\Vh}{2},b\At{1{-}\frac{\Vh}{2}}}; \Vv\Gets \bot$, that chooses box $\Vh$ uniformly, and then draws a ball $\Vv$ from that Box~$\Vh$: from the description above (and the code) we can see that with probability  $\Vh/2$ the ball is white, and with probability $1{-}\Vh/2$ it is black. Then the ball is replaced. A typical security concern is ``How much information about $\Vh$ is revealed by its assignments to $\Vv$?''

We use this program, and that question, to motivate our program syntax and semantics, to make Program $S$ the above program precise and to provide the framework for asking --and answering-- such security questions.

We begin by introducing \emph{distribution} notation, generalising the notations for na{\"i}ve set theory.

\subsection{Distributions: explicit, implicit and expected values over them}\label{s1729}

We write function application as $f.x$, with ``$.$'' associating to the left. Operators without their operands are written between parentheses, as $(\ERef)$ for example. 
\emph{Set} comprehensions are written as $\Set{s\In S}{G}{E}$ meaning the set formed by instantiating bound variable $s$ in the expression $E$ over those elements of $S$ satisfying formula $G$.\,%
\footnote{This is a different order from the usual notation $\{E \mid s{\in}S \land G\}$, but we have good reasons for using it: calculations involving both sets and quantifications are made more reliable by a careful treatment of bound variables and by arranging that the order $S/G/E$ is the same in both comprehensions and quantifications (as in $\ForAll{s\In S}{G}{E}$ and $\Exists{s\In S}{G}{E}$).}

By $\Dist{S}$ we mean the set of \emph{discrete sub-distributions} on set $S$ that sum to no more than one, and $\TDist{S}$ means the \emph{full} distributions that sum to one exactly. The \emph{support} $\Supp{\delta}$ of (sub-)distribution $\delta \In \Dist{S}$ is those elements $s$ in $S$ with $\delta.s{\neq}0$, and the \emph{weight} $\sum\delta$ of a distribution is $\Sum{s\In\Supp{\delta}}{}{\delta.s}$, so that full distributions have weight 1. Distributions can be scaled and summed according to the usual pointwise extension of arithmetic to real-valued functions, so that $(c{*}\delta).s$ is $c{*}(\delta.s)$ for example; the \emph{normalisation} of a (sub-)distribution $\delta$ is defined $\Norm{\delta}\Defs\delta/\sum\delta$.

Here are our notations for \emph{explicit} distributions (cf.\ set enumerations):
\begin{description}
\item[multiple] We write $\PSet{}{}{x\At{p}, y\At{q},\cdots, z\At{r}}$ for the distribution assigning probabilities $p,q,\cdots,r$ to elements $x,y,\cdots,z$ respectively, with $p{+}q{+}\cdots{+}r\leq1$.
\item[uniform] When explicit probabilities are omitted they are uniform: thus $\PSet{}{}{x}$ is the point distribution $\PSet{}{}{x\At{1}}$, and $\PSet{}{}{x,y,z}$ is $\PSet{}{}{x\Att{1}{3},y\Att{1}{3},z\Att{1}{3}}$. And $\delta_1{\PC{}}\delta_2$ is $\delta_1{\PC{\frac{1}{2}}}\delta_2$.
\end{description}

In general, we write $\Exp{d\In\delta}{E}$ for the \emph{expected value} $\Sum{d\In\Supp{\delta}}{}{\delta.d*E}$ of expression $E$ interpreted as a random variable in $d$ over distribution $\delta$.\footnote{It is a dot-product between the distribution and the random variable as state-vectors.}
If however $E$ is Boolean, then it is taken to be 1 if $E$ holds and 0 otherwise: thus in that case $\Exp{d\In\delta}{E}$ is the combined probability in $\delta$ of all elements $d$ that satisfy $E$.

We write \emph{implicit} distributions (cf.\ set comprehensions) as $\PSet{d\In\delta}{R}{E}$, for distribution $\delta$, real expression $R$ and expression $E$, meaning
\begin{equation}\label{e1221}
\Exp{d\In\delta}{R * \Point{E}} ~/~ \Exp{d\In\delta}{R}
\end{equation}
where, first, an expected value is formed in the numerator by scaling and adding point-distribution $\PSet{}{}{E}$ as a real-valued function: this gives another distribution. The scalar denominator then normalises to give a distribution yet again. A missing $E$ is implicitly $d$ itself. If $R$ is missing, however, then $\PSet{d\In\delta}{}{E}$ is just $\Exp{d\In\delta}{\Point{E}}$ --- in that case we do not multiply by $R$ in the numerator, nor do we divide (by anything).

Thus $\PSet{d\In\delta}{}{E}$ \emph{maps} expression $E$ in $d$ over distribution $\delta$ to make a new distribution on $E$'s type.  When $R$ is present, and Boolean, it is converted to 0,1; thus in that case $\PSet{d\In\delta}{R\,}{}$ is $\delta$'s \emph{conditioning} over formula $R$ as predicate on $d$.

Finally, for \emph{Bayesian belief revision} we let $\delta$ be an a-priori distribution over some $D$, and we let expression $R$ for each $d$ in $D$ be the probability of a certain subsequent result if that $d$ is chosen. Then $\PSet{d\In\delta}{R}{}$ is the a-posteriori distribution over $D$ when that result actually occurs. Thus in the three-box program $S$ let the value first assigned to $\Vv$ be $\hat{v}$. The a-priori distribution over $\Vh$ is uniform, and the probability that the chosen ball is white, that $\hat{v}{=}w$, is therefore $1/3*(0/2+1/2+2/2) = 1/2$. But the a-posteriori distribution of $\Vh$ \emph{given that $\hat{v}{=}w$} is $\PSet{h\In \delta}{h/2}{}$, which from \Eqn{e1221} we can evaluate
{\small\[
\newcommand{\Dii}[1]{\mbox{\Large{$\frac{#1}{2}$}}}
 =~~~
 \Exp{h\In\PSet{}{}{0,1,2}}{\Dii{h} * \Point{h}} ~/~ \Exp{h\In\PSet{}{}{0,1,2}}{\Dii{h}}
 \Wide{=}
 \PSet{}{}{1\Att{1}{6},2\Att{1}{3}} ~/~ \Dii{1} ~,
\]}
that is $\PSet{}{}{1\Att{1}{3},2\Att{2}{3}}$, to calculate our way to the conclusion that if a white ball is drawn ($\hat{v}{=}w$) then the chance it came from Box 2 is $2/3$, the probability of $\Vh{=}2$ in the a-posteriori distribution.

\subsection{Program denotations over a visible/hidden ``split'' state-space}
We account for the \emph{visible} and \emph{hidden} partitioning of the finite state space $\VV{\times}\HH$ in our new model by building \emph{split-states} of type $\VV{\times}\TDist{\HH}$, whose typical element $(v,\delta)$ indicates that we know $\Vv{=}v$ exactly, but that all we know about $\Vh$ --which is not directly observable-- is that it takes value $h$ with probability $\delta.h$. 

Programs become functions $(\VV{\times}\TDist{\HH}) \Fun \TDist{(\VV{\times}\TDist{\HH})}$ from split-states to distributions over them, called \emph{hyper-distributions} since they are distributions with other distributions inside them: the outer distribution is directly visible but the inner distribution(s) over $\HH$ are not. Thus for a program $P$ with semantics $\Sem{P}$, the application $\Sem{P}.(v,\delta)$ is the distribution of final split-states produced from initial $(v,\delta)$. Each $(v',\delta')$ in the support of that outcome, with probability $p$ say in the outer- (left-hand) $\TDist$ in $\TDist{(\VV{\times}\TDist{\HH})}$, means that with probability $p$ an attacker will observe that $\Vv$ is $v'$ and simultaneously will be able to deduce (via the explicit observation of $v$ and $v'$ and other implicit observations) that $\Vh$ has distribution $\delta'$.

When applied to hyper-distributions, addition, scaling and probabilistic choice ($\PC{p}$) are to be interpreted as operations on the outer distributions (as explained in \Sec{s1729}).

\subsection{Program syntax and semantics}\label{s47563}

The programming language semantics is given in \Fig{f1228}. In this presentation we do not treat loops and, therefore, all our programs are terminating. 

When we refer to \emph{classical} semantics, we mean the interpretation of a program without distinguishing its visible and hidden variables, thus as a ``relation'' of type $(\VV{\times}\HH)\Fun\TDist(\VV{\times}\HH)$.\,%
\footnote{Classical relational and \emph{non-probabilistic} semantics over a state-space $\VV{\times}\HH$ is strictly speaking $(\VV{\times}\HH){\Rel}(\VV{\times}\HH)$ or equivalently $\Pow((\VV{\times}\HH)^2)$. Further formulations include however both $(\VV{\times}\HH){\Fun}\Pow(\VV{\times}\HH)$ and $\VV{\Fun}\HH{\Fun}\Pow(\VV{\times}\HH)$. Because all these are essentially the same, we call $(\VV{\times}\HH){\Fun}\TDist(\VV{\times}\HH)$ a ``relational'' semantics.}
\begin{Figure}[ht!]{f1228} 
\begin{displaymath}
 \begin{array}{p{10em}@{~}l@{~}lr}
  Program type &\HangRight{Program text $P$} &\multicolumn{1}{r}{\textrm{Semantics $\Sem{P\,}.(v,\delta)$}} \\ \hline \\
  Identity & \Skip & \Point{~(v, \delta)~}  
  & \HangRight{$~\star$} \\
  Assign to visible & \Vv\Gets E.\Vv.\Vh & 
   \PSet{~h\In\delta}{}{(E.v.h, \PSet{h'\In\delta}{E.v.h'{=}E.v.h}{})~}  
   & \HangRight{$~\star$} \\
  Assign to hidden & \Vh\Gets E.\Vv.\Vh & 
  \Point{~(v, \PSet{h\In\delta}{}{E.v.h})~} 
  & \HangRight{$~\star$} \\
  Choose prob.\ visible & \Vv\PGets D.\Vv.\Vh &  
   \PSet{~v'\In \Exp{h\In \delta}{D.v.h}}{}{(v', \PSet{h'\In\delta}{D.v.h'.v'}{})~} 
  & \HangRight{$~\star$} \\
  Choose prob.\ hidden & \Vh\PGets D.\Vv.\Vh & 
   \Point{~(v, \Exp{h\In \delta}{D.v.h})~} 
   & \HangRight{$~\star$} \\[1.5ex]
  Composition & P_1; P_2 & 
   \Exp{(v',\delta')\In\Sem{P_1}.(v,\delta)}{~\Sem{P_2}.(v',\delta')} \\  
  General prob.\ choice & P_1 \PC{q.\Vv.\Vh} P_2
                     & \hspace{3em} p*\Sem{P_1}.(v,\PSet{h\In\delta}{q.v.h}{})
                     \hspace{1em}\textrm{\tiny $p$ is $\Exp{h\In\delta}{q.v.h}$} \\
                     & 
                     & \makebox[3em][r]{$+$~} 
                                   (1{-}p)*\Sem{P_2}.(v,\PSet{h\In\delta}{1{-}q.v.h}{}) 
                    \\[1.5ex]
  Probabilistic choice & P_1 \PC{p} P_2 & 
   p*\Sem{P_1}.(v,\delta) + (1{-}p)*\Sem{P_2}.(v,\delta)
   \hspace{1em}\textrm{\tiny $p$ is constant} \\[1.5ex]
  Conditional choice & \makebox[0pt][l]{$\If~ G.\Vv.\Vh ~\Then~ P_t$} 
                     & \hspace{3em} p*\Sem{P_t}.(v,\PSet{h\In\delta}{G.v.h}{})
                     \hspace{1em}\textrm{\tiny $p$ is $\Exp{h\In\delta}{G.v.h}$} \\
                     & \makebox[0pt][l]{$\Else~ P_f ~\Fi$}
                     & \makebox[3em][r]{$+$~} (1{-}p)*\Sem{P_f}.(v,\PSet{h\In\delta}{\neg G.v.h}{})
\end{array}
\end{displaymath}
For simplicity let $\VV$ and $\HH$ have the same type $\cal X$. Expression $E.\Vv.\Vh$ is then of type ${\cal X}$, distribution $D.\Vv.\Vh$ is of type $\TDist{\cal X}$ and expression $G.\Vv.\Vh$ is Boolean. Expressions $p$ and $q.\Vv.\Vh$ are of type $[0,1]$.

\medskip
The syntactically atomic commands marked $\star$ have semantics calculated by taking the classical meaning and then applying \Def{d0855}. The third column for $\star$'d commands is the result of doing that.

\medskip
Further, the Assign-to semantics are special cases of the Choose-prob.\ semantics, obtained by making the distribution $D$ equal to the point distribution $\Point{E}$. And the (simple) probabilistic choice is a special case of the general prob.\ choice, taking $q.v.h$ to be the constant function always returning $p$. Finally, conditional choice is the special case of general prob.\ choice obtained by taking $q.v.h$ to be 1 when $G.v.h$ holds and 0 otherwise.

\medskip

For distributions in \emph{program texts} we allow the more familiar infix notation $\PC{p}$, so that we can write $\Vh\Gets 0{\PC{\frac{1}{3}}}1$ for $\Vh\From\PSet{}{}{0\Att{1}{3},1\Att{2}{3}}$ and $\Vh\Gets 0{\PC{}}1$ for the uniform $\Vh\From\PSet{}{}{0,1}$. The degenerate cases $\Vh\Gets0$ and $h\From\PSet{}{}{0}$ are then equivalent, as they should be.

\caption{Split-state semantics of commands}
\end{Figure}

\subsubsection{Atomic commands}

Syntactically \emph{atomic} program (fragments), noted $\star$ in \Fig{f1228}, are first interpreted with respect to their classical probabilistic semantics, and are then embedded into the split-state model. To emphasise that they are syntaxtically atomic, we call them ``$A$'' (rather than ``$P$'') in this section.

\newcommand\VProj {\textsf{vProj}}
\newcommand\VCond {\textsf{vCond}}

Thus the first step is to interpret an atomic program $A$ as a function from $\VV{\times}\HH$ -pairs to distributions $\TDist(\VV{\times}\HH)$ of them \cite{Kozen:85,McIver:05a} --- call that classical interpretation $\SemC{A}$ so that for an initial $(v,h)$ program $A$ produces a final distribution $\SemC{A}.(v,h)$, that is some distribution $\delta'{\in}\TDist(\VV{\times}\HH)$.

Given such a distribution $\delta'$, define its $\Vv$-projection $\VProj.\delta'$ to be given by $\PSet{(v,h)\In\delta'}{}{v}$, that is the distribution over $\VV$, alone, that $\delta'$ defines if we ignore (and aggregate) the $h$-components for each distinct $v$.

Then define for $\delta'$ its $v'$-conditioning $\VCond.\delta'.v'$, that is the distribution $\PSet{(v,h)\In\delta'}{v{=}v'}{h}$ over $\HH$ that we get by concentrating on a particular value $v'$.

With these two preliminaries, the distribution over $\VV{\times}\TDist\HH$ we get by interpreting $\delta'$ atomically is defined
\[
 \Hide.\delta' \Wide{\Defs} \PSet{v'\In\VProj.\delta'~}{}{~(v',\VCond.\delta'.v')} ~,
\]
which is in essence just the ``grouping together'' of all elements $(v',h')$ in $\delta'$ that have the same $v'$.

There are two routine steps left to finish off the embedding of whole programs; and they are given here in \Def{d0855}:

\begin{Definition}{Induced secure semantics for atomic programs}{d0855}
Given a \emph{syntactically atomic} program $A$ we define its \emph{induced secure semantics} $\Sem{A}$ via
\begin{equation}\label{e1028}
 \Sem{A}.(v,\delta) \Wide{\Defs} \Hide.\Exp{h\In \delta~}{~\SemC{A}.(v,h)}~.
\end{equation}
Thus $A$ is applied to the incoming distribution $(v,\delta)$ by applying its classical meaning $\SemC{A}$ to each $(v,h)$-pair separately, noting that pair's implied weight, and then using those weights to combine the resulting $(v',h')$-distributions into a single distribution $\delta'$ of type $\TDist(\VV{\times}\HH)$. That distribution $\delta'$ is then embedded into the split-state model as above.

The effect overall is that an embedding imposes the largest possible ignorance of $h'$ that is consistent with seeing $v'$ and knowing the classical semantics $\SemC{A}$.
\end{Definition}

We illustrate the definitions in \Fig{f1228} by looking at some simple examples.

Program $\Skip$ modifies neither $\Vv$ nor $\Vh$, nor does it change an attacker's knowledge of $\Vh$. 
Assignments to $\Vv$ or $\Vh$ can use an expression $E.\Vv.\Vh$ or a distribution $D.\Vv.\Vh$; and assignments to $\Vv$ might reveal information about $\Vh$. For example, from \Fig{f1228} we can explore various assignments to $\Vv$:
\newcounter{z}\begin{enumerate}[(i)]
\item A direct assignment of $\Vh$ to $\Vv$ reveals everything about $\Vh$: \\
  \hspace*{\fill}
  $\Sem{\Vv\Gets \Vh}.(v,\delta) = \PSet{h\In \delta}{}{(h,\PSet{}{}{h})}$
\item Choosing $\Vv$ from a distribution independent of $\Vh$ reveals nothing about $\Vh$: \\
  \hspace*{\fill}
  $\Sem{\Vv\Gets 0\PC{1/3}1}.(v,\delta) = \PSet{}{}{(0,\delta)\Att{1}{3},(1,\delta)\Att{2}{3}}$
\item Partially $\Vh$-dependent assignments to $\Vv$ might reveal something about $\Vh$:\\
  \hspace*{\fill} 
  $\Sem{\Vv \Gets \Vh\Mod 2}.(v,\Uniform{0,1,2}) =  
  \PSet{}{}{(0,\Uniform{0,2})\Att{2}{3}, (1,\Uniform{1})\Att{1}{3}}$
\setcounter{z}{\value{enumi}}\end{enumerate}
As a further illustration, we calculate the effect of the first assignment to $\Vv$ in Program $S$ as follows:
\begin{Reason}
\Step{}{
\Sem{\Vv\From \PSet{}{}{w\Att{\Vh}{2},b\At{1{-}\frac{\Vh}{2}}}}.(v,\Uniform{0,1,2})
}
\Space
\StepR{$=$}{Choose prob.\ visible}{
\PSet{~v'\In 1/3{*}(\Uniform{b} + \Uniform{w,b} + \Uniform{w})}{}{
\\ \hspace{6em}~
(v',~ \PSet{h'\In\Uniform{0,1,2}}{
\PSet{}{}{w\Att{h'}{2},b\At{1{-}\frac{h'}{2}}}.v'}{})~}
}
\Space
\WideStepR{$=$}{simplify the summation}{
\PSet{~v'\In \Uniform{w,b}}{}{
(v', \PSet{h'\In\Uniform{0,1,2}}{
\PSet{}{}{w\Att{h'}{2},b\At{1{-}\frac{h'}{2}}}.v'}{})~}
}
\Space
\WideStepR{$=$}{evaluate outer comprehension}{
\Uniform{~(w, \PSet{h'\In \Uniform{0,1,2}}{\frac{h'}{2}}{}),~ 
          (b, \PSet{h'\In \Uniform{0,1,2}}{1{-}\frac{h'}{2}}{})~}
}
\Space
\StepR{$=$}{evaluate conditional distributions}{
\Uniform{~(w,\PSet{}{}{1\Att{1}{3},2\Att{2}{3}}),
         ~(b,\PSet{}{}{0\Att{2}{3},1\Att{1}{3}})~}~.
}
\end{Reason}

As for assignments to $\Vh$, we see that they affect $\delta$ directly; thus Choosing hidden $\Vh$ might
\begin{enumerate}[(i)]\setcounter{enumi}{\value{z}}
\item increase our uncertainty of $\Vh$:
 \hspace*{\fill}
 $\Sem{\Vh\Gets 0 {\PC{}} 1 {\PC{}} 2}.(v,\Uniform{0,1}) = \Uniform{(v,\Uniform{0,1,2})}$
\item or reduce it: 
 \hspace*{\fill}
 $\Sem{\Vh\Gets 0{\PC{}} 1}.(v, \Uniform{0,1,2}) = \Uniform{(v,\Uniform{0,1})}$ \label{e4446}
\item or leave it unchanged:
 \hspace*{\fill}
 $\Sem{\Vh\Gets 2{-}\Vh}.(v,\Uniform{0,1,2}) = \Uniform{(v,\Uniform{2,1,0})}$
\end{enumerate}

In all of the above, we saw that the assignment statements were atomic --- an attacker may not directly witness the evaluation of their right-hand sides. For instance, the atomic probabilistic choice $\Vv\Gets \Vh {\PC{}} \neg \Vh$  does \emph{not} reveal which of the equally likely operands of $(\PC{})$ was used.

\subsubsection{Non-atomic commands}
The first, \emph{Composition} $P_1;P_2$, gives an attacker \textbf{perfect recall} after $P_2$ of the visible variable $\Vv$ as it was after $P_1$, even if $P_2$ overwrites $\Vv$.%
\footnote{It is effectively the Kleisli composition over the outer distribution.}
To see the effects of this, we compare the three-box Program $S$ from the start of \Sec{s1121}, that is 
\begin{equation*}
\Vh\Gets 0{\PC{}}1{\PC{}}2; \Vv\From \PSet{}{}{w\Att{\Vh}{2},b\At{1{-}\frac{\Vh}{2}}}; \Vv\Gets \bot ~,
\end{equation*}
with the simpler Program $I_1$ defined $\Vh \Gets 0{\PC{}}1{\PC{}}2; \Vv\Gets \bot$ in which no ball is drawn: the final hyper-distributions are respectively
\medskip\par\noindent
\hfill\(
 \PSet{}{}{~(\bot,\PSet{}{}{1\Att{1}{3},2\Att{2}{3}}),~(\bot,\PSet{}{}{0\Att{2}{3},1\Att{1}{3}})~}
\)\hfill\makebox[0pt][r]{($\Delta'_S$)}
\par\noindent
\hfill\(
 \makebox[0pt][r]{and\hspace{7em}}
 \PSet{}{}{~(\bot,\PSet{}{}{0,1,2})~}\makebox[0pt][l]{~.}
\)\hfill\makebox[0pt][r]{($\Delta'_{I_1}$)}%
\medskip\par\noindent
We calculated $\Delta_S'$ as follows:
\begin{Reason}
\Step{}{
\Sem{\Vh\Gets 0{\PC{}}1{\PC{}}2;
     \Vv\From \PSet{}{}{w\Att{\Vh}{2},b\At{1{-}\frac{\Vh}{2}}};
     \Vv\Gets\bot}.(v,\delta)
}
\Space
\StepR{$=$}{Choose hidden; Composition}{
\Sem{\Vv\From \PSet{}{}{w\Att{\Vh}{2},b\At{1{-}\frac{\Vh}{2}}};\Vv\Gets\bot}.(v,\PSet{}{}{0,1,2})
}
\Space
\StepR{$=$}{Composition}{
\Exp{(\hat{v},\hat{\delta})\In 
\Sem{\Vv \PGets \PSet{}{}{w\Att{\Vh}{2},b\At{1{-}\frac{\Vh}{2}}}}.(v,\Uniform{0,1,2})
}{\Sem{\Vv\Gets\bot}.(\hat{v},\hat{\delta})}
}
\Space
\WideStepR{$=$}{assignment $\Vv\Gets\bot$ independent of $\Vh$}{
\Exp{(\hat{v},\hat{\delta})\In 
\Sem{\Vv \PGets \PSet{}{}{w\Att{\Vh}{2},b\At{1{-}\frac{\Vh}{2}}}}.(v,\Uniform{0,1,2})
}{\Uniform{(\bot,\hat{\delta})}}
}
\Space
\WideStepR{$=$}{Choose prob.\ visible (see earlier calculation)}{
\Exp{(\hat{v},\hat{\delta})\In 
\Uniform{(w,\PSet{}{}{1\Att{1}{3},2\Att{2}{3}}), 
         (b,\PSet{}{}{0\Att{2}{3},1\Att{1}{3}})}
}{ \Uniform{(\bot,\hat{\delta})} }
}
\Space
\StepR{$=$}{evaluate expected value}{
\Uniform{
(\bot,\PSet{}{}{1\Att{1}{3},2\Att{2}{3}}), 
(\bot,\PSet{}{}{0\Att{2}{3},1\Att{1}{3}}) }~.
}
\end{Reason}

In neither case $\Delta'_S$ nor $\Delta'_{I_1}$ does the final value $\bot$ of $\Vv$ reveal anything about $\Vh$. But $\Delta'_{I_1}$ is a point (outer) distribution (thus concentrated on a single split-state), whereas $\Delta'_S$ is a uniform distribution over two split-states each of which recalls implicitly the observation of an intermediate value $\hat{v}$ of $\Vv$ that was made during the execution leading to that state. Generally, if two split-states $(v',\delta_1')$ and $(v',\delta_2')$ occur with $\delta'_1{\neq}\delta'_2$ then it means an attacker can deduce whether $\Vh$'s distribution is $\delta_1'$ or $\delta_2'$ even though $\Vv$ has the same final value $v'$ in both cases. Although the direct evidence $\hat{v}$ has been overwritten, the distinct split-states preserve the attacker's deductions from it.

The meaning of \emph{General prob.\ choice} $P_1 \PC{p.\Vv.\Vh} P_1$ --of which both \emph{Probabilistic choice} and \emph{Conditional choice} are specific instances-- makes it behave like $\Sem{P_1}$ with probability $p.\Vv.\Vh$ and $\Sem{P_2}$ with the remaining probability. The definition allows an attacker to observe which branch was taken and, knowing that, she might be able to deduce new facts about $\Vh$. Thus unlike for \Eqn{e4446} above we have $\Sem{\Vh\Gets0\PC{}\Vh\Gets{1}}.(v,\delta) = \PSet{}{}{(v,\PSet{}{}{0}),(v,\PSet{}{}{1})}$, which is an example of \textbf{implicit flow}. 

A similar implicit information flow in any Conditional choice with guard $G.\Vv.\Vh$ makes it possible for an attacker to deduce the value of the guard exactly.

For General prob.\ choice $P_1 \PC{p.\Vv.\Vh}P_2$ however, the implicit flow might only partially reveal the value of the expression $p.\Vv.\Vh$. For example, suppose we execute the probabilistic assignment $\Vh\Gets \frac{1}{4}\PC{}\frac{1}{2}$, which establishes that $\Vh$ is either $\frac{1}{4}$ or $\frac{1}{2}$ with equal probability of each: its output is $\PSet{}{}{(v, \Uniform{\frac{1}{4}, \frac{1}{2}})}$. Then we execute program $\Skip \PC{\Vh} \Skip$ from there, and we find that we do not \emph{entirely} discover the value of $\Vh$. But still we do discover something: we find that
\[
 \Sem{\Skip \PC{\Vh} \Skip}.(v, \Uniform{\IIFrac{1}{4}, \IIFrac{1}{2}})
 \Wide{=} 
 \PSet{}{}{
  (v, \PSet{}{}{\IIFrac{1}{4}\Att{1}{3}, \IIFrac{1}{2}\Att{2}{3}})\Att{3}{8},
  (v, \PSet{}{}{\IIFrac{1}{4}\Att{3}{5}, \IIFrac{1}{2}\Att{2}{5}})\Att{5}{8}
 }~,
\]
and see that indeed the chance of guessing $\Vh$'s value has increased, though we still do not know it for certain. Our probability initially of guessing $\Vh$ is $1/2$. But after the choice we will guess $\Vh{=}\frac{1}{2}$ when we see the choice went left, which happens with probability $3/8$; but if we saw the choice going right we will guess $\Vh{=}\frac{1}{4}$, which happens with probability $5/8$. Our average chance of guessing $\Vh$ is thus $(2/3){*}(3/8)+(3/5){*}(5/8) = 5/8$, which is more than the $1/2$ it was initially: that increased knowledge is what was revealed by the $(\PC{\Vh})$.

\section{The Bayes-Risk based elementary testing order}\label{s1649}

The elementary testing order comprises functional- and security characteristics. 

Say that two programs are \emph{functionally equivalent} iff from the same input they produce the same \emph{overall} output distribution \cite{Kozen:85,McIver:05a}, defined for hyper-distribution $\Delta'$ to be $\Ft.\Delta' \Defs \PSet{(v',\delta')\In \Delta'; h'\In\delta'}{}{(v',h')}$.\,%
\footnote{Two program \emph{texts} $P_{\{1,2\}}$ denote functionally equivalent secure programs just when their classical denotations agree, that is when $\SemC{P_1}{=}\SemC{P_2}$. The function $\Ft$ expresses that semantically, and the connection is thus that $\SemC{P_1}{=}\SemC{P_2}$ just when $\Ft.(\Sem{P_1}.(v,\delta)){=}\Ft.(\Sem{P_2}.(v,\delta))$ for all $(v,\delta)$.}
We consider state-space $\VV{\times}\HH$ jointly, i.e.\ not $\VV$ alone, because differing distributions over $\Vh$ alone can be revealed by the context $\SeqCC{\Vv\Gets\Vh}$ that appends an assignment $\Vv\Gets\Vh$.

We measure the \emph{security} of a program with ``Bayes Risk''~\cite{Smith:07,Chatzikokolakis:07b,Braun:08,Braun:09}, which determines an attacker's chance of guessing the final value of $\Vh$ in one try. The most effective such attack is to determine which split-state $(v',\delta')$ in a final hyper-distribution actually occurred, and then to guess that $\Vh$ has some value $h'$ that maximises $\delta'$, i.e. so that $\delta'\kern-.25em.h' = \UMax{}\delta'.$ For a whole hyper-distribution we average the attacks over its elements, weighted by the probability it gives to each, and so we we define the \textbf{Bayes Vulnerability} of $\Delta'$ to be $\Et.\Delta'\Defs\Exp{(v',\delta')\In\Delta'}{\UMax{}\delta'}$.\footnote{We use vulnerability rather than risk because ``greatest chance of leak'' is more convenient than the dual ``least chance of no leak.'' Our definition corresponds to Smith's \emph{vulnerability}~\cite{Smith:07}.}

For Program $S$ the vulnerability is the chance of guessing $\Vh$ by remembering $\Vv$'s intermediate value, say $\hat{v}$, and then guessing that $\Vh$ at that point had the value most likely to have produced that $\hat{v}$: when $\hat{v}{=}w$ (probability $1/2$), guess $\Vh{=}2$; when $\hat{v}{=}b$, guess $\Vh{=}0$. Via $\Et.\Delta'_S$ that vulnerability is $1/2{*}2/3 + 1/2{*}2/3 = 2/3$. For $I_1$, however, there is no ``leaking'' $\hat{v}$, and so it is less vulnerable, having $\Et.\Delta'_{I_1} = 1/3$.

The elementary testing order on hyper-distributions is then defined $\Delta_S{\ERef}\Delta_I$ iff $\Ft.\Delta_S{=}\Ft.\Delta_I$ and $\Et.\Delta_S{\geq}\Et.\Delta_I$, and it extends pointwise to the \textbf{elementary testing order} on whole programs. That is, we say that $S{\ERef}I$ just when for corresponding inputs (i) $S,I$ are functionally equivalent and (ii) the vulnerability of $I$ is no more than the vulnerability of $S$. Thus $S{\ERef}I_1$ because they are functionally equivalent and the vulnerabilities of $S,I_1$ are $2/3,1/3$ resp.

The direction of the inequality $(\ERef)$ corresponds to increasing security (and thus decreasing vulnerability). This agrees with other notions of security that increase with increasing entropy of the hidden distribution.

\section{Non-compositionality of the elementary testing order}\label{s1201}
Although $S{\NotERef}I$ is an (elementary) failure of implementation, the complementary $S{\ERef}I$ is not necessarily a success: it is quite possible, in spite of that, that there is a context $\CC$ with $\CC(S){\NotERef}\CC(I)$. That is, simply having $S{\ERef}I$ does not mean that $I$ is safe to use in place of $S$ in general.

Thus for stepwise development we require more than just $S{\ERef}I$: we must ensure that $\CC(S){\ERef}\CC(I)$ holds for \emph{all} contexts $\CC(\cdot)$ in which $S,I$ might be placed --- and we do not know in advance what those contexts might be.

Returning to the boxes, we consider now another variation Program $I_2$ in which both Boxes 0,1 have two black balls: thus the program code becomes $\Vh\Gets0{\PC{}}1{\PC{}}2;\Vv\From\PSet{}{}{w\At{(\Vh\div2)},b\At{1{-}(\Vh\div2)}};\Vv\Gets\bot$ with final hyper-distribution
\begin{EQN}{$\Delta'_{I_2}$}
$\PSet{}{}{~(\bot,\PSet{}{}{2})\Att{1}{3}~, ~(\bot,\PSet{}{}{0,1})\Att{2}{3}~}\makebox[0pt][l]{~.}$
\end{EQN}
The vulnerability of $I_2$ is $1/3{*}1+2/3{*}1/2$, again $2/3$ so that $S{\ERef}I_2$. Now if context $\CC$ is defined $\SeqCC{\Vh\Gets\Vh{\div}2}$, the vulnerability of $\CC(S)$ is $1/2{*}2/3+1/2{*}1 = 5/6$: it is more than for $S$ alone because there are fewer final $\Vh$-values to choose from. But for $\CC(I_2)$ it is greater still, at $1/3{*}1+2/3{*}1=1$.

Thus $S{\ERef}I_2$ but $\CC(S){\NotERef}\CC(I_2)$, and so $(\ERef)$ is not compositional. This makes $(\ERef)$ unsuitable, on its own, for secure-program development of any size; and its failure of compositionality is the principal problem we solve.

\section{The refinement order, and compositional closure}
The compositional closure of an ``elementary'' partial order over programs, call it $(\leq_E)$, is the largest subset of that order that is preserved by composition with other programs, that is with being placed in a program context. Call that closure $(\leq_C)$.

The utility of $(\leq_C)$ is first that $A{\leq_C}B$ implies $A{\leq_E}B$, so that $A{\leq_C}B$ suffices if $A{\leq_E}B$ is all that we want: but it implies further that $\CC(A){\leq_E}\CC(B)$ for all contexts $\CC$, as well. Its being the \emph{greatest} such subset of $(\leq_E)$ means that it relates as many programs as possible, never claiming that $A{\not\leq_C}B$ unless there is some context $\CC$ that forces it to do so because in fact $\CC(A){\not\leq_E}\CC(B)$.

Thus to address the non-compositionality exposed in \Sec{s1201}, we seek the \emph{compositional closure} of $(\ERef)$, the unique \emph{refinement} relation $(\Ref)$ such that (\emph{soundness}) if $S{\Ref}I$ then for all $\CC$ we have $\CC(S){\ERef}\CC(I)$; and (\emph{completeness}) if $S{\NotRef}I$ then for some $\CC$ we have $\CC(S){\NotERef}\CC(I)$. Soundness gives refinement the property (\Sec{s1201}) we need for stepwise development; and completeness makes refinement as liberal as possible consistent with that.

We found above that $S{\NotRef}I_2$; we show later (\Sec{s1213}) that we do have $S{\Ref}I_1$.

\section{Constructive definition of the refinement order}\label{s1457}

Although saying thet $(\Ref)$ is the compositional closure of $(\ERef)$ does define it completely, it is of little use if to establish $S{\Ref}I$ in practice we have to evaluate and compare $\CC(S){\ERef}\CC(I)$ for all contexts $\CC$. Instead we seek an explicit construction that is easily verified for specific cases. We give a detailed example to help introduce our definition.

For integers $x,n$, let $x\Rnd n$ be a distribution over the multiple(s) of $n$ closest to $x$: usually there will be exactly two such multiples, one on either side of $x$ and, in that case, the probabilities of each are inversely proportional to their distance from $x$. Thus $1\Rnd4$ is $\PSet{}{}{0\Att{3}{4},4\Att{1}{4}}$ and $2\Rnd4$ is $\PSet{}{}{0\Att{1}{2},4\Att{1}{2}}$ and $3\Rnd4$ is $\PSet{}{}{0\Att{1}{4},4\Att{3}{4}}$. If however $x$ happens to be an integer multiple of $n$ then the outcome is definite, a point distribution: thus $0\Rnd4=\PSet{}{}{0}$ and $4\Rnd4=\PSet{}{}{4}$.

Now consider the two programs
\begin{equation}\label{e1930}
 \begin{array}{lrl}
                    & P_2 \Defs~~~ & \Vh\Gets1{\PC{}}2{\PC{}}3;~ \Vv \From \Vh\Rnd2;~ \Vv \Gets \Vh\Mod2 \\                    
  \textrm{and}\quad 
                    & P_4 \Defs~~~ & \Vh\Gets1{\PC{}}2{\PC{}}3;~ \Vv\From \Vh\Rnd4;~\Vv \Gets \Vh\Mod2 ~.
\end{array}
\end{equation}
Both reveal $\Vh\Mod2$ in $\Vv$'s final value $v'$, but each $P_n$ also reveals in the overwritten visible $\hat{v}$, say, something about $\Vh\Rnd n$; and intuition suggests that $P_n{\Ref} P_m$ for $n{\leq}m$ only. Yet in fact the vulnerability is $5/6$ for both $P_{2,4}$, which we can see from their final hyper-distributions; they are $\Delta'_{P_2}$ and $\Delta'_{P_4}$ given by
%
\medskip\par\noindent
\hfill\(
 \PSet{}{}{~~(0,\PSet{}{}{2})\Att{1}{3},~
       (1,\PSet{}{}{1})\Att{1}{6},
       (1,\PSet{}{}{1,3})\Att{1}{3},
       (1,\PSet{}{}{3})\Att{1}{6}~~}
\)\hfill\makebox[0pt][r]{($\Delta'_{P_2}$)}
\par\noindent
\hfill\(
 \PSet{}{}{~~(0,\PSet{}{}{2})\Att{1}{3},
       \underbrace{(1,\PSet{}{}{1\Att{3}{4},3\Att{1}{4}})\Att{1}{3}}
       _{\makebox[0pt]{\hspace{3em}\parbox{33em}{\scriptsize
         With overall probability $1/3{*}3/4+1/3{*}1/4=1/3$ the final $v'$ will be 1 and $\hat{v}$ will be 0;
         since $v'$ is 1 then $\Vh$ must be 1 or 3;
         but if $\hat{v}$ was 0 that $\Vh$ is three times
         as likely to have been 1.}}},
       (1,\PSet{}{}{1\Att{1}{4},3\Att{3}{4}})\Att{1}{3}~~}
\)\hfill\makebox[0pt][r]{($\Delta'_{P_4}$)}%
\medskip\par\noindent
so that e.g.\ $1/3{*}1+1/3{*}3/4+1/3{*}3/4 = 5/6$ for $P_4$.
The overall distribution of $(v',h')$ is $\PSet{}{}{(0,2),(1,1),(1,3)}$ in both cases, so that $P_{2,4}$ are functionally equivalent; but they have different residual uncertainties of $\Vh$.

\subsection{Hyper-distributions as partitions of fractions}

In our definition of refinement we will consider the hyper-distributions corresponding to each value of $\Vv$ separately. 

In the example above, if we consider just the $\Vh$-distributions associated with $v'{=}1$ then we can, by multiplying through their associated probabilities from the hyper-distributions, present them as a collection of \emph{fractions}, that is sub-distributions over $\HH$. We call such collections \emph{partitions} and here they are given for $P_2$ and $P_4$ respectively by
\footnote{Strictly speaking, partitions are multisets of fractions, i.e.\ without order but possibly having repeated elements.}
\begin{equation}\label{e1948}
 \textrm{when $v'{=}1$}\quad\left\{\quad\quad
 \begin{array}{r@{\hspace{3em}}l}
  \Pi'_{P_2}\In&\List{\PSet{}{}{1\Att{1}{6}}, \PSet{}{}{1\Att{1}{6},3\Att{1}{6}}, \PSet{}{}{3\Att{1}{6}}} \\
  \Pi'_{P_4}\In&\List{\PSet{}{}{1\Att{1}{4},3\Att{1}{12}}, \PSet{}{}{1\Att{1}{12},3\Att{1}{4}}} ~.
 \end{array}
 \right.
\end{equation}

In general, let the function $\Fracs.\Delta.v$ for hyper-distribution $\Delta$ and value $v$ give the partition of fractions extracted from $\Delta$ for $\Vv{=}v$, as we extracted $\Pi'_{\{2,4\}}$ from $\Delta'_{\{2,4\}}$ and $v'{=}1$ at \Eqn{e1948} above.

\subsection{Operations on fractions and partitions}

Distribution operations such as support $(\Supp{\cdot})$ and weight $(\sum)$ and normalise $\Norm{\cdot}$ apply to fractions, and for example we have that $\PSet{}{}{1\Att{1}{6}} + \PSet{}{}{1\Att{1}{6},3\Att{1}{6}}$ is $\PSet{}{}{1\Att{1}{3},3\Att{1}{6}}$ and $\sum\PSet{}{}{1\Att{1}{3},3\Att{1}{6}}$ is $1/2$ and $\Norm{\PSet{}{}{1\Att{1}{3},3\Att{1}{6}}}$ is $\PSet{}{}{1\Att{2}{3},3\Att{1}{3}}$.
For partitions $\Pi$ we write $\sum \Pi$ as shorthand for $\List{\Sum{\pi \In \Pi}{}{\pi}}$, so that 
\[
 \sum\List{\PSet{}{}{1\Att{1}{6}},\PSet{}{}{1\Att{1}{6},3\Att{1}{6}}}
 \WIDERM{is}
 \List{\PSet{}{}{1\Att{1}{3},3\Att{1}{6}}}~.
\]
Note that the sum of a partition is still a partition, albeit always with only a single fraction in it.
Scaling, when applied partition is applied pointwise to each of its fractions.
An empty partition is written $\List{}$, and a zero(-weight) fraction is written $\PSet{}{}{}$; thus $\List{\PSet{}{}{}}$ is a zero-weight partition containing exactly one fraction. 

Finally, the Bayes Vulnerability of a partition $\Et.\Pi$ is $\Sum{\pi\In \Pi}{}{\UMax{}\pi}$, and the Bayes Vulnerability of a hyper-distribution may be equivalently expressed using partitions as $\Sum{v\In \VV}{}{\Et.(\Fracs.\Delta.v)}$.

\subsection{Relationships between fractions and partitions}

Say that two non-zero fractions {$\pi_{\{1,2\}}$ are \emph{similar}, written $\pi_1{\Similar}\pi_2$, just when their normalisations are equal, that is when $\Norm{\pi_1}{=}\Norm{\pi_2}$ so that they are multiples of each other: this is an equivalence relation. For example we have $\PSet{}{}{1\Att{1}{3},2\Att{2}{3}}\Similar\PSet{}{}{1\Att{1}{4},2\Att{1}{2}}$ because both normalise to the former. 

Say that a partition is \emph{reduced} just when it contains no two similar fractions, and no zero fractions at all.\,%
\footnote{Allowing zero fractions, in the unreduced case, simplifies some proofs.}
For any hyper-distribution $\Delta$ and value $v$, we have that $\Fracs.\Delta.v$ is in reduced form by construction.
Thus partitions are more expressive than hyper-distributions.

The \emph{reduction} of a partition is obtained by by adding-up all its similar fractions and removing its all-zero fractions, that is by \emph{reducing} it, and we} say that two partitions $\Pi_{\{1,2\}}$ are \emph{similar}, written $\Pi_1{\Similar}\Pi_2$, just when they have the same reduction. Thus for example we have
\[
 \List{\PSet{}{}{},\PSet{}{}{0\Att{1}{3}},\PSet{}{}{1\Att{1}{6},2\Att{1}{6}},\PSet{}{}{1\Att{1}{6},2\Att{1}{6}}}
 \Wide{\Similar}
 \List{\PSet{}{}{0\Att{1}{3}},\PSet{}{}{1\Att{1}{9},2\Att{1}{9}},\PSet{}{}{1\Att{2}{9},2\Att{2}{9}}}~,
\]
because both reduce to $\List{\PSet{}{}{0\Att{1}{3}},\PSet{}{}{1\Att{1}{3},2\Att{1}{3}}}$.
If two partitions are similar then. for any distribution $\delta$ over $\HH$, the probability that an attacker may deduce that $\Vh$ is distributed according to $\delta$ is the same in either partition.

Say that one partition $\Pi_1$ is \emph{as fine as} another $\Pi_2$, written $\Pi_1{\Finer}\Pi_2$, just when $\Delta_2$ can be obtained by adding-up one or more groups of fractions in $\Delta_1$. Thus for example we have
\[
 \List{\PSet{}{}{0\Att{1}{3}},\PSet{}{}{1\Att{1}{9},2\Att{2}{9}},\PSet{}{}{1\Att{2}{9},2\Att{1}{9}}}
 \Wide{\Finer}
 \List{\PSet{}{}{0\Att{1}{3}},\PSet{}{}{1\Att{1}{3},2\Att{1}{3}}}
\]
by adding-up the second and third fractions on the left. For as-fine-as the added-up fractions do not have to be similar: if however they \emph{are} similar, then we have $\Pi_1{\Similar}\Pi_2$ as well as $\Pi_1{\Finer}\Pi_2$; if they are not similar, we can write $\Pi_1{\StrictFiner}\Pi_2$.

Combining two dissimilar fractions in a partition represents removal of the implicit observations that distinguished them. Hence if $\Pi_1{\StrictFiner}\Pi_2$ then  partition $\Pi_2$ conceals $\Vh$ strictly better than $\Pi_1$ does.

Note that in both cases $\Pi_1{\Similar}\Pi_2$ and $\Pi_1{\Finer}\Pi_2$ we have $\sum\Pi_1=\sum\Pi_2$, i.e.\ that neither relation allows a change in the overall probability assigned to each of the elements.

\subsection{Constructive definition of refinement}\label{s1213}

We use the relations $(\Similar)$ and $(\Finer)$ between partitions to define refinement.

\begin{Definition}{Secure refinement}{d1424}
We say that hyper-distribution $\Delta_{S}$ is securely-refined by $\Delta_{I}$, written $\Delta_{S}\Ref\Delta_{I}$, just when for every $v$ there is some intermediate partition $\Pi$ of fractions
so that first (i) $\Fracs.\Delta_{S}.v$ is similar to $\Pi$ and then (ii) $\Pi$ is as fine as $\Fracs.\Delta_{I}.v$.%
\footnote{In our earlier \emph{qualitative} work \cite{Morgan:07} refinement reduces to taking unions of equivalence classes of hidden values, so-called ``Shadows.'' K{\"o}pf et al.\ observe similar effects \cite{Kopf:07}.}
That is, we have
\[
 \Delta_{S}\Ref\Delta_{I}
 \WideRm{~~iff~~}
 \Fracs.\Delta_{S}.v ~\Similar~ \Pi ~\Finer~ \Fracs.\Delta_{I}.v
 \quad\textrm{for some partition $\Pi$.}
\]
The fractions of $\Delta_S$ are first split-up into similar sub-fractions; and then some of those sub-fractions are rejoined to create the fractions of $\Delta_I$.
 
Refinement of hyper-distributions extends pointwise to the programs that produce them.
\end{Definition}
Note that since both $(\Similar)$ and $(\Finer)$ preserve partition-sum, we have that $(\Ref)$ from \Def{d1424} implies functional equality. Informally speaking, refinement may not change the functional behaviour of a secure program, but it may reduce the implicit observations available to an attacker, and hence the deductions an attacker can make about $\Vh$.

We return to $\Delta'_S$ for an example, getting $\List{\PSet{}{}{1\Att{1}{6},2\Att{1}{3}},\PSet{}{}{0\Att{1}{3},1\Att{1}{6}}}$ for $\Fracs.\Delta'_S.\bot$ by multiplying through. For $\Delta'_{I_1}$ we get $\List{\PSet{}{}{0\Att{1}{3},1\Att{1}{3},2\Att{1}{3}}}$ similarly for $\Fracs.\Delta'_{I_1}.\bot$. The two fractions of the former sum to the single fraction of the latter, and so $S{\Ref}I_1$ according to our definition \Def{d1424} of secure refinement.

For the more detailed $\Delta'_{P_2}{\Ref}\Delta'_{P_4}$ and $v'{=}1$, we need the intermediate partition $\Pi\Defs\,\List{\PSet{}{}{1\Att{1}{6}}, \PSet{}{}{1\Att{1}{12},3\Att{1}{12}}, \PSet{}{}{1\Att{1}{12},3\Att{1}{12}}, \PSet{}{}{3\Att{1}{6}}}$, whose middle two fractions turn out to be equal, thus certainly similar: summing them gives the middle $\PSet{}{}{1\Att{1}{6},3\Att{1}{6}}$ of $\Pi'_{P_2}$, so that $\Pi'_{P_2}{\Similar}\Pi$. On the other hand, summing the first two fractions of $\Pi$ gives $\PSet{}{}{1\Att{1}{4},3\Att{1}{12}}$, the first fraction of $\Pi'_{P_4}$, and summing the last two give the second fraction of $\Pi'_{P_4}$; thus $\Pi{\Finer}\Pi'_{P_4}$. Partition $\List{\PSet{}{}{2\Att{1}{3}}}$ deals trivially with $v'{=}0$, and so indeed we have $P_2{\Ref}P_4$ altogether. In \App{s9347} we show however that $P_4{\NotRef}P_2$.

\subsection{Properties of refinement}\label{s9345}
The refinement relation $(\Ref)$ is a partial order (hence it is transitive), and program contexts preserve it (thus it is monotonic). Consequently, we can reason incrementally and compositionally about refinement relation between large programs.
\begin{Theorem}{Partial order}{t1042}
The refinement relation $(\Ref)$ is a partial order over the set of hyper-distribut\-ions; and so, by extension, it is a partial order over programs.
\begin{Proof} See \App{a2937}. \end{Proof}
\end{Theorem}
\begin{Theorem}{Monotonicity of refinement}{t1120}
If $S{\Ref}I$ then $\CC(S){\Ref}\CC(I)$ for all contexts $\CC$ built from programs as defined in \Fig{f1228}.
\begin{Proof} See \App{a9475}. \end{Proof}
\end{Theorem}

Furthermore, we define \emph{strict} refinement such that $S \StrictRef I$ when  $S \Ref I$ but $I \NotRef S$.

\section{Refinement $(\Ref)$ is the compositional closure of $(\ERef)$}\label{s1724}

In this proof we will manipulate partitions, sequential composition, refinement and Bayes Vulnerability in terms of matrices, as follows.

\subsection{Matrix representation and manipulation of partitions}\label{s1552}

\subsubsection{Partitions as matrices}

Assume \Wlog\ that $\HH$ is the integers $1..H$. For a particular input $(v,\delta)$ and a chosen visible output $v'$, a program $P$ will produce as output a partition $\Pi = \Fracs.(\Sem{P}.(v,\delta)).v'$ over hidden values containing some number $F$ of fractions that we index $1..F$. Each fraction on its own is a vector of length $H$ of probabilities; if we put them together as rows, we get an $F{\times}H$-matrix that represents the partition as a whole. For example, we have from \Eqn{e1948} the following matrix representations of partitions output from Programs $P_{\{2,4\}}$ for $v'{=}1$:
\begin{equation}\label{e1520}
 \Pi_{P_2}'{:}\quad
  \left (\begin{array}{c@{~}c@{~}c}
          \IFrac{1}{6} & 0 &  0 \\
          \IFrac{1}{6} & 0 & \IFrac{1}{6} \\
          0 & 0 & \IFrac{1}{6}
         \end{array}
  \right)
 \hspace{5em}
 \Pi_{P_4}'{:}\quad
  \left (\begin{array}{c@{~}c@{~}c}
          \IFrac{1}{4}  & 0 & \IFrac{1}{12} \\
          \IFrac{1}{12} & 0 & \IFrac{1}{4}\\
         \end{array}
  \right)~.
\end{equation}
There are three possible values of $\Vh$ in each case, so that $H{=}3$; and $P_2$'s partition has 3 fractions, so that $F_2{=}3$ and thus it generates a $3{\times}3$ matrix. Program $P_4$'s partition has only 2 fractions, so that $F_4{=}2$ and it generates a $2{\times}3$ matrix.

For simplicity in the proof, we will arrange that $H{=}F$ so that all matrices are of the same (square) dimension $N{\times}N$. This is without loss of generality, since we can extend $\HH$ with extra, unused values; and we can extend  our partitions with extra, zero fractions. For instance $\Pi'_{P_4}$ becomes a $3{\times}3$ matrix, as $\Pi'_{P_2}$ is already, if we add an extra row underneath (representing an all-zero fraction):
\begin{equation}\label{e64957}
 \Pi_{P_4}'{:}\quad
  \left (\begin{array}{c@{~}c@{~}c}
          \IFrac{1}{4}  & 0 & \IFrac{1}{12} \\
          \IFrac{1}{12} & 0 & \IFrac{1}{4} \\
          0 & 0 & 0
         \end{array}
  \right)~.
\end{equation}

\subsubsection{(A) Sequential composition as matrix multiplication}

In our completeness proof, our program-differentiating context $\CC$ will post-compose a probabilistic assignment $\Vh\From D.\Vh$ so that, for each of its incoming values $h$, the output value $h'$ will be chosen from the distribution $D.h$, thus with probability $D.h.h'$. In effect the context redistributes variable $\Vh$ in a way that depends on its current value.

We can consider $D$ itself to be an $N{\times}N$ matrix whose value in row $h$ and column $h'$ is just $D.h.h'$. If we do that, then the output partition $\Pi'$ that results from executing $\Vh\From D.\Vh$ on input partition $\Pi$ is just $\Pi{\times}D$, where ($\times$) is matrix multiplication. For example, suppose our post-composed context were
\begin{equation}\label{e1329}
 \Vh\From~(\,\PSet{}{}{1\Att{1}{2}, 2\Att{1}{4}, 3\Att{1}{4}}
           ~~\If~\Vh{=}1~\Else~~
           \PSet{}{}{2\Att{1}{2}, 3\Att{1}{2}}\,)~,           
\end{equation}
so that matrix $D$ would be
\[
 \left(
  \begin{array}{c@{~}c@{~}c}
   \IFrac{1}{2} & \IFrac{1}{4} & \IFrac{1}{4} \\
   0 & \IFrac{1}{2} & \IFrac{1}{2} \\
   0 & \IFrac{1}{2} & \IFrac{1}{2}
  \end{array}
 \right)~.
\]
From \Eqn{e1520} we take the incoming partition $\Pi$ to the post-composed context \Eqn{e1329} to be the outgoing partition $\Pi'_{P_2}$ from Program $P_2$, and so determine the outgoing partition $\Pi'$ from $(P'_2; \Vh\From D.\Vh)$ overall to be
\[
  \left (\begin{array}{c@{~}c@{~}c}
          \IFrac{1}{6} & 0 &  0 \\
          \IFrac{1}{6} & 0 & \IFrac{1}{6} \\
          0 & 0 & \IFrac{1}{6}
         \end{array}
  \right)
 \times
 \left(
  \begin{array}{c@{~}c@{~}c}
   \IFrac{1}{2} & \IFrac{1}{4} & \IFrac{1}{4} \\
   0 & \IFrac{1}{2} & \IFrac{1}{2} \\
   0 & \IFrac{1}{2} & \IFrac{1}{2}
  \end{array}
 \right)
 \Wide{=}
  \left(\begin{array}{c@{~\,}c@{\,~}c}
        \IFrac{1}{12} & \IFrac{1}{24} &  \IFrac{1}{24} \\
        \IFrac{1}{12} & \IFrac{1}{8} & \IFrac{1}{8} \\
        0 & \IFrac{1}{12} & \IFrac{1}{12}
       \end{array}
  \right)~.
\]

\subsubsection{(B) Refinement as matrix multiplication}

Also refinement can be formulated as matrix multiplication, since it is essentially a rearranging of fractions within a partition that, therefore, boils down to rearrangement of rows within a matrix. For example, from \Sec{s1213} we recall that to refine $\Pi'_{P_2}$ into $\Pi'_{P_4}$ we split the middle fraction of the former into two equal pieces, and add them to the other two, and that is achieved by the left-hand matrix in the pre-multiplication shown here:
\[
  \left(\begin{array}{c@{~}c@{~}c}
        1 & \IFrac{1}{2} & 0 \\
        0 & \IFrac{1}{2} & 1 \\
        0 & 0 & 0
  \end{array}\right)
 \times
  \left (\begin{array}{c@{~}c@{~}c}
          \IFrac{1}{6} & 0 &  0 \\
          \IFrac{1}{6} & 0 & \IFrac{1}{6} \\
          0 & 0 & \IFrac{1}{6}
         \end{array}
  \right)
 \Wide{=}
  \left (\begin{array}{c@{~}c@{~}c}
         \IFrac{1}{4}  & 0 & \IFrac{1}{12}\\
         \IFrac{1}{12} & 0 & \IFrac{1}{4} \\
         0 & 0  & 0
        \end{array}
  \right)~.
\]

In general a partition $\Pi_S$ is refined by $\Pi_I$ iff there exists a \emph{refinement matrix} $R$, a matrix whose columns are non-negative and one-summing, such that $R{\times}\Pi_S$ equals $\Pi_I$. Entry $(r,c)$ of such a refinement matrix describes what proportion of the $c^{\mathit th}$ fraction (row) of $\Pi_S$ is to contribute by addition to the $r^{\mathit th}$ fraction of $\Pi_I$.

\subsubsection{(C) Bayes Vulnerability as matrix multiplication}

Finally we bring Bayes Vulnerability into the matrix algebra as well. For a partition $\Pi$ as a matrix, the vulnerability is found by taking the individual row maxima and adding them together: the result is a scalar. Thus for $\Pi'_{P_4}$, for example, we have the matrix
\[
  \left (\begin{array}{c@{~}c@{~}c}
          \mbox{\boldmath$\IFrac{1}{4}$}  & 0 & \IFrac{1}{12} \\
          \IFrac{1}{12} & 0 & \mbox{\boldmath$\IFrac{1}{4}$} \\
          0 & 0 & 0
         \end{array}
  \right)
\WideRm{with maxima selected by the \emph{strategy} matrix $G$:}
  \left (\begin{array}{c@{~}c@{~}c}
         1  & 0 & 0\\
         0 & 0 & 1\\
         0 & 0 & 1
        \end{array}
  \right)
\]
whose maxima have been set in bold and are selected by the 1 entries in the matrix $G$ at right. Note that strategy matrices have the same shape as the matrix from which they select, and that they are 0/1 matrices with exactly one 1 per row.\,%
\footnote{Of course in an all-zero row it makes no difference which entry is selected.}

To determine the vulnerability associated with the $\Pi$, we calculate
\begin{equation}\label{e1605}
 \General{\MAX}{\,\textrm{strategy matrices $G$}}{}{\Tr.(\Transpose{G}{\times}\Pi)}
\end{equation}
in general, where $\Transpose{(\cdot)}$ is matrix transpose and $\Tr$ takes the \emph{trace} of a square matrix, i.e.\ the sum of its diagonal. Note that the maximum is actually attained, for some $G$, since there are only finitely many of them. In this particular case we use the $G$ above to calculate $\Tr.(\Transpose{G}{\times}\Pi'_{P_4})$, and have therefore
\[
  \left (\begin{array}{c@{~}c@{~}c}
         1 & 0 & 0 \\
         0 & 0 & 0 \\
         0 & 1 & 1
        \end{array}
  \right)
  \times
  \left (\begin{array}{c@{~}c@{~}c}
          \mbox{\boldmath$\IFrac{1}{4}$}  & 0 & \IFrac{1}{12} \\
          \IFrac{1}{12} & 0 & \mbox{\boldmath$\IFrac{1}{4}$} \\
          0 & 0 & 0
         \end{array}
  \right)
 \Wide{=}
  \left (\begin{array}{c@{~}c@{~}c}
         \mbox{\boldmath$\IFrac{1}{4}$} & 0 & \IFrac{1}{12} \\
         0 & \mbox{\boldmath$0$} & 0 \\
         \IFrac{1}{12} & 0 & \mbox{\boldmath$\IFrac{1}{4}$} 
        \end{array}
  \right) ~,
 \]
whose trace is $1/4 + 0 + 1/4 = 1/2$ to give the Bayes Vulnerability of $\Pi'_{P_4}$.

\subsubsection{(D) The connection between strategy matrices and refinement}

For any strategy matrix $G$, the transpose $\Transpose{G}$ has exactly one 1 in each column, and thus can be regarded as  a \emph{simple} refinement matrix, one of those which (when pre-multiplied with a partition) merges only whole fractions.
If we denote the set of $N{\times}N$ strate$\StratMatrices$y matrices by $\StratMatrices_N$, the set of $N{\times}N$ $\RefMatrices$efinement matrices by $\RefMatrices_N$, and the si$\FinerMatrices$ple subset of these (having only one non-zero entry per row) by $\FinerMatrices_N$, we thus have that 
\begin{equation*}
\Set{G\In \StratMatrices_N}{}{\Transpose{G}} 
\Wide{=} \FinerMatrices_N 
\Wide{\subseteq} \RefMatrices_N ~.
\end{equation*}
Furthermore, it can be shown that the complete set of refinement matrices $\RefMatrices_{N}$ is in fact the convex closure of its simple subset:
\begin{equation}\label{e29374}
\RefMatrices_{N} \Wide{=} \CCL.(\FinerMatrices_N)~.
\end{equation}
From \Eqn{e1605} and by linearity of matrix operations multiplication and trace, we thus have for any $N{\times}N$-dimensional $\Pi$ that the Bayes Vulnerability is given by
\begin{equation}\label{e8564}
\Et.\Pi
\Wide{=}
\General{\MAX}{G\In\StratMatrices_N}{}{\Tr.(\Transpose{G}{\times}\Pi)}
\Wide{=}
\General{\MAX}{R\In\RefMatrices_N}{}{\Tr.(R{\times}\Pi)} ~,
\end{equation}
because the extra elements in $\RefMatrices_N$ but not $\Transpose{\StratMatrices_N\kern-.2em}$ are only interpolations, and so cannot increase the maximum of a linear expression. (Recall from above that this maximum is attained for some $R$.)

Additionally, $\RefMatrices_{N}$ 
forms a monoid under matrix multiplication, that is
\begin{equation}\label{e58762}
(\RefMatrices_{N}, {\times}, \UnitMatrix_{N}) \Wide{\textrm{is a monoid,}}
\end{equation}
where $\UnitMatrix_{N}$ is the $N{\times}N$ unit of matrix multiplication.\,%
\footnote{Note that it is not a group because only the matrices in $\RefMatrices_{N}$ that permute --but do not combine-- fractions have inverses.}
We refer to \App{a39475} for a proof of Properties \Eqn{e29374} and \Eqn{e58762}.

\subsection{Soundness}\label{s1353}

Here from $S{\Ref}I$ we must show that $\CC(S){\ERef}\CC(I)$ for all contexts $\CC$. From monotonicity (\Thm{t1120}) it suffices to show that $S{\Ref}I$ implies $S{\ERef}I$.

Fix an initial split-state and construct the output hyper-distributions $\Delta'_{\{S,I\}}$ that result from $S,I$ respectively. Then since we assume $S{\Ref}I$ we must have $\Delta'_S{\Ref}\Delta'_I$. We now show that this implies $\Delta'_S{\ERef}\Delta'_I$.

Since $S{\Ref}I$ trivially guarantees that $\Ft.\Delta'_S = \Ft.\Delta'_I$ --recall \Def{d1424}-- we need to show that the Bayes-Vulnerability condition in the elementary testing order is satisfied. Since $\Et.\Delta = \Sum{v\In\VV}{}{\Et.(\Fracs.\Delta.v)}$, it is enough to show that for each $v\In \VV$ the vulnerability of $\Pi'_S \Defs \Fracs.\Delta'_{S}.v$ is no less than that of $\Pi'_I \Defs \Fracs.\Delta'_{I}.v$. 

For any such $\Pi'_{\{S,I\}}$ assume \Wlog\ that they are represented as $N{\times}N$ matrices.
We then have that
\begin{Reason}
\Step{}{
  \Et.\Pi'_S
}
\StepR{$=$}{from (D), Property \Eqn{e8564}}{
  \General{\MAX}{R \In \RefMatrices_{N}}{}{\Tr.(R{\times}\Pi'_S)} ~.
}
\StepR{$=$}{from (D), Property \Eqn{e58762}}{
  \General{\MAX}{R_1,R_2 \In \RefMatrices_{N}}{}{\Tr.(R_1{\times}R_2{\times}\Pi'_S})
}
\StepR{$\geq$}{for any $\widehat{R_2}$}{
  \General{\MAX}{R_1 \In \RefMatrices_{N}}{}{\Tr.(R_1{\times}\widehat{R_2}{\times}\Pi'_S})
}
\StepR{$=$}{from (B), choose $\widehat{R_2}$ so that $\widehat{R_2}{\times}\Pi'_S = \Pi'_I$}{
  \General{\MAX}{R_1 \In \RefMatrices_{N}}{}{\Tr.(R_1{\times}\Pi'_I)} ~.
}
\StepR{$=$}{from (D), Property \Eqn{e8564}}{
  \Et.\Pi'_I ~.
}
\end{Reason}
That gives us
\begin{Theorem}{Refinement is sound for Bayes Risk}{t1241}\quad
If $S{\Ref}I$ then $\CC(S){\ERef}\CC(I)$ for all contexts $\CC$.
\end{Theorem}

\subsection{Completeness}\label{s1634}

Here from $S{\NotRef}I$ we must discover a context $\CC$ such that $\CC(S){\NotERef}\CC(I)$. (The proof here is self-contained; but as background we give a fully worked example in \App{s9347}).

Since $S{\NotRef}I$, there must be an initial split-state $(v,\delta)$ from which $S,I$ yield hyper-distributions $\Delta'_{\{S,I\}}$ with $\Delta'_S{\NotRef}\Delta'_I$. We can assume however that $\Delta'_{S,I}$ give equal overall probabilities to \emph{visible} variables since, if they did not, they would be functionally different, giving $S{\NotERef}I$ immediately. This being so, we can assume that for some final $v'$ we have that partition $\Pi_S\Defs \Fracs.\Delta'_S.v'$ cannot be transformed into partition $\Pi_I\Defs \Fracs.\Delta'_I.v'$ via the two steps (i), (ii) in \Def{d1424}. That is, we have $\Pi_S \NotRef \Pi_I$.

We will define a distribution $D$ such that the context $\SeqCC{C}$ where $C$ is 
\begin{equation*}
\If~\Vv{=}v'~\Then~\Vh\From D.\Vh~\Else~\Vh\Gets0~\Fi
\end{equation*}
can be used to differentiate $S$ from $I$ using elementary testing.

We dispose of the simple case first: if $v''{\neq}v'$ then $\Fracs.(\Sem{S;C}.(v,\delta)).v''$ equals $\Fracs.(\Sem{I;C}.(v,\delta)).v''$  since, first, hyper-distributions $\Delta_{\{S,I\}}'$ give equal probabilities to that $v''$ and, second, the final value $h'$ of $\Vh$ is zero for both $S;C$ and $I;C$ in that case. The vulnerability associated with these partitions is therefore the same. To establish $\Sem{S;C} \NotERef \Sem{I;C}$ for our chosen $C$, it is thus enough to show that the vulnerability of $\Fracs.(\Sem{I;C}.(v,\delta)).v'$ is strictly greater than for $\Fracs.(\Sem{S;C}.(v,\delta)).v'$. 
Treating $\Pi_{\{S,I\}}$ as $N{\times}N$ matrices, we calculate
\begin{Reason}
\Step{}{
 \textrm{``Bayes Vulnerability of $\Pi_S;C$\,''}
}
\StepR{$=$}{(A) above; definition of $C$ based on $D$\,}{
 \textrm{``Bayes Vulnerability of $\Pi_S{\times}D$\,''}
}
\StepR{$=$}{\Eqn{e8564} in (D) above; for some maximising $\widehat{R}{\in}\RefMatrices_{N}$}{
 \Tr.(\widehat{R}{\times}\Pi_S{\times}D)
}
\StepR{$=$}{(B) above; for refinement $\widehat{\Pi} = \widehat{R}{\times}\Pi_S$ of $\Pi_S$}{
 \Tr.(\widehat{\Pi}{\times}D)
}
\Space
\StepR{$<$}{$D$ was chosen in advance, using the \emph{Separating} \\ \emph{Hyperplane Lemma}, 
and does not depend on $\widehat{\Pi}$: see below}{
 \Tr.(\Pi_I{\times}D)
}
\Space
\StepR{$=$}{identity}{
 \Tr.(\IdM{}{\times}\Pi_I{\times}D)
}
\StepR{$\leq$}{$\IdM{}{\in}\RefMatrices_{N}$}{
 \General{\MAX}{R\In\RefMatrices_{N}}{}{\Tr.(R{\times}\Pi_I{\times}D)}
}
\StepR{$=$}{\Eqn{e8564} in (D) above}{
 \textrm{``Bayes Vulnerability of $\Pi_I{\times}D$\,''}
}
\StepR{$=$}{(A) above; definition of $C$\,}{
 \textrm{``Bayes Vulnerability of $\Pi_I;C$\,''}~.
}
\end{Reason}

The structure of the argument is basically a reformulation on the $S$ side, an appeal to the separation property of the ``pre-selected'' matrix $D$, and then a complementary un-reformulation on the $I$ side. Thus for ``see below'' we argue as follows.

To prepare $D$ we consider all possible refinements of $\Pi_S$ together.  These refinements $\PSet{R\In \RefMatrices_N}{}{R {\times} \Pi_S}$ comprise a \emph{convex} set of $N{\times}N$ matrices, (where convexity follows from \Eqn{e29374} and linearity of matrix multiplication).
Since $\Pi_I$ is not a refinement of $\Pi_S$, we know $\Pi_I$ is not in that set. If we ``flatten out'' the matrices into vectors of length $N^2$, say by glueing their rows together, then we have a ``point'' $\Pi_I$ in Euclidean space that is strictly outside of that convex set 
and by the \emph{Separating Hyperplane Lemma} \cite{Trustrum:71} there must be a plane 
with normal $X$ that strictly separates that whole set of refinements (including $\widehat{\Pi} = \widehat{R}{\times}\Pi_S$) from the single point $\Pi_I$. The point $X$ too will be a vector of length $N^2$ and, written with matrices, the strict-separation condition is then that
\[
 \Tr.(\widehat{\Pi}{\times}\Transpose{X})
 \Wide{<}
 \Tr.(\Pi_I{\times}\Transpose{X})
 \hspace{3em}
 \textrm{for all $\widehat{\Pi}$ refining $\Pi_S$}
\]
since the dot-product of two $N^2$-vectors $A,B$ written as matrices of size $N{\times}N$ is just $\Tr.(A{\times}\Transpose{B})$. This is precisely what we required above; and so our $D$ is made by taking the direction numbers of the separating hyperplane in Euclidean $N^2$-space and turning them back into a matrix, and transposing the result.

We admit that there is no guarantee that the $D$ constructed as above will have one-summing rows. However, we can choose $D$ to have all non-negative coefficients because $\Pi_I$ and all the refinements $\widehat{\Pi}$ of $\Pi_S$ have the same weight, and thus we can add any constant to all elements of $D$ without affecting its separating property; similarly we can scale it by any positive number. Thus we can assume \Wlog\ that $D$ is non-negative and that all its rows sum to no more than 1. 
To then make each row of $D$ sum to one exactly we can extend it with an extra  ``column zero'' whose entries are chosen just for that purpose. We then need to guarantee --as a technical detail-- that neither the Bayes Vulnerability strategy matrix for $\Pi_S$ or $\Pi_I$ chooses $\Vh$ to be zero. 
We do that, if necessary, by adding a second context program that acts as \Skip\ when $\Vh{\neq}0$; but when $\Vh{=}0$ it executes a large probabilistic choice over $\Vh$ to distribute the 0 value over enough new values $-1,-2\ldots$ to make sure none of them individually will have a large enough probability to attract a maximising choice.\,%
\footnote{At most $N$ new values will be required for such a context.}
That gives us 
\begin{Theorem}{Refinement is complete for Bayes Risk}{t1240}\quad If $S{\NotRef}I$ then $\CC(S){\NotERef}\CC(I)$ for some context $\CC$.
\end{Theorem}

\subsection{Maximal discrimination of the Bayes-Risk elementary order}
In this section only, we write ``$\BRef$'' for the Bayes-Risk based elementary testing order $(\ERef)$, and we write ``$\ORef{1}$'' etc.\ to stand generically for any similar order based on one of the four alternative entropy measures set out in \Sec{s1028}.

The problem discussed in \Sec{s1028} was that one could have $A{\StrictORef{1}}B$ and yet $B{\StrictORef{2}}A$ for programs $A,B$ and competing elementary orders $(\ORef{1})$ and $(\ORef{2})$. Similarly, for any of the four $(\StrictORef{1})$ including $(\StrictBRef)$ itself, it's easy to manufacture examples where we have $A{\StrictORef{1}}B$ but there is a context $\CC$ that reverses the comparison, so that $\CC(B){\StrictORef{1}}\CC(A)$. This seems a hopelessly confused situation.

Luckily it turns out (\App{s3448}) that refinement $(\Ref)$ is sound not only for $(\BRef)$ but for the other three orders as well and --since $(\Ref)$ is complete for Bayes Risk-- that gives us
\begin{Theorem}{Bayes Risk is maximally discriminating}{t1054}
With context, Bayes Risk is maximally discriminating among the orders of \Sec{s1028}: that is if $(\ORef{1})$ is an order derived from one of the entropies of \Sec{s1028}, then whenever for two programs $S,I$ and all contexts $\CC$ we have $\CC(S){\BRef}\CC(I)$ we also have $\CC(S){\ORef{1}}\CC(I)$ for all $\CC$.

Equivalently, if two programs $A,B$ are distinguished by any $(\ORef{1})$ from \Sec{s1028}, that is $A{\NotORef{1}}B$, then there is a context $\CC$ such that $(\BRef)$ in particular distinguishes $\CC(A)$ and $\CC(B)$, that is such that $\CC(A){\NotBRef}\CC(B)$.
\begin{Proof}
The equivalence of the first and second formulations is straightforward;
\footnote{First implies second:
\begin{quote}
If $A{\NotORef{1}}B$ then, appealing to the identity context in the conclusion of the first formulation, for some $\CC$ we have $\CC(A){\NotBRef}\CC(B)$.
\end{quote}
Second implies first:
\begin{quote}
Assume $\CC(S){\NotORef{1}}\CC(I)$ for some $\CC$, whence immediately from the second formulation we have $\DD(\CC(S)){\NotBRef}\DD(\CC(I))$ for some context $\DD(\CC(\cdot))$.
\end{quote}
}
we prove the second, reasoning
\begin{Reason}
\Step{}{
 A{\NotORef{1}}B
}
\StepR{$\Implies$}{soundness of $(\Ref)$ for $(\ORef{1})$, see \App{s3448}}{
 A{\NotRef}B
}
\StepR{$\Implies$}{completeness of $(\Ref)$ for $(\BRef)$; some context $\CC$}{
 \CC(A){\NotBRef}\CC(B) ~.
}
\end{Reason}
\end{Proof}
\end{Theorem}

It's the completeness result for $(\BRef)$ that makes it maximal, i.e.\ that seems to single it out from among the other orders. 
Whether or not the other orders are also complete is an open problem.

\section{Case study: The \emph{Three Judges} protocol}\label{s1727}

The motivation for our case study is to suggest and illustrate techniques for reasoning compositionally from specification to implementation of noninterference \cite{Morgan:07,McIver:09b,PSG}. Our previous examples include (unboundedly many) Dining Cryptographers \cite{Chaum:88}, Oblivious Transfer \cite{Rivest:99} and Multi-Party Shared Computation \cite{Yao:82}. All of them however used our \emph{qualitative} model for compositional noninterference \cite{Morgan:07,McIver:09b}; here of course we are using instead a \emph{quantitative} model.

The example is as follows. Three judges $A,B,C$ are to give a majority decision, innocent or guilty, by exchanging messages but concealing their individual votes $\Va,\Vb$ and $\Vc$, respectively.\,%
\footnote{Though this is similar to the (generalised) Dining Cryptographers, it is more difficult: we do not reveal anonymously the total number of guilty votes; rather we reveal only whether that total is a majority \cite[\texttt{Morgan:09a}]{PSG}.}

We describe this protocol with a program fragment, a specification which captures exactly the functional and security properties we want. Its variables are Boolean, equivalently $\{0,1\}$ and, including some notational conventions explained below, it evaluates $(\Va{+}\Vb{+}\Vc \geq 2)$ atomically, and reveals the value of the expression to everyone:
\begin{NumMini}{e1035}\begin{Reason}
\Step{}{
 \hspace{-1em}
  \Vis_{A}~\Va;~ \Vis_{B}~\Vb;~ \Vis_{C}~\Vc;
  & \hspace{3em}\textrm{$\leftarrow$\small\ These are global variables.} \\
 \Reveal~(\Va{+}\Vb{+}\Vc \geq 2) ~.
  & \hspace{3em}\textrm{$\leftarrow$\small\ Atomically evaluate}\\
  & \hspace{3em}\textrm{\hspace{1em}\small\ and reveal expression.}
}
\end{Reason}\end{NumMini}%
Note that this specification is \emph{not} noninterference-secure in the usual sense: for example when $\Va$ judges ``not guilty'' (\False) and yet the defendant is found guilty by majority, Agent $A$ learns that both $\Vb,\Vc$ must have judged ``guilty '' --- and that is a release of information. This allows a similar behaviour in the implementation, strictly speaking a declassification: but we need no special measures to deal with it.

We interpret the specification as follows. The system comprises four agents: the judges $A,B,C$ and (say) some Agent $X$ as an external observer. The participating agents ($A,B,C$) are distributed, each with its own state-space; and the external observer has no state. The annotations $\Vis_{\{A,B,C\}}$ above indicate that the variables $\Va,\Vb,\Vc$ are located with the agents $A,B,C$ respectively and are visible only to them: that is, only Agent $A$ can see variable $\Va$ etc.\,%
\footnote{In principle we could have separate annotations for visibility and for location, allowing thus variables located at $A$ that however $A$ cannot see, and (complementarily) variables located at $B$ that $A$ can see. But in this example we do not need that fine control, and so we use $\Vis$ for both.}

The $\Reveal$ command (explained in more detail below) publishes its argument for all agents to see.

The \emph{location} of a variable has no direct impact on semantics (in our treatment here); but it does affect our judgement of what is directly executable and what is not. In particular, an expression is said to be \emph{localised} just when all its variables are located at the same agent, and only localised expressions can be directly executed (by that agent, thus). Thus $\Va{+}\Vb{+}\Vc \geq 2$ is not localised, in spite of its being meaningful in the sense of having a well defined value; and it is precisely because it is not localised that we must develop the specification further. Assignment statements $\Va\Gets E$, where $\Va$ is in Agent $A$, say, and $E$ is localised in Agent $B$, are implemented by $B$'s calculating $E$ and then sending its value in a message to $A$.

The \emph{visibility} of a variable does affect semantics. A variable annotated $\Vis_{A}$, for example, is treated as if it were simply annotated $\Vis$ when we are reasoning from Agent $A$'s point of view; from Agents' $B,C$ points of view, it is treated as if it were annotated $\Hid$; and the same applies analogously to the other agents. Thus in the example we will treat three agents $A,B,C$ each with her own view: variables visible to one (declared $\Vis$) will be hidden from another (declared $\Hid$) --- and vice versa. The ``extra'' Agent $X$ (mentioned above) sees none of $\Va,\Vb,\Vc$, but does observe the $\Reveal$. This simple approach is possible for us because we are not dealing with agents whose actions can be influenced by other agents' knowledge.

In principle the \Vis-subscripting convention means that protocol development, e.g.\ as in \Sec{s1324}ff.\ to come, will require a separate proof for each observer (since the patterns of variables' visibility might differ); but in practice we can usually find a single chain of reasoning each of whose steps is valid for two or even all three observers at once.

Before incrementally developing \Eqn{e1035} into an implementation in order to localise its expressions, we introduce some further extensions, including the $\Reveal$ statement mentioned above \cite{McIver:08a}, that will be used in the subsequent program derivation.

\subsection{Further program-language extensions} \label{s9374}
\subsubsection{Multiple- and local variables}
To this point we have had just two variables, visible $\Vv$ and hidden $\Vh$, and a split-state $\VV{\times}\TDist\HH$ to describe their behaviour. In practice each of $\VV,\HH$ will each comprise many variables, represented in the usual Cartesian way. Thus if we have variables $\Va\In{\cal A},\Vb\In{\cal B},\Vc\In{\cal C},\Vd\In{\cal D}$ with the first two $\Va,\Vb$ visible and the last two $\Vc,\Vd$ hidden, then $\VV$ is ${\cal A}{\times}{\cal B}$ and $\HH$ is ${\cal C}{\times}{\cal D}$ so that the state-space is ${\cal A}{\times}{\cal B}{\times}\TDist({\cal C}{\times}{\cal D})$. Assignments and projections are handled as normal.

We allow local variables, both visible and hidden, which are treated (also) as normal: within the scope of a visible local-variable declaration $\Begin~\Vis~\Vx\In{\cal X}\cdots~\End$, the $\VV_{\rm local}$ used is ${\cal X}{\times}\VV_{\rm global}$. Hidden variables are treated similarly.\footnote{Implicitly local variables are assumed to be initialised by a uniform choice over their finite state space. In our examples however, we always initialize local variables explicitly, to avoid confusion.}

\subsubsection{Revelations}\label{s1423}
Command $\Reveal~E$  publishes expression $E$ for all to see: it is equivalent to the local block
\begin{equation}\label{e1253}
 \Begin~\Vis~\Vv;~ \Vv\Gets E~\End ~,
\end{equation}
but it avoids the small extra complexity of declaring the temporary visible-to-all variable $\Vv$ and the having to introduce the scope brackets as \Eqn{e1253} does. The attraction of this is that the $\Reveal$ command has a simple algebra of its own, including for example that $\Reveal~E = \Reveal~F$ just when $E$ and $F$ are interdeducible given the values of (other) visible variables \cite{McIver:08a,McIver:09a}. Thus for example (and slightly more generally) we have
\[
 \Reveal~\Va{\Xor}\Vb;~\Reveal~\Vb{\Xor}\Vc
 \Wide{=}
 \Reveal~\Vb{\Xor}\Vc;~\Reveal~\Vc{\Xor}\Va ~,
\]
using $(\Xor)$ to denote exclusive-or, because from $\Va{\Xor}\Vb$ and $\Vb{\Xor}\Vc$ an observer can deduce both $\Vb{\Xor}\Vc$ and $\Vc{\Xor}\Va$, and vice versa.

\subsubsection{Bulk atomicity}\label{s1044}

In \Fig{f1228} we introduced the semantics of commands and remarked that for syntactically atomic commands the secure semantics is given by \Def{d0855}, based on the classical semantics of the same command. With atomicity brackets $\Atomic{\cdot}$ we make \emph{groups} of commands atomic ``by fiat,'' so that \Def{d0855} applies to them as well.
We have
\begin{Definition}{Secure semantics atomicity brackets}{d0855a}
Given any program $P$ we define
\begin{equation}\label{e1028a}
 \Sem{\Atomic{P}}.(v,\delta) \Wide{\Defs} \Hide.\Exp{h\In \delta~}{~\SemC{P}.(v,h)}~.
\end{equation}
The effect overall, as earlier, is to impose the largest possible ignorance of $h'$ that is consistent with seeing $v'$ and knowing the classical semantics $\SemC{P}$ of the program between brackets. In particular, perfect recall and implicit flow are both suppressed by $\Atomic{\cdot}$.
\end{Definition}
A comparison of Defs.~\ref{d0855}~and~\ref{d0855a} shows immediately that for any syntactically atomic command $A$ we have $A = \Atomic{A}$, just as one would expect.\,%
\footnote{Note that although $\Reveal~E$ looks syntactically atomic, it is via \Eqn{e1253} actually an abbreviation of a compound command: thus in fact $\Atomic{\Reveal~E}\neq\Reveal~E$ in general. Actually $\Atomic{\Reveal~E}=\Skip$ in all cases, whereas $\Reveal~E=\Skip$ only when $E$ is visible.} With groups of commands of course the equality does not hold in general: for example we cannot reason
\begin{Reason}
\Step{}{
 \Vv\Gets \Vh;~\Vv\Gets0
}
\StepR{$=$}{syntactically atomic, both}{
 \Atomic{\Vv\Gets \Vh};~\Atomic{\Vv\Gets0}
}
\StepR{\HangLeft{\textrm{\small?}}$=$}{\emph{invalid step}}{
 \Atomic{\Vv\Gets \Vh;~\Vv\Gets0}
}
\StepR{$=$}{classical equality}{
 \Atomic{\Vv\Gets0}
}
\StepR{$=$}{syntactically atomic}{
 \Vv\Gets 0~,
}
\end{Reason}
because --as we have often stressed-- an assignment of $\Vh$ to $\Vv$  \emph{does} reveal $\Vh$ to an observer, even of $\Vv$ is immediately overwritten. The invalid step violates the conditions of \Lem{l1222} immediately below, which gives an important special situation in which we do have distribution of atomicity inwards:
\begin{Lemma}{Distribution of atomicity}{l1222}
Given is a sequential composition of two programs $P;Q$. If by observation of the visible variable $\Vv$ before the execution of $P$ and after the execution $Q$ it is always possible to determine the value $\Vv$ had \emph{between} $P$ and $Q$, then we do have
\[
 \Atomic{P;Q} \Wide{=} \Atomic{P}; \Atomic{Q} ~.
\]
\begin{Proof} 
(sketch) The full proof is given in \App{a29375}.

It can be shown that the left- and right-hand sides' classical effect on $\Vv$ and $\Vh$ are the same, and so the only possible difference between the two can be the degree to which $\Vh$ is hidden. On the left, variable $\Vh$ must be maximally hidden since that is the (defined) effect of the atomicity brackets $\Atomic{\cdot}$. Thus we need only argue that $\Vh$ is maximally hidden on the right as well.

Since $\Vh$ is maximally hidden after $\Atomic{P}$, the only way $\Vh$ can fail to be maximally hidden after the subsequent $\Atomic{Q}$ is if there are two (or more) distinct values of $\Vv$ after $P$, say $\hat{v}_{{\{0,1\}}}$ each with its associated hidden distribution $\delta_{\{0,1\}}$ of $\Vh$, that are brought together to the \emph{same} final value $v'$ by execution of $Q$. For that would mean that after $Q$ we could have two distinct distributions $\delta'_{\{0,1\}}$ of $\Vh$ associated with that single $v'$, which is precisely what it means not to be maximally hidden. Each $\delta'_{\{0,1\}}$ would have been derived from the corresponding $\delta_{\{0,1\}}$ in between.

That scenario cannot occur if for any particular starting $v$ before $P$ that leads to two (or more) values $\hat{v}_{\{0,1\}}$ between $P$ and $Q$, we never have $Q$ bringing those values back together again to a single final value $v'$. That amounts to being able to determine the intermediate value $\hat{v}$ of $\Vv$ from its values before ($v)$  and after ($v'$).
\end{Proof}
\end{Lemma}

In fact our invalid step $\Atomic{\Vv\Gets \Vh;\Vv\Gets0}\neq\Atomic{\Vv\Gets \Vh};\Atomic{\Vv\Gets0}$ above shows off the condition exactly. Although $\Vv$'s intermediate value is indeed determined by the initial $\Vh$, that is not good enough because we cannot see that $\Vh$: we have access only to the initial $\Vv$. And $\Vv$'s final value is always 0, again hiding $\Vv$'s intermediate value from us. Knowing $\Vv$ before and after, in this example, does not tell us its intermediate value (which is in fact $\Vh$).\,%

By definition, semantic equivalence of $P$ and $Q$ in the classical model entails semantic equivalence of $\Atomic{P}$ and $\Atomic{Q}$ --- that is why \emph{within} atomicity brackets we can use classical equality reasoning.

\subsection{Subprotocols: qualitative vs.\ quantitative reasoning}

Rather than appeal constantly to the basic semantics (\Fig{f1228}) instead we have accumulated, with experience, a repertoire of identities --a program algebra-- which we use to reason at the source level. Those identities themselves are proved directly in the semantics but, after that, they become permanent members of the designer's toolkit. One of the most common is the \emph{Encryption Lemma}.

\subsubsection{The Encryption Lemma}
Let statement $(\Vv{\Xor}\Vh)\Gets E$ set Booleans $\Vv,\Vh$ so that their exclusive-or $\Vv{\Xor}\Vh$ equals Boolean $E$: there are exactly two possible ways of doing so. In our earlier work \cite{Morgan:07}, we proved that when the choice is made demonically, on a single run nothing is revealed about $E$; in our refinement style we express that as
\begin{equation}\label{e1105}
\Skip
\Wide{=}
\Begin~\Vis~\Vv; \Hid~\Vh;~(\Vv{\Xor}\Vh)\Gets E ~\End ~.
\end{equation}
In our current model we can prove that \emph{exactly the same identity holds} provided the choice of possible values for $\Vv$ and $\Vh$ is made uniformly:
\begin{Lemma}{The Encryption Lemma}{l1335}
For any Boolean expression $E$ we have that the following block is equal to \Skip, and so reveals nothing:
\[
 \Begin~\Vis~\Vv;\Hid~\Vh;~ (\Vv{\Xor}\Vh)\Gets E~\End~.
\]
For this we require that the implicit choice in $(\Vv{\Xor}\Vh)\Gets E$ is made uniformly.
\begin{Proof} We calculate
\begin{Reason}
\Step{}{
 \Begin~\Vis~\Vv;\Hid~\Vh;~ (\Vv{\Xor}\Vh)\Gets E~\End
}
\StepR{$=$}{syntactically atomic}{
 \Begin~\Vis~\Vv;\Hid~\Vh;~ \Atomic{(\Vv{\Xor}\Vh)\Gets E}~\End
}
\StepR{$=$}{classical equality~($\dagger$)}{
 \Begin~\Vis~\Vv;\Hid~\Vh;~ \Atomic{\Vv\Gets\True{\PC{}}\False; \Vh\Gets \Vv{\Xor}E}~\End
}
\StepR{$=$}{\Lem{l1222}}{
 \Begin~\Vis~\Vv;\Hid~\Vh;~ \Atomic{\Vv\Gets\True{\PC{}}\False}; \Atomic{\Vh\Gets \Vv{\Xor}E}~\End
}
\StepR{$=$}{syntactically atomic}{
 \Begin~\Vis~\Vv;\Hid~\Vh;~ \Vv\Gets\True{\PC{}}\False; \Vh\Gets \Vv{\Xor}E~\End
}
\StepR{$=$}{$\Hid~\Vh$ does not capture}{
 \Begin~\Vis~\Vv;~\Vv\Gets\True{\PC{}}\False;~\Begin~\Hid~\Vh;~ \Vh\Gets \Vv{\Xor}E~\End~\End
}
\StepR{$=$}{assignment to local hidden is \Skip}{
 \Begin~\Vis~\Vv;~\Vv\Gets\True{\PC{}}\False~\End
}
\StepR{$=$}{value assigned to local $\Vv$ is known already ($\ddagger$)}{
 \Skip~.
}
\end{Reason}
\end{Proof}
\end{Lemma}

The crucial step in the proof above was the classical equality at ($\dagger$), and we note that other variations are possible: for example we also have the classical equality
\begin{equation}\label{e1359}
 (\Vv{\Xor}\Vh)\Gets E
 \WIDE{=}
 \Vh\Gets\True{\PC{}}\False;~\Vv\Gets E{\Xor}\Vh ~,
\end{equation}
which suggests the operational procedure of ``flipping a private coin $\Vh$\,'' and then revealing (via assignment to local $\Vv$) the exclusive-or of that private coin with some expression $E$. The above reasoning shows that also to be equal to \Skip.

Finally, we recall that $(\PC{})$ means choose \emph{uniformly}, and we now show that it is essential for the ($\dag$) step and for the equality \Eqn{e1359}: if for example we had $\Vh\Gets\True\PC{p}\False;~\Vv\Gets E{\Xor}\Vh$ in \Eqn{e1359} on the right-hand side, but with $p{\neq}1/2$, it would not be possible to rewrite that in the form $\Vv\Gets\True\PC{1/2}\False; \Vh\Gets \Vv{\Xor}E$ as we had at ($\dagger$) but with the $1/2$ here exposed. Instead we'd have
\[
 \Vv\Gets\neg E{\PC{p}}E;~~\Vh\Gets \Vv{\Xor}E
\]
and the last step ($\ddagger$) would then be invalid if $E$ contained hidden variables (as it usually would). The role of $p{=}1/2$ is thus that the equality
\[
 \Vv\Gets\neg E\PC{1/2}E \WIDE{=} \Vv\Gets\True\PC{1/2}\False
\]
holds no matter what expression $E$ is, and in particular even if it contains hidden variables --- but only (in general) when the choice is with probability 1/2.

\Lem{l1335} means that extant q\underline{ual}itative source-level proofs that rely only on ``upgradeable identities'' like \Eqn{e1105} can be used \emph{as is} for q\underline{uant}itative results provided the demonic choices involved are converted to uniform choice. And that is the case with our current example.

Beyond the Encryption Lemma, we use \emph{Two-party Conjunction} \cite{Yao:82} and \emph{Oblivious Transfer} \cite{Rivest:99} in our implementation.
Just as for the Encryption Lemma, the algebraic proofs of their implementations \cite{Morgan:07,McIver:09b} apply quantitatively provided we interpret the (formerly) demonic choice as uniform. We now look briefly at those subprotocols.

\subsubsection{Two-Party Conjunction} \label{s1223}
In the Two-Party Conjunction subprotocol,  the conjunction of two privately held Booleans is published without revealing either Boolean separately. It is an instance of Yao's \emph{Multi-party Computation} technique \cite{Yao:82} and we have given a formal derivation of it elsewhere \cite{McIver:09b}. Its specification is
\begin{NumMini}{e1058}\begin{Reason}
\Step{}{
 \hspace{-1em}\textit{Two-Party Conjunction} \\[0.5em]
 \hspace{-1em}\Vis_{B}~\Vb; \Vis_{C}~\Vc;
  & \hspace{3em}\textrm{$\leftarrow$\small\ These are global variables.} \\
 \Reveal~\Vb\land \Vc~,
}
\end{Reason}\end{NumMini}%
and its similarity to \Eqn{e1035} is clear: a compound outcome $\Vb\land \Vc$ is published without revealing the components $\Vb,\Vc$ --- except that, just as before, if for example the revealed outcome is \False\ but $\Vb$ is \True, then $B$ can deduce that $\Vc$ must have been \False\ (and similar).

We develop an implementation of \Eqn{e1058} in several steps, as follows. Note that for some steps the \emph{justification} varies depending on the agent although we have arranged that the claimed equality is valid for all of them. We have
\begin{Reason}
\StepR{\HangLeft{\Eqn{e1058}} $=$}{identity}{
 \Skip; \\
 \Reveal~\Vb\land \Vc
}
\Space
\StepR{$=$}{Encryption Lemma for $A,C$; \\ obvious for $B$; see below ($\ddagger$).}{
 \Begin~\Vis_{B}~\Vb_{0},\Vb_{1};~ (\Vb_{0}{\Xor}\Vb_{1})\Gets \Vb;~ \Reveal~\Vb_{0} \End; \\
 \Reveal~\Vb\land \Vc
}
\Space
%
%
\StepR{$=$}{Revelation algebra: in this context $\Vb\land \Vc \equiv \Vb_{0}{\Xor}\Vb_{\Vc}$\HangRight{\quad$\dagger$} \\where $\Vb_{\Vc} \Defs (\Vb_1 ~\If~ \Vc ~\Else ~\Vb_0)$.}{
 \InQuadL{\Begin}\Vis_{B}~\Vb_{0},\Vb_{1}; \\
 \quad (\Vb_{0}{\Xor}\Vb_{1})\Gets \Vb;~ \Reveal~\Vb_{0}; \\
 \quad \Reveal~\Vb_{\Vc} \\
 \End
}
\Space
\StepR{$=$}{Delegate second revelation to Agent $C$.}{
 \InQuadL{\Begin}\Vis_{B}~\Vb_{0},\Vb_{1}; \\
 \quad (\Vb_{0}{\Xor}\Vb_{1})\Gets \Vb;~ \Reveal~\Vb_{0}; \\
 \quad\InQuadL{\Begin}\Vis_{C}~\Vc_{0};~ \Vc_{0}\Gets \Vb_{\Vc};~ \Reveal~\Vc_{0}~\End \\
 \End
}
\Space
\StepR{$=$}{Rearrange declarations; clean up.}{
 \InQuadL{\Begin}\Vis_{B}~\Vb_{0},\Vb_{1};~ \Vis_{C}~\Vc_{0}; \\[0.5em]
 \quad (\Vb_{0}{\Xor}\Vb_{1})\Gets \Vb;~\Reveal~\Vb_{0};
  & \hspace{1em}\textrm{$\leftarrow$\small\ This done by Agent $B$.} \\
 \quad \Vc_{0}\Gets \Vb_{\Vc};
  & \hspace{1em}\textrm{$\leftarrow$\small\ An ``Oblivious Transfer'' between $B,C$.} \\
 \quad\Reveal~\Vc_{0}
  & \hspace{1em}\textrm{$\leftarrow$\small\ This done by Agent $C$.} \\
 \End~.
}
\end{Reason}
At ($\ddagger$) we find a case where the same equality applies to all agents, although in fact the reasons for its validity use agent-specific reasoning. For example, for Agents $A,C$ the fragment is effectively
\[
 \Begin~\Hid~\Vb_{0},\Vb_{1};~ (\Vb_{0}{\Xor}\Vb_{1})\Gets \Vb;~ \Reveal~\Vb_{0} \End~,
\]
which is a version of the Encryption Lemma in which $\Vb_{0}$, being revealed, takes the role of the local visible variable. Variable $\Vb_{1}$ is the local hidden, and (hidden) variable $\Vb$ is the expression $E$ (on which there are no restrictions). On the other hand, for Agent $B$ the fragment is
\[
 \Begin~\Vis~\Vb_{0},\Vb_{1};~ (\Vb_{0}{\Xor}\Vb_{1})\Gets \Vb;~ \Reveal~\Vb_{0} \End
\]
with $\Vb$ a global visible: this is trivially equal to \Skip\ because all variables are visible.

At ($\dagger$) we use the revelation algebra mentioned in \Sec{s1423} to reason that once $\Vb_{0}$ is revealed, going on to reveal $\Vb\land \Vc$ is equivalent to revealing just $\Vb_{\Vc}$ since --knowing $\Vb_{0}$-- we can calculate each of $\Vb\land \Vc$ and $\Vb_{\Vc}$ from the other.\,%
\footnote{We have
\[
 \Vb_{0}{\Xor}\Vb_{\Vc}
 ~\equiv~
 (\Vb_{0}{\Xor}\Vb_{1}~\If~\Vc~\Else~\Vb_{0}{\Xor}\Vb_{0})
 ~\equiv~
 (\Vb~\If~\Vc~\Else~\False)
 ~\equiv~
 \Vb\land \Vc ~.
\]
}

More interesting than any of that, however, is that in the last step we appeal to a further subprotocol by including the \emph{specification} of the ``Oblivious Transfer Protocol'' \cite{Rabin:81,Rivest:99}. Here Agent $C$ has a private $\Set{}{}{0,1}$-valued variable $\Vc$ and obtains from Agent $B$ either $\Vb_{0}$ or $\Vb_{1}$, depending on $\Vc$. But Agent $B$ does not discover what $\Vc$ is, and Agent $C$ does not discover $\Vb_{\neg \Vc}$. We give a rigorous implementation of the protocol elsewhere \cite{Morgan:07}; an informal explanation may be found in \App{a1446}.

Finally, to emphasise our earlier point about declassification, we suppose $\Vb$ is \True\ but $\Vc$ is \False\ and thus that $B$ learns $\Vc$ by noting that \False\ is revealed overall; note that this is a property of the \emph{specification}. Now, in the implementation, we can see how this happens: when $\Vb$ is true the local variables $\Vb_{0,1}$ will be complementary and so --in spite of not learning $\Vc$ while the Oblivious Transfer is carried out-- Agent $B$ will still learn $\Vc$ afterwards by comparing $\Vc_{0}$ with her own $\Vb_{0,1}$.

\subsection{The Three-Judges implementation: first attempt}\label{s1324}

We begin with an implementation attempt that fails, because this will illustrate two things. The first is that our model prevents incorrect developments, that is it stops us from constructing implementations less secure than their specifications: arguably this ``negative'' aspect of a method is its most important property, since it would be trivial to describe a method that allowed secure refinements\ldots\ and all others as well. The key is what is \emph{not} allowed.

The second thing illustrated here is that a conditional $\If~E~\cdots~\Fi$ should be considered to reveal its condition $E$ implicitly. This \emph{implicit flow} is a property forced upon us by our advocacy of program algebra and our use of compositionality: since the \Then- and the \Else\ branch of a conditional can be developed differently \emph{after} the conditional has been introduced, we must expect that those differences might reveal to an attacker which branch is being executed (and hence the condition implicitly). This is exactly what we are about to see.

We start with some Boolean algebra
\[
 (\Va{+}\Vb{+}\Vc \geq 2)
 \Wide{\equiv}
 \Va\land(\Vb\lor \Vc) ~\lor~ \Vb\land \Vc
 \Wide{\equiv}
 (\Vb\lor \Vc ~\If~\Va~\Else~ \Vb\land \Vc) ~,
\]
and that suggests the first development steps
\begin{NumMini}{e1303}\begin{Reason}
\Step{}{
 \Reveal~(\Va{+}\Vb{+}\Vc \geq 2)
}
\Space
\WideStepR{$\stackrel{\textrm{?}}{=}$}{$P \stackrel{\textrm{?}}{=} \If~E~\Then~P~\Else~P~\Fi$}{
 \makebox[4.1em][l]{$\If~\Va~\Then$}~ \Reveal~ (\Va{+}\Vb{+}\Vc \geq 2) \\
 \makebox[4.1em][l]{\hfill\Else}~ \Reveal~ (\Va{+}\Vb{+}\Vc \geq 2) \\
 \Fi
}
\Space
\StepR{$=$}{After \Then\ we can assume $a$; \\ after \Else\ we can assume $\neg a$.}{
 \If~\Va \\
 \quad\makebox[2.3em][l]{\Then}~ \Reveal~ \Vb\lor \Vc \\
 \quad\makebox[2.3em][l]{\Else}~ \Reveal~ \Vb\land \Vc \\
 \Fi\ldots
}
\end{Reason}\end{NumMini}

Now we can deal immediately with the \Else-part by adapting the Two-Party Conjunction Protocol of \Sec{s1223} so that it reveals $\Vb\land \Vc$ only to Agent $A$; we introduce a pair of $A$-private local variables for that purpose. The result is
\begin{Reason}
\StepR{\HangLeft{\ldots}$=$}{Adapting \Sec{s1223}}{
 \multicolumn{2}{@{}l}{\InQuadL{\Begin} \Vis_{A}~\Va_{B},\Va_{C};~\Vis_{B}~\Vb_{0},\Vb_{1};~\Vis_{C}~\Vc_{0};} \\[0.5em]
 \quad\If~\Va \\
 \quad\quad\makebox[2.3em][l]{\Then}~ \Reveal~ \Vb\lor \Vc \\
 \quad\quad\makebox[2.3em][l]{\Else}
                          (\Vb_{0}{\Xor}\Vb_{1})\Gets \Vb;
                           & \hspace{1em}\textrm{$\leftarrow$\small\ Done privately by Agent $B$.} \\
 \quad\quad\hspace{2.3em} \Va_{B}\Gets \Vb_{0};
                           & \hspace{1em}\textrm{$\leftarrow$\small\ Message $B{\rightarrow}A$.} \\
 \quad\quad\hspace{2.3em} \Vc_{0}\Gets \Vb_{\Vc};
                           & \hspace{1em}\textrm{$\leftarrow$\small\ Oblivious Transfer $B{\rightarrow}C$.} \\
 \quad\quad\hspace{2.3em} \Va_{C}\Gets \Vc_{0};
                           & \hspace{1em}\textrm{$\leftarrow$\small\ Message $C{\rightarrow}A$.} \\
 \quad\quad\hspace{2.3em} \Reveal~\Va_{B}{\Xor}\Va_{C}
                           & \hspace{1em}\textrm{$\leftarrow$\small\ Agent $A$ announces majority verdict.} \\
 \quad\Fi \\
 \End\ldots
}
\end{Reason}

For the \Then-part we write $\Vb\lor \Vc$ as $\neg(\neg \Vb \land \neg \Vc)$ and adapt the \Else-part accordingly; the effect overall turns out to be replacing the initial $\Vb$ by $\neg \Vb$ and changing the following assignment. Once we factor out the common portion of the conditional, we have
\begin{Reason}
\StepR{\HangLeft{\ldots}$=$}{Using \emph{de Morgan}}{
 \multicolumn{2}{@{}l}{\InQuadL{\Begin} \Vis_{A}~\Va_{B},\Va_{C};~\Vis_{B}~\Vb_{0},\Vb_{1};~\Vis_{C}~\Vc_{0};} \\[1em]
 \quad\If~\Va~\Then~(\Vb_{0}{\Xor}\Vb_{1})\Gets \neg \Vb; \Va_{B}\Gets \Vb_{1}
             ~\Else~(\Vb_{0}{\Xor}\Vb_{1})\Gets \Vb; \Va_{B}\Gets \Vb_{0}~\Fi; \\
 \quad \Vc_{0}\Gets \Vb_{\Vc};~ \Va_{C}\Gets \Vc_{0}; \\
 \quad \Reveal~\Va_{B}{\Xor}\Va_{C} \\
 \End ~.
}
\end{Reason}

Now we see that the problem with going further is that Agent $A$ must somehow arrange that $B$ carries out either $(\Vb_{0}{\Xor}\Vb_{1})\Gets \neg \Vb$ or $(\Vb_{0}{\Xor}\Vb_{1})\Gets \Vb$, with that arrangement depending on the value of $A$'s private variable $\Va$. Since $B$'s two potential computations are different, there is no way this can occur without $B$'s learning the value of $\Va$ in the process: this code is already incorrect.

Thus we must abandon this attempt, and admit that the questionable step at \Eqn{e1303} above was indeed wrong.
In order to allow us to develop distributed implementations, we make the (reasonable) assumption that each agent knows the code it is instructed to execute, with those instructions coming possibly from another agent. In this case Agent $B$ must execute either $(\Vb_{0}{\Xor}\Vb_{1})\Gets \neg \Vb; \Va_{B}\Gets \Vb_{1}$ or $(\Vb_{0}{\Xor}\Vb_{1})\Gets \Vb; \Va_{B}\Gets \Vb_{0}$, depending on the value of $a$ which is supposed to be visible only to Agent $A$.

Our semantics recognises implicit flow, and does not allow in general the transformation of $P$ into $\If~E~\Then~P~\Else~P~\Fi$, for exactly this reason.\,%
\footnote{In our related work for noninterference with demonic choice and without probability \cite{Morgan:06,Morgan:07}, we give further arguments for this point of view, but based directly on program algebra. The extra feature there is that even \emph{classical} programs have a non-trivial refinement relation; here, we have proper refinement only for secure programs.}
Similarly, a fragment $\Va\Gets \Vb; \Va\Gets \Vc$ represents two messages, one $B{\rightarrow}A$ and then a second one $C{\rightarrow}A$; with perfect recall we recognise that $A$ can learn $\Vb$ by examining $\Va$ after the first message has arrived, but before the second.

\subsection{The Three-Judges implementation: second attempt (sketch)}\label{s1203}

An ``obvious'' remedy for \Sec{s1324}'s problem, that Agent $B$'s is aware of which procedure she must follow, is to make $B$ follow \emph{both} procedures, speculatively: she does not know which one $A$ will actually use.

The difficulty is now with Agent $A$, who learns both $\Vb\land \Vc$ and $\Vb\lor \Vc$. Although those two values do not (always) determine $\Vb$ and $\Vc$ themselves, they do provide strictly more information to $A$ than her knowing $\Va$ and $(\Va{+}\Vb{+}\Vc \geq 2)$ would have provided on their own.\,%
\footnote{If $\Va$ and $(\Va{+}\Vb{+}\Vc \geq 2)$ are both $\False$, then Agent $A$ concludes $\neg(\Vb{\land}\Vc)$, for which there are the three possibilities $\False/\False, \True/\False,\False/\True$. Agent $A$'s additionally knowing $\Vb\lor \Vc$ would eliminate at least one of those three.}
Thus this approach fails also.

Our attention is therefore drawn to arranging for $B$ (and $C$) to do both two-party calculations, but then for $A$ to get the results of only one of them. That leads naturally to the approach of the next section, a combination of two two-party computations (letting Agents $B,C$ do both calculations) and two (more) oblivious transfers (letting Agent $A$ learning about only one of them.)\,%
\footnote{The ``more'' refers to the fact that the two-party computations have oblivious transfers inside of them.}

\subsection{The Three-Judges implementation: successful development}\label{s1323}

To repair the problem we encountered above we must arrange that Agents $B,C$ as far as possible carry out procedures independent of $A$'s variable $\Va$, in particular so that calculations relating to $\Vb\lor \Vc$ and to $\Vb\land \Vc$ both occur, irrespective of which result $A$ actually needs.

To achieve this we need a slightly more general form of two-party computation. We begin by introducing the specification of such a two-party conjunction, with its variables made local so that the introduced code is equivalent to \Skip:
\begin{Reason}
\Step{}{
 \Reveal~(\Va{+}\Vb{+}\Vc \geq 2)
}
\Space
\StepR{$=$}{Two-party conjunction}{
 \Begin~\Vis_{B}~\Vb_{0}; \Vis_{C}~\Vc_{0};~ (\Vb_{0}{\Xor}\Vc_{0})\Gets \Vb\land \Vc; ~\End; \\
 \Reveal~(\Va{+}\Vb{+}\Vc \geq 2)
}
\end{Reason}
From Agent $A$'s point of view, the introduced statement is trivially equivalent to \Skip: all assignments are to local variables that $A$ cannot see. From Agent $B,C$'s points of view, it is equivalent to \Skip\ because it is an instance of the Encryption Lemma: each of those two agents can see only one of the two variables assigned-to, and so learns nothing about the expression $\Vb\land \Vc$.\,%
\footnote{The following bogus counter-argument is an example of what having a careful definition of equality and refinement helps us to avoid.

``Agent $B$ might know that $\Vb$ is \False, and then perhaps receive \False\ also in $\Vb_{\land}$. She concludes that $\Vc_{\land}$ is also \False, which is a leak since $\Vc_{\land}$ is supposed to be private to $C$, invisible to $B$.''

In fact this is not a leak, because to judge it so we must refer to the \emph{specification} of this fragment. But that is simply $\Skip$ and there is no $\Vc_{\land}$ declared there: the revealed variable is local to the implementation only.

That is, publishing the value of a hidden variable declared only in the implementation might \emph{look} like a leak in the conventional interpretation --consider $\Begin~\Vis_{C}~\Vc;~ \cdots; \Reveal~\Vc~\End$ for example-- but it is actually a leak only if that variable $\Vc$ has come to contain information (via assignments say in the ``$\cdots$'' portion) about other, more global hidden variables that were present in the specification, originally. Our semantics checks for that automatically.}

The statement $(\Vb_{0}{\Xor}\Vc_{0})\Gets \Vb\land \Vc$ we have introduced is a more general form of two-party conjunction than the $\Reveal~\Vb\land \Vc$ we illustrated earlier in \Sec{s1223} --- that is because the conjunction is not actually revealed, not yet; instead it is split into two ``shares,'' one belonging to each party $B,C$. Since each party has only one share, the conjunction is not revealed at all at this stage. But those shares can be used as inputs to further two-party computations, while preserving the security, and the contribution of the conjunction to a larger computation is revealed at a later point.

The extra generality introduced by the shares does not cause us extra work here, since we are using only the specification for our reasoning and that (we will see) suffices. When we come to \emph{implement} the general two-party conjunction in more primitive terms, however, we would then have further work to do. We have given such an implementation elsewhere \cite{McIver:09b}.

With exactly the same reasoning as above we can introduce two-party disjunction and, with both conjunction and disjunction present, perform some reorganisation:
\begin{Reason}
\StepR{\HangLeft{\ldots}$=$}{Two-party disjunction}{
 \Begin~\Vis_{B}~\Vb_{0}; \Vis_{C}~\Vc_{0};~ (\Vb_{0}{\Xor}\Vc_{0})\Gets \Vb\land \Vc; ~\End; \\
 \Begin~\Vis_{B}~\Vb_{1}; \Vis_{C}~\Vc_{1};~ (\Vb_{1}{\Xor}\Vc_{1})\Gets \Vb\lor \Vc; ~\End; \\
 \Reveal~(\Va{+}\Vb{+}\Vc \geq 2)
}
\Space
\StepR{$=$}{Reorganise declarations and scoping}{
 \InQuadL{\Begin} \Vis_{B}~\Vb_{0},\Vb_{1}; \Vis_{C}~\Vc_{0}, \Vc_{1}; \\[.5em]
 \quad (\Vb_{0}{\Xor}\Vc_{0})\Gets \Vb\land \Vc; \\
 \quad (\Vb_{1}{\Xor}\Vc_{1})\Gets \Vb\lor \Vc; \\
 \quad \Reveal~(\Va{+}\Vb{+}\Vc \geq 2) \\
 \End
}
\Space
\StepR{$=$}{Boolean algebra}{
 \InQuadL{\Begin} \Vis_{B}~\Vb_{0},\Vb_{1}; \Vis_{C}~\Vc_{0}, \Vc_{1}; \\[.5em]
 \quad (\Vb_{0}{\Xor}\Vc_{0})\Gets \Vb\land \Vc; \\
 \quad (\Vb_{1}{\Xor}\Vc_{1})\Gets \Vb\lor \Vc; \\
 \quad \Reveal~\Vb_a{\Xor}\Vc_a \\
 \End\ldots
}
\end{Reason}

Now since $\Vb_a{\Xor}\Vc_a$ is revealed to everyone, and thus to $A$ in particular, it does no harm first to capture that value in variables of $A$, and then to have Agent $A$ reveal those instead:
\begin{Reason}
\StepR{\HangLeft{\ldots}$=$}{Introduce local private variables of $A$}{
 \InQuadL{\Begin} \Vis_{A}~\Va_{B}, \Va_{C}; \\
 \quad \Vis_{B}~\Vb_{0},\Vb_{1}; \\
 \quad \Vis_{C}~\Vc_{0}, \Vc_{1}; \\[.5em]
 \quad (\Vb_{0}{\Xor}\Vc_{0})\Gets \Vb\land \Vc; \\
 \quad (\Vb_{1}{\Xor}\Vc_{1})\Gets \Vb\lor \Vc; \\
 \quad (\Va_{B}{\Xor}\Va_{C})\Gets \Vb_a{\Xor}\Vc_a; \\
 \quad \Reveal~\Va_{B}{\Xor}\Va_{C} \\
 \End\ldots
}
\end{Reason}
The point of using two variables $\Va_{\{B,C\}}$ rather than one is to be able to split the transmission of information $B,C{\rightarrow}A$ into two separate oblivious transfers $B{\rightarrow}A$ and $C{\rightarrow}A$.

Thus the protocol boils down to three two-party computations: a conjunction $\Vb\land \Vc$, a disjunction $\Vb\lor \Vc$ and an exclusive-or $\Vb_a{\Xor}\Vc_a$. The \RHS\ of the last is actually within an atomic conditional on $\Va$, that is $\ITE{(\Vb_0{\Xor}\Vc_0)}{\Va{=}0}{(\Vb_1{\Xor}\Vc_1)}$.

\subsection{Two-party exclusive-or}\label{s1045}

Our final step is to split the two-party exclusive-or into two separate assignments. This is achieved by introducing a local shared variable $\Vh$ that is visible to $B,C$ only, i.e.\ not to $A$, and encrypting both hidden variables with it. Thus we take the step
\begin{Reason}
\Step{}{
 (\Va_{B}{\Xor}\Va_{C})\Gets \Vb_{\Va}{\Xor}\Vc_{\Va}
}
\Space
\Step{$=$}{
  \InQuadL{\Begin}\Vis_{B,C}~\Vh; \\
  \quad \Vh\Gets\True\PC{}\False; \\
  \quad \Va_B\Gets \Vb_a {\Xor}\Vh; \\
  \quad \Va_C\Gets \Vc_a {\Xor}\Vh\\
  \End ~,
}
\end{Reason}
justified trivially for $B,C$ since the only assignments of non-constants are to variables visible only to $A$. For $A$ the justification comes from the the use of classical equality reasoning within a temporary atomicity block (refer \Lem{l1222}): the effect of the two fragments above on $\Va_B,\Va_C$ is identical, and there are no overwritten visible values.

We will now show that in fact the extra variable $\Vh$ is not necessary: by absorbing it into earlier statements, and with some rearrangement of scopes we can rewrite our code at the end of \Sec{s1323} as
\begin{Reason}
\Step{\HangLeft{\ldots}$=$}{
 \InQuadL{\Begin} \Vis_{A}~\Va_{B}, \Va_{C}; ~\Vis_{B}~\Vb_{0},\Vb_{1}; ~\Vis_{C}~\Vc_{0}, \Vc_{1}; ~\Vis_{B,C}~ \Vh; \\[.5em]
 \quad \Vh\Gets \True\PC{}\False; \\
 \quad (\Vb_{0}{\Xor}\Vc_{0})\Gets \Vb\land \Vc; \\
 \quad (\Vb_{1}{\Xor}\Vc_{1})\Gets \Vb\lor \Vc; \\[0.5em]
 \quad \Va_B\Gets \Vb_a{\Xor}\Vh; \\
 \quad \Va_C\Gets \Vc_a{\Xor}\Vh; \\
 \quad \Reveal~\Va_{B}{\Xor}\Va_{C} \\
 \End\ldots
}
\end{Reason}
where in fact we have moved the declaration and initialisation of $\Vh$ right to the beginning. We now absorb it into the earlier two-party computations by introducing temporarily variables $\Vb'_{\{0,1\}}$ and $\Vc'_{\{0,1\}}$ which correspond to their unprimed versions except that they, too, are encrypted with $\Vh$. That gives
\begin{Reason}
\StepR{\HangLeft{\ldots}$=$}{Boolean reasoning}{
 \InQuadL{\Begin} \Vis_{A}~\Va_{B}, \Va_{C}; \\
 \quad \Vis_{B}~\Vb_{0},\Vb_{1},\Vb'_{0},\Vb'_{1}; \\
 \quad \Vis_{C}~\Vc_{0}, \Vc_{1},\Vc'_{0}, \Vc'_{1}; \\
 \quad \Vis_{B,C}~ \Vh; \\[.5em]
 \quad \Vh\Gets \True\PC{}\False; \\
 \multicolumn{2}{l}{\quad (\Vb_{0}{\Xor}\Vc_{0})\Gets \Vb\land \Vc;\quad \Vb'_0,\Vc'_0\Gets \Vb_0{\Xor}\Vh, \Vc_0{\Xor}\Vh;} \\
  \multicolumn{2}{l}{\quad (\Vb_{1}{\Xor}\Vc_{1})\Gets \Vb\lor \Vc;\quad \Vb'_1,\Vc'_1\Gets \Vb_1{\Xor}\Vh, \Vc_1{\Xor}\Vh;} \\[0.5em]
 \quad \Va_B\Gets \Vb'_a;
   & \hspace{1em}\textrm{$\leftarrow$\small\ Note primes, justified by earlier.} \\
 \quad \Va_C\Gets \Vc'_a;
   & \hspace{1em}\textrm{$\leftarrow$\small\ Assignments to $\Vb'_{\{0,1\}},\Vc'_{\{0,1\}}$.} \\
 \quad \Reveal~\Va_{B}{\Xor}\Va_{C}~\End\ldots
}
\end{Reason}
where we have replaced the $\Vb_a{\Xor}\Vh$ and $\Vc_a{\Xor}\Vh$ at the end of the code with their simpler, primed versions where the encryption in built-in. Now we can rearrange the statements using $\Vh$ so that not only $\Vh$ but also the unprimed $\Vb_{\{0,1\}}$ and $\Vc_{\{0,1\}}$ become auxiliary; that is, we have for the conjunction
\begin{Reason}
\Step{}{
  (\Vb_{0}{\Xor}\Vc_{0})\Gets \Vb\land \Vc; ~~
   \Vb'_0,\Vc'_0\Gets \Vb_0{\Xor}\Vh, \Vc_0{\Xor}\Vh
}
\StepR{$=$}{introduce atomicity}{
  \InQuadL{\AtomicOpen}(\Vb_{0}{\Xor}\Vc_{0})\Gets \Vb\land \Vc; ~~
   \Vb'_0,\Vc'_0\Gets \Vb_0{\Xor}\Vh, \Vc_0{\Xor}\Vh ~\AtomicClose
}
\StepR{$=$}{classical equality}{
  \InQuadL{\AtomicOpen}(\Vb'_{0}{\Xor}\Vc'_{0})\Gets \Vb\land \Vc; ~~
   \Vb_0,\Vc_0\Gets \Vb'_0{\Xor}\Vh, \Vc'_0{\Xor}\Vh ~\AtomicClose
}
\StepR{$=$}{remove atomicity}{
  (\Vb'_{0}{\Xor}\Vc'_{0})\Gets \Vb\land \Vc; ~~
  \Vb_0,\Vc_0\Gets \Vb'_0{\Xor}\Vh, \Vc'_0{\Xor}\Vh ~,
}
\end{Reason}
and similarly for the disjunction. Removing the auxiliaries, and then applying a trivial renaming to get rid of the primes, we end up with
\begin{Reason}
\StepR{\HangLeft{\ldots}$=$}{Consolidating the above}{
 \multicolumn{2}{l}{\InQuadL{\Begin} \Vis_{A}~\Va_{B}, \Va_{C};~ \Vis_{B}~\Vb_{0},\Vb_{1};~ \Vis_{C}~\Vc_{0}, \Vc_{1};} \\[.5em]
 \quad (\Vb_{0}{\Xor}\Vc_{0})\Gets \Vb\land \Vc;
  & \hspace{1em}\textrm{$\leftarrow$\small\ Two-Party Conjunction (contains Oblivious Transfer).} \\
 \quad (\Vb_{1}{\Xor}\Vc_{1})\Gets \Vb\lor \Vc;
  & \hspace{1em}\textrm{$\leftarrow$\small\ Two-Party Disjunction (contains Oblivious Transfer).} \\
 \quad \Va_B\Gets \Vb_a;
  & \hspace{1em}\textrm{$\leftarrow$\small\ Oblivious Transfer.} \\
 \quad \Va_C\Gets \Vc_a;
  & \hspace{1em}\textrm{$\leftarrow$\small\ Oblivious Transfer.} \\[0.5em]
 \quad \Reveal~\Va_{B}{\Xor}\Va_{C}~\End
}
\end{Reason}
which is precisely what we sought.

In \Fig{f1640} we give the code with the (two) two-party computations instantiated. In \Fig{f1641} we instantiate one of the (four) oblivious transfers.
\begin{Figure}{f1640}
\[
 \begin{array}{l}
  \InQuadL{\Begin} \Vis_{A}~\Va_{B}, \Va_{C};~ \Vis_{B}~\Vb_{0},\Vb_{1};~ \Vis_{C}~\Vc_{0}, \Vc_{1}; \\
  \quad \Vb_0\Gets\True\PC{}\False; \\
  \quad \Vb_1\Gets\True\PC{}\False; \\\\
  \quad\left.\begin{array}{@{}ll}
             \Vc_0\Gets& (\Vb{\Xor}\Vb_0~\If~\Vc~\Else~\Vb_0); \\
             \Vc_1\Gets& (\neg \Vb_1~\If~\Vc~\Else~\Vb{\Xor}\Vb_1); \\
             \Va_B\Gets& (\Vb_1~\If~\Va~\Else~\Vb_0); \\
             \Va_C\Gets& (\Vc_1~\If~\Va~\Else~\Vc_0);
            \end{array}
            ~~\right\}\textit{Four oblivious transfers.}
            \\\\
  \quad \Reveal~\Va_{B}{\Xor}\Va_{C}~\End
 \end{array}
\]

We replace the two Two-Party 'junctions by their implementations as oblivious transfers: each becomes two statements instead of one. The random flipping of bits $\Vb_{\{0,1\}}$ is then collected at the start.

\medskip
The preservation of correctness is guaranteed by the compositionality of the security semantics.
\caption{Three-Judges Protocol assuming Oblivious Transfers as primitives}
\end{Figure}
\begin{Figure}[ht!]{f1641}
Starting from \Fig{f1640}, we replace the specification of the first of its four oblivious transfers $\Vc_0\Gets (\Vb{\Xor}\Vb_0~\If~\Vc~\Else~\Vb_0)$ by an implementation in elementary terms \cite{McIver:09b}:
\[
 \begin{array}{ll} 
  \multicolumn{2}{l}{\InQuadL{\Begin} \Vis_{A}~\Va_{B}, \Va_{C};~ \Vis_{B}~\Vb_{0},\Vb_{1};~ \Vis_{C}~\Vc_{0}, \Vc_{1};} \\
  \quad \Vb_0\Gets\True\PC{}\False; \\
  \quad \Vb_1\Gets\True\PC{}\False; \\\\

  \multicolumn{2}{l}{\quad\InQuadL{\Begin} \Vis_B~ \Vm'_0,\Vm'_1; \Vis_C~\Vc',\Vm';} \\
  \quad\quad \Vc'\Gets\True\PC{}\False; \\
  \multicolumn{2}{l}{\quad\quad \Vm'_0\Gets\True\PC{}\False;~\Vm'_1\Gets\True\PC{}\False;} \\
  \quad\quad \Vm'\Gets \Vm'_{\Vc'}; 
    & \hspace{1em}\textrm{$\leftarrow$\small\ Done in advance by trusted third party.~$\dagger$} \\\\
   \quad\quad \Vis_{ABC}~\Vx,\Vy_0,\Vy_1;
    & \hspace{1em}\textrm{$\leftarrow$\small\ Note these are visible to all three agents.} \\
   \quad\quad \Vx\Gets \Vc{\Xor}\Vc'; \\
   \quad\quad \Vy_0\Gets \Vb_0{\Xor}\Vm'_{\Vx}; \\
   \quad\quad \Vy_1\Gets \Vb{\Xor}\Vb_0{\Xor}\Vm'_{\neg \Vx}; \\
   \quad\quad \Vc_{0}\Gets \Vy_c{\Xor} \Vm' 
   & \hspace{1em}\textrm{$\leftarrow$\small\ Although $\Vy_{\Vc}$ is public, only Agent $C$ knows $\Vm'$.~$\ddagger$} \\
  \quad\End; \\\\
  \multicolumn{2}{l}{
  \quad\left.\begin{array}{@{}ll}
              \Vc_1\Gets& (\neg \Vb_1~\If~\Vc~\Else~\Vb{\Xor}\Vb_1); \\
              \Va_B\Gets& (\Vb_1~\If~\Va~\Else~\Vb_0); \\
              \Va_C\Gets& (\Vc_1~\If~\Va~\Else~\Vc_0);
             \end{array}
             ~~\right\}~{\begin{tabular}{l}\textit{Three more oblivious transfers}, \\ each one to be \\ expanded as above.\end{tabular}}}
   \\\\
  \quad \Reveal~\Va_{B}{\Xor}\Va_{C}~\End
 \end{array}
\]

Each of the other three transfers would expand to a similar block of code, making about 40 lines of code in all. The Oblivious Transfer is formally derived elsewhere \cite{Morgan:07}; an informal explanation is given in \App{a1446}.

\medskip
The preservation of correctness, under expansion, is again guaranteed by the compositionality of the security semantics.

\medskip
Note that aside from the statement marked $\dagger$ (and its three other instances within the three other, unexpanded oblivious transfers), all messages are wholly public because of the declarations $\Vis~\Vx,\Vy_0,\Vy_1$; that is, all the privacy needed is provided already by the exclusive-or'ing with hidden Booleans, as $\ddagger$ shows. 

The only private communications ($\dagger\times4$) are done with the aid of a trusted third party. As explained by Rivest \cite{Rivest:99,Morgan:07} this party's involvement occurs only \emph{before} the protocol begins, and it is trusted not to observe any data exchanged subsequently between the agents; alternatively, the subsequent transfers can themselves be encrypted without affecting the protocol's correctness. (A trusted third party without these limitations would implement trivially any protocol of this kind, simply by collecting the secret data, processing it and then distributing the result.)

\caption{Three-Judges Protocol in elementary terms}
\end{Figure}

\section{Conclusion: a challenge and an open problem}\label{s1731}
We have investigated the foundations for probabilistic non-interference security by proposing a semantics, and a refinement order between its programs, which we have demonstrated has connections with existing entropy-based measures. Especially it is related to Bayes Risk and we have given a soundness and completeness result that establishes compositional closure.

Our approach has a  general goal: to justify practical methods which support accurate analysis of programs operating in a context of probabilistic uncertainty. Abstraction underlies tractable analysis, but the results of such analyses become relevant \emph{only if} the method of abstraction aptly preserves the properties intended for examination. The impact of this research is to show firstly that our refinement order aptly characterises Bayes Risk, and secondly that the former discrepancies between Bayes Risk and other information orders can be rationalised by taking contexts into account.

By taking a fresh point of view, we have related entropies that were formerly thought to be inconsistent.
Furthermore, we highlight the \emph{similarities} between non-interference (as defence against an adversary) and large-scale structuring techniques (such as stepwise refinement and its associated information hiding \cite{Parnas:72}) for probabilistic systems. Both require a careful distinction between what data can be observed and what data must be protected; by observing that distinction in the definition of abstraction, we allow the tractable analysis of properties which rely on ``secrecy" (on the one hand) or ``probabilistic local state'' (on the other). This unified semantic foundation opens up the possibility for a uniform approach to the specification of security properties, along with other safety-critical features, during system design \cite{McIver:09f}.

These positive results now present a challenge and an open problem. The \underline{challen}g\underline{e} is to find a model where all three features, probability, nondeterminism and hidden state, can reside together, and then an equivalence between semantic objects which respects an appropriate definition of testing. The presence of nondeterminism would then include a treatment of distributed systems with schedulers having a restricted view of the state \cite{Chatzikokolakis:10}; that is because nondeterminism can be interpreted either as underspecification, or as a range of decisions presented to a scheduler.  Within such a model we would be increasing the power of the adversary  to harvest information about the hidden state by increasing the expressivity of the contexts she can create. It is an \underline{o}p\underline{en} p\underline{roblem} whether that increased power is sufficient to  make the various information-theoretic orders (Bayes Risk, Shannon Entropy, Marginal Guesswork etc.) equivalent or whether they remain truly distinct.

\subsection*{Related techniques}

The use of information orders, such as those summarised in \Sec{s1028}, to determine the extent to which programs leak their secrets is widespread. Early work that took this approach includes  \cite{Millen:87,Wittbold:90,Gray:91}, and more recently it has been employed in \cite{Smith:07,Kopf:07,Braun:08,Clarkson:09,Malacaria:10}.
One of the contributions of this paper is to show how those evaluations can be related by taking a refinement-oriented perspective. Compositionality plays a major role in our definition of refinement and we note that other orders between probability distributions such as  the ``peakedness'' introduced by Dubois and H\"ullermeir  \cite{Dubois:05} appear not to be compositional when generalised to our hyperdistributions.

More significant than the particular information order is the way that it is used in the analysis of programs. Our approach uses specifications to characterise permitted leaks, and a refinement order which ensures that for our chosen information order (i.e. Bayes Risk), the implementation is at least as secure as its specification. An alternative mode is taken by Braun et al.\ \cite{Braun:08}. Rather than restricting  the elementary \emph{testing-relation} $(\ERef)$ to a compositional subset $(\Ref)$, they identify the  \emph{safe} contexts $\CC_{\rm safe}$ such that $I\mathrel{\mbox{``$\ERef$''}}\CC_{\rm safe}(I)$. With our emphasis on implementations $I$ \emph{and} their specifications $S$, by analogy we would be looking for $S{\ERef}I$ implies $\CC_{\rm safe}(S){\ERef}\CC_{\rm safe}(I)$.

Building on the theoretical approaches, others have investigated the use of automation to evaluate the quantitative weaknesses in programs. Heusser and Malacaria \cite{Heusser:10}, for example, have automated a technique based on Shannon entropy. Andr{\'e}s et al. \cite{Andres:10} similarly consider efficient calculation of information leakage, which can provide diagnostic feedback to the designer.

In some ways our semantics is related in structure to \emph{Hidden Markov Models} \cite{Jurafsky:00} suggesting that, in the future, the algorithmic methods developed in that field might apply to the special concerns of program development. A Hidden Markov Model considers a system partitioned into hidden states (our $\Vh$) and observable outputs (similar to our $\Vv$). The $\Vh$-state evolves according to a Markov Chain, in our terms repeated execution of a fragment $\Vh\From D.\Vh$ in which the probability of the next state $h'$ is given by a fixed ``matrix'' $D$ as $D.h.h'$ where $h$ is the current state. Associated with each transition is an observation, in our terms execution of a fragment $\Vv\From E.\Vh$. Put together, therefore, the \HMM\ evolves according to repeated executions of the fragment
\begin{equation}\label{e1411}
 \Vh\From D.\Vh;~\Vv\From E.\Vh ~,
\end{equation}
which fragment is a special case of our probabilistic-choice statements since the distributions on the right in \Eqn{e1411} do not depend on $\Vv$, whereas in \Fig{f1228} they can.

The canonical problems associated with \HMM's are (in the terms above)
\begin{enumerate}
\item Given the source code (that is, the matrices $D,E$), compute the probability of observing a given sequence of values assigned to $\Vv$.
\item Given a sequence of output values, determine the most likely values of $D,E$.
\item Given the source code and a particular sequence of values assigned to $\Vv$, calculate the sequence of values assigned to $\Vh$ that was most likely to have occurred.
\end{enumerate}

The first of those is basically the classical semantics \cite{Kozen:85,McIver:05a}, but projected onto $\Vv$ since we are not interested in $\Vh$'s values. The second we do not treat at all --- it is tantamount to trying to guess a program's source code (in a limited repertoire) given the outputs it produces. The third is closest to our security concerns, since it is in a sense trying to guess $\Vh$ from observation of $\Vv$.

But in fact we address none of the three problems directly, since even in the third case we have a different concern: in \HMM\ terms we are comparing \emph{two} systems  $D_S,E_S$ and $D_I,E_I$, asking  whether --according to certain entropy measures-- the entropy of the a-posteriori distribution of the final value of $\Vh$ is at least as secure in system $D_I,E_I$ as it is in $D_S,E_S$. Furthermore, our concern with compositionality would in \HMM\ terms relate to the question of embedding each of $D_S,E_S$ and $D_I,E_I$ ``inside'' another system $D,E$.

The application of \HMM\ techniques to our work would in the first instance probably be in the efficient calculation of whether $D_S,E_S$, the specification, was secure enough for our purposes: once that was done, the refinement relation would ensure that the implementation $D_I,E_I$ was also secure enough, without requiring a second calculation. The advantage of this is that the first calculation, over a smaller and more abstract system, is likely to be much simpler than the second would have been.

\bibliographystyle{plain}
\bibliography{ProbsNew}

\newpage
\appendix

\section{Proofs for partition-based matrix representations}\label{a39475}

We give here the proofs for properties we relied on in \Sec{s1724}.

\subsubsection{Property \ref{e29374} (in \Sec{s1552}): Convex closure of refinement matrices}

To show that the set of $N{\times}N$ refinement matrices is the convex closure of the transpose of the set of $N{\times}N$ strategy matrices, i.e.\ that
\begin{equation*} 
\RefMatrices_{N} \Wide{=} \CCL.(\FinerMatrices_{N})~, 
\end{equation*}
we first observe $(\supseteq)$ that every element in $\CCL.(\FinerMatrices_{N})$ is trivially non-negative and column-one-summing (that is, it is a refinement matrix). It remains to show $(\subseteq)$ that any refinement matrix $R$ in $\RefMatrices_{N}$ can be expressed as an interpolation of matrices from $\FinerMatrices_{N}$.

We argue as follows. Fix $R$, and identify a non-zero minimum element
in each of its columns; let $c$ be the minimum of those column-minima;
select the $M$ in $\FinerMatrices_{N}$ that has 1's in the column-minima
positions exactly; and subtract $c M$ from $R$ to give some $R'$.

Now $R'$ has at least one more 0 entry than $R$ did, and yet the columns of $R'$ still have equal sums, now $1{-}c$. Continue this process from $R'$ onwards: it must stop, since the number of 0's increases each time; and when it does stop it must be because there is an all-0 column, in which case \emph{all} columns must be all-0, since the column sums have remained the same all the way through.\par
The collection of $M$'s and their associated $c$'s is the interpolation we had to find: for example, in three steps the procedure generates the interpolation
\begin{displaymath}
 \left(\begin{array}{c@{~~}c} 1/3 & 3/4 \\ 2/3 & 1/4 \end{array}\right)
 \Wide{=}
   1/4\left(\begin{array}{cc} 1&0\\0&1\end{array}\right)
 + 1/12\left(\begin{array}{cc} 1&1\\0&0\end{array}\right)
 + 2/3\left(\begin{array}{cc} 0&1\\1&0\end{array}\right) ~.
\end{displaymath}

\subsubsection{Property \Eqn{e58762} (in \Sec{s1552}): refinement matrices form a monoid}

Since matrix multiplication is associative and the identity $\UnitMatrix_{N}$ is an element of $\RefMatrices_N$, we need only demonstrate that $\RefMatrices_N$ is closed under multiplication. That can be checked by direct calculation.

\section{Secure semantics via matrices} \label{a394759}

In \Sec{s1724} we appealed to matrix representations of partitions to construct our proof that $(\Ref)$ is the compositional closure of $(\ERef)$. Here we we project the rest of our semantics into matrix algebra, giving matrix representations of split-states, hyper-distributions, programs, and refinement. These representations are used to verify both monotonicity (\Thm{t1120} from \Sec{s9345}) in \App{a9475} and the Atomicity Lemma (\Lem{l1222} from \Sec{s9374}) in \App{a29375}.

\subsection{Notation}

For $i$ taken from some ordered index set $I$ we will write 
$\General{\Above}{i\In I}{R}{M_i}$ 
for the vertical concatenation of matrices $M_i$ for those $i$ satisfying $R$, taken in $I$-order: for this to be well defined, the column-count must be the same for all $M$'s; but their row-counts may differ.  In the same way we will write 
$\General{\Beside}{i\In I}{R}{M_i}$ 
for horizontal matrix concatenation (in which case the row-counts must agree).

For a given dimension $N$ and expression $E.\Vn$ we write $\DiagM{E.\Vn}$ for the $N{\times}N$ diagonal matrix whose value at the element (doubly) indexed $n$ is $E.n$. Thus for example we have $\IdM{}=\DiagM{1}$.

\subsection{Split-states as single-column matrices}

Let $N$ be the cardinality of $\VV{\times}\HH$. A split-state has type $\VV{\times}\TDist\HH$ and can be written as an $1{\times}N$ matrix, a row of probabilities in some agreed-upon index order of $\VV{\times}\HH$ where the element at (column) index $(v,h)$ gives the probability $\delta.h$ associated with that pair.

Naturally the row sums to 1 but --more than that-- each such representation of a split-state will have nonzero entries only in columns whose first index-component is the $v$ appearing in $(v,\delta)$. We say that such a row is \emph{$\VV$-unique} and that it has \emph{characteristic} $v$.

Write $\IdM{v}$ for the $N{\times}N$ diagonal matrix $\DiagM{1~\If~ \Vv{=}v~\Else~0}$ having ones only at positions whose row- (or equivalently column-) index has that $v$ as its first component; elsewhere in the diagonal (and everywhere off the diagonal) the entries are zero. The row-matrix representation $\MatSem{v,\delta}$ of split-state $(v,\delta)$ then satisfies $\MatSem{v,\delta}= \MatSem{v,\delta} {\times} \IdM{v}$ because it has characteristic $v$, so that the multiplication by $\IdM{v}$ sets to zero only elements that were zero already.

\subsection{Hyper-distributions as matrices}

In \Sec{s1552} we interpreted whole partitions as matrices, with each row (fraction) giving a possible distribution over $\HH$ for some fixed $v$. Here we proceed similarly, but we do not fix $v$, so that a hyper-distribution $\Delta$ whose support has cardinality $F$ is represented as an $F{\times}N$ matrix $\MatSem{\Delta}$ each of whose rows is $\VV$-unique, as above, thus independently representing some split-state $(v,\delta)$. 
Extending the matrix representation of individual split-states, we can represent whole hyper-distributions according to
\begin{equation}\label{e1729}
 \MatSem{\Delta}
 \Wide{\Defs}
 \General{\Above}{(v,\delta)\In \Support{\Delta}}{}{\Delta.(v,\delta) * \MatSem{v,\delta}}
\end{equation}
where, as in \Sec{s1552}, with the multiplier $\Delta.(v,\delta)$ we are scaling the rows so that the total weight of each gives the probability of that split-state in the hyper-distribution overall; the distribution the split-state actually contains (the $\delta$ in the $(v,\delta)$ that the row represents) is as usual recoverable by normalising. Because each of the rows is $\VV$-unique we say that the matrix as a whole, also, is $\VV$-unique; but note that it is possible to have several rows with the same characteristic $v$. $\VV$-uniqueness means that no two distinct $v$'s appear with non-zero probability in the \emph{same} row.

As for partitions, in such matrices we define \emph{similarity} between rows and say that a hyper-distribution is in \emph{reduced} matrix representation if all its similar rows have been added together, and all its all-zero rows have been removed. We say that two hyper-distribution matrices are \emph{similar} $(\Similar)$ if their reductions are equivalent up to a reordering of rows. Similarity is a congruence for matrix multiplication on the right, but not on the left; vertical concatenation $(\Above)$ respects similarity on both sides.

While the column-order of $\MatSem{\Delta}$ is fixed by our (arbitrary) ordering of $\VV{\times}\HH$, the row-order might vary since there is no intrinsic order on fractions. We therefore regard $\MatSem{\Delta}$ as determined only up to similarity, and our reasoning below will be restricted to operations for which similarity is a congruence. In particular we have that $\MatSem{\Delta_1}{\Similar}\MatSem{\Delta_2}$ implies $\Delta_1{=}\Delta_2$, i.e.\ that $\MatSem{\cdot}$ is injective up to similarity.

The operation $\MatSem{{\cdot}}$ on (sub-)hyper-distributions is linear in the sense that
\begin{equation}\label{e1503}
 \begin{array}{p{4em}rcl}
      & \MatSem{p{*}\Delta} &\Wide{=}& p*\MatSem{\Delta} \\
   and & \MatSem{\Delta_1{+}\Delta_2} &\Wide{\Similar}& \MatSem{\Delta_1}\Above\MatSem{\Delta_2} ~.
 \end{array}
\end{equation}

\subsection{Classical commands as matrices}

We recall from \Sec{s47563} that the classical ``relational'' semantics  $\SemC{P}$ of a program $P$ is a function $\VV{\times}{\HH} \Fun \TDist{(\VV{\times}\HH)}$ and may hence be treated (just as $D.\Vv.\Vh$ from \Sec{s1552} was) as an $N{\times}N$ matrix written $\MatSemC{P}$ whose value in row $(v,h)$ and  column $(\Vpk,h')$ is just $\SemC{P}.(v,h).(\Vpk,h')$. 
\footnote{Note that operation $\MatSemC{{\cdot}}$ applies to texts, i.e.\ syntax but $\MatSem{\cdot}$ applies to hyper-distributions, i.e.\ semantics.}%

Sequential composition between classical commands is then represented by matrix multiplication, in the usual Markov style, so that we have
\begin{equation}\label{e1606}
\MatSemC{P_1; P_2} \Wide{=} 
\MatSemC{P_1} \times \MatSemC{P_2} ~.
\end{equation}

\subsection{Secure commands as matrices}

We will establish that for any secure program $P$ there is an $I$-indexed set of $N{\times}N$ matrices such that 
\begin{equation}\label{e1612}
\MatSem{\,\Sem{P}.(v,\delta)\,} 
\Wide{\Similar}
\General{\Above}{i\In I}{}{\MatSem{v,\delta} \times M_i}
\end{equation}
for any split-state $(v,\delta)$. We think of the matrices as giving a \emph{normal form} for $P$. Using the normal form, we will be able to represent the lifting of $P$'s secure semantics using matrix operations, since then
\begin{equation}\label{l9745}
\MatSem{\,\Exp{(v,\delta)\In \Delta}{\Sem{P}.(v,\delta)}\,}
\Wide{\Similar}
\General{\Above}{i\In I}{}{\MatSem{\Delta} \times M_i}
\end{equation}
can be established by the calculation
\begin{Reason}
\Step{}{
  \MatSem{\,\Exp{(v,\delta)\In \Delta}{\Sem{P}.(v,\delta)}\,}
}
\StepR{$=$}{definition expected value \Sec{s1729}}{
  \MatSem{\,\General{\sum}{(v,\delta)\In \Support{\Delta}}{}{
    \Delta.(v,\delta) * \Sem{P}.(v,\delta)\,}
  }
}
\StepR{$\Similar$}{from \Eqn{e1503}}{
  \General{\Above}{(v,\delta)\In \Support{\Delta}}{}{
    \Delta.(v,\delta) * \MatSem{\Sem{P}.(v,\delta)}
  }
}
\StepR{$\Similar$}{normal form \Eqn{e1612}}{
  \General{\Above}{(v,\delta)\In \Support{\Delta}}{}{
    \Delta.(v,\delta) * 
    \General{\Above}{i\In I}{}{\MatSem{v,\delta}\times M_i}
  }
}
\StepR{$=$}{distribute multiplication}{
  \General{\Above}{(v,\delta)\In \Support{\Delta}; i\In I}{}{ 
    (\Delta.(v,\delta) * \MatSem{v,\delta}) \times M_i
  }
}
\Space
\StepR{$\Similar$}{rearrange rows; distribute\\ post-multiplication}{
  \General{\Above}{i\In I}{}{ 
    \General{\Above}{(v,\delta)\In \Support{\Delta}}{}{
        \Delta.(v,\delta) * \MatSem{v,\delta}
    }
    \times M_i
  }
}
\Space
\StepR{$\Similar$}{from \Eqn{e1729} defining $\MatSem{\Delta}$}{
  \General{\Above}{i\In I}{}{ 
     \MatSem{\Delta} \times M_i
  } ~.
}
\end{Reason}
We now show by structural induction how embedded classical commands, general choice, sequential composition (and hence all of our secure commands) can be translated into this normal form.

\subsubsection{Embedded classical commands}

In \Def{d0855a} from \Sec{s1044} we gave the semantics $\Sem{\Atomic{P}}$ of a program $P$ considered as an atomic unit; we now do the same here in matrix style.

If we were to execute an atomic program $\Atomic{P}$ from a split-state $(v,\delta)$, in the matrix style we would begin by calculating $\MatSem{v,\delta} {\times} \MatSemC{P}$, giving again a single row; but that row might not be $\VV$-unique, in which case a further step would be needed. We'd split its possibly non-unique rows into (maximally) $\VV$-unique portions, an operation that corresponds roughly to the $\Hide$ funtion used in \Def{d0855}.

Given a row matrix $R$ that is $\VV{\times}\HH$-indexed by column (such as the output $\MatSem{v,\delta} {\times} \MatSemC{P}$ from just above) the splitting of its possibly non-unique row is achieved via
\begin{equation}\label{e1951}
 \Hide.R \Wide{\Defs}\General{\Above}{v'\In \VV}{}{R \times \IdM{v'}} ~,
\end{equation}
in which each of the values $v'$ in $\VV$ is used, in turn, to construct a row matrix of characteristic $v'$ projected from $R$ by zeroing all other entries: those characteristic-$v'$ projections are then stacked on top of each other with $(\Above$) to make a single (possibly quite tall!)\ matrix that is derived from $R$ but now is $\VV$-unique.\,%
\footnote{For example, if the row $R$ is $\VV$-unique already then $R'\Defs\Hide.R$ will stack up a great many all-zero rows. But still we will have $R{\Similar}R'$, so no damage is done.}
With that apparatus, we have
\begin{equation}\label{e1558}
 \MatSem{\,\Sem{\Atomic{P}.(v,\delta)}\,}
 \Wide{\Similar}
 \General{\Above}{v'\In\VV}{}{
\MatSem{v,\delta} \times \MatSemC{P} \times \IdM{v'}}~,
\end{equation}
thereby giving the $\VV$-unique matrix representation (up to similarity) of the hyper-distribution output by $\Atomic{P}$ if executed from incoming split-state $(v,\delta)$.\,%
\footnote{Note the algebra of similarity here: if we have $R{\Similar}\MatSem{v,\delta}$ for some $R$, then $\General{\Above}{v'\In\VV}{}{R \times \MatSemC{P} \times \IdM{v'}}$ is similar to the right-hand side above.}

\subsubsection{General choice}

For both general choice and sequential composition we assume inductively that the semantics of subprograms $P_1$ and $P_2$ can be written in matrix normal form so that for each split-state $(v,\delta)$ we have
\begin{equation*}
\MatSem{\,\Sem{P_{i}}.(v,\delta)\,} \Wide{\Similar}
\General{\Above}{j_i\In J_{i}}{}{\MatSem{v,\delta} \times M_{i,j_i}} ~.
\end{equation*}

To show that general choice can be expressed in matrix normal form, we use the following identity which expresses the conditioning of a split-state $(v,\delta)$ by expression $E.\Vv.\Vh$ in terms of matrix operations:
\begin{equation}\label{e4573}
\Exp{h\In \delta}{E.v.h} * 
\MatSem{v,\PSet{h\In\delta}{E.v.h}{}}
\Wide{=}
\MatSem{v,\delta}\times \DiagM{E.\Vv.\Vh} ~.
\end{equation}%

We then have
\begin{Reason}
\Step{}{
  \MatSem{\,\Sem{P_1 \PC{q.\Vv.\Vh} P_2}.(v,\delta)\,}
}
\Space
\WideStepR{$=$}{general choice from \Fig{f1228}; $p \Defs \Exp{h\In\delta}{q.v.h}$}{
  \MatSem{\,p*\Sem{P_1}.(v,\PSet{h\In\delta}{q.v.h}{}) + {}
  (1{-}p)*\Sem{P_2}.(v,\PSet{h\In\delta}{1{-}q.v.h}{})\,}
}
\Space
\WideStepR{$\Similar$}{from \Eqn{e1503}}{
  p * \MatSem{\,\Sem{P_1}.(v,\PSet{h\In\delta}{q.v.h}{})\,} \AboveRel {}
  (1{-}p)*\MatSem{\,\Sem{P_2}.(v,\PSet{h\In\delta}{1{-}q.v.h}{})\,}
}
\Space
\WideStepR{$\Similar$}{inductive assumption: matrix normal form of $P_1$ and $P_2$}{
\begin{array}{lrl}
  &
  p*{}&
  \General{\Above}{j_1\In J_{1}}{}{
    \MatSem{v,\PSet{h\In\delta}{q.v.h}{}} \times M_{1,j_1} 
  }\\
  \AboveRel &
  (1{-}p)*{}&
  \General{\Above}{j_2\In J_{2}}{}{     
    \MatSem{v,\PSet{h\In\delta}{1{-}q.v.h}{}} \times M_{2,j_2} 
  }
\end{array}
}
\Space
\WideStepR{$=$}{distribute scalar multiplications}{
  &
  \General{\Above}{j_1\In J_{1}}{}{
    (p * \MatSem{v,\PSet{h\In\delta}{q.v.h}{}}) \times M_{1,j_1} 
  }\\
  \AboveRel &
  \General{\Above}{j_2\In J_{2}}{}{
    ((1{-}p)*\MatSem{v,\PSet{h\In\delta}{1{-}q.v.h}{}}) \times M_{2,j_2} 
  }
}
\Space
\WideStepR{$=$}{recall $p \Defs \Exp{h\In\delta}{q.v.h}$; from \Eqn{e4573}}{
  &
  \General{\Above}{j_1\In J_{1}}{}{
    \MatSem{v,\delta} \times \DiagM{q.\Vv.\Vh} \times M_{1,j_1}
  }\\
  \AboveRel &
  \General{\Above}{j_2\In J_{2}}{}{
    \MatSem{v,\delta} \times \DiagM{1{-}q.\Vv.\Vh} \times  M_{2,j_2}
  }
}
\Space
\WideStepR{$=$}{Let $p_1.\Vv.\Vh \Defs q.\Vv.\Vh$ and 
                    $p_2.\Vv.\Vh \Defs 1{-}q.\Vv.\Vh$}{
  \General{\Above}{i\In\{1,2\}; j\In J_i}{}{
     \MatSem{v,\delta} \times \DiagM{p_i.\Vv.\Vh} \times M_{i,j}
   } ~.
}
\end{Reason}

\subsubsection{Sequential composition}

For sequential composition of $P_1$ and $P_2$ we have
\begin{Reason}
\Step{}{
  \MatSem{\,\Sem{P_1;P_2}.(v,\delta)\,}
}
\StepR{$=$}{Composition from \Fig{f1228}}{
  \MatSem[.9em]{\,\Exp{(\Vpk,\delta')\In\Sem{P_1}.(v,\delta)}{\Sem{P_2}.(\Vpk,\delta')}\,}
}
\StepR{$\Similar$}{\Eqn{l9745}; matrix normal form of $P_2$}{
  \General{\Above}{j_2\In J_2}{}{
    \MatSem{\Sem{P_1}.(v,\delta)}
    \times M_{2,j_2}
  }
}
\StepR{$\Similar$}{matrix normal form of $P_1$}{
  \General{\Above}{j_1\In J_1; j_2\In J_2}{}{
     \MatSem{v,\delta} \times M_{1,j_1}\times M_{2,j_2}
  } ~.
}
\end{Reason}

\subsection{Refinement as matrix multiplication}

In \Sec{s1552} we showed how refinement between partitions could be defined using matrix multiplication. We can promote this to hyper-distributions by dealing with each $v$ separately: we have that hyper-distribution $\Delta_S$ is refined by $\Delta_I$ just when for each $v{\in}\VV$ there exists a refinement matrix (i.e. a non-negative, column one-summing matrix) $R_v$ such that
\begin{equation}\label{e0231}
R_v \times \MatSem{\Delta_S} \times \IdM{v} 
\Wide{\Similar}
\MatSem{\Delta_I} \times \IdM{v} ~.
\end{equation}
The effect of requiring similarity for each $v$ separately is to prevent rows with differing $v$'s from being added together.

\section{Proofs for the refinement relation}

\subsection{Secure programs are partially ordered by $(\Ref)$}\label{a2937}

We show (\Thm{t1042} in \Sec{s9345}) that the refinement relation $(\Ref)$ defines a partial order over hyper-distribut\-ions. It follows by extension that it is a partial order over secure programs.

\subsubsection{Reflexivity} For any hyper-distribution $\Delta$ reflexivity holds trivially since, for each $v$, the intermediate partition $\Fracs.\Delta.v$ is both similar to and as fine as itself.

\subsubsection{Transitivity} Assume that $\Delta_1{\Ref}\Delta_2$ and $\Delta_2{\Ref}\Delta_3$. It is enough to show that for each $v$ we have $\Fracs.\Delta_1.v\Ref\Fracs.\Delta_3.v$.
For each $i$ let $\Pi_{i}$ be the $N{\times}N$ matrix representation (\Sec{s1552}) of $\Fracs.\Delta_i.v$ for some $N$.
To prove $\Pi_1{\Ref}\Pi_3$, we need to find a refinement matrix $R_{31}$ such that $\Pi_3$ is $R_{31}{\times}\Pi_{1}$.

From above there are refinement matrices $R_{32},R_{21}$ with $\Pi_3 = R_{32} {\times} \Pi_2$ and $\Pi_2 = R_{21} {\times} \Pi_1$. Thus $R_{31}$ defined $R_{32} {\times} R_{21}$ satisfies $\Pi_3 = R_{31} {\times} \Pi_1$, and it is a refinement matrix by Property \Eqn{e58762} from \Sec{s1552}.

\subsubsection{Antisymmetry}

Assume that both $\Delta_1{\Ref}\Delta_2$ and $\Delta_2{\Ref}\Delta_1$ but $\Delta_1 \neq \Delta_2$. From the first and third assumptions, with \Lem{l2121} (\App{s3628}) we have that the Shannon Entropy 
of $\Delta_1$ is strictly less than that of $\Delta_2$; from the second and third, we have the opposite --- thus a contradiction.

\subsection{Monotonicity of secure programs w.r.t.\ $(\Ref)$} \label{a9475}

We use the following technical results to verify that $(\Ref)$ is monotonic with respect to secure program contexts (\Thm{t1120} from \Sec{s9345}). They are verified using the matrix algebra from \App{a394759} above.

\begin{Lemma}{}{l35679}
For any indexed set of matrices $\Set{i\In I}{}{M_i}$ each of dimension $F{\times}N$ and corresponding refinement matrices $R_i$ each having $F_i$ rows and $F$ columns, there exists a refinement matrix $R$ such that 
\begin{equation}\label{l59867}
\General{\Above}{i\In I}{}{R_i \times M_i} 
\Wide{=}
R \times \General{\Above}{i\In I}{}{M_i} ~.
\end{equation}
\begin{Proof}
Refinement matrix $R$ can be given directly as
\begin{equation*}
\General{\Above}{i\In I}{}{
  \General{\Beside\,}{i'\In I}{}{(~R_i~\If~i=i'~\Else~ \ZeroMatrix_{F_i{\times}F}~)}
} ~.
\end{equation*}
That $R$ is a refinement matrix (i.e. it has non-negative entries and is column-one-summing) follows from its definition and the fact that each $R_i$ is a refinement matrix. It can be established by matrix multiplication that \Eqn{l59867} holds.\,%
\footnote{A sketch of the block matrices helps to see the pattern.}
\end{Proof}
\end{Lemma}

\begin{Lemma}{Additive monotonicity of hyper-distributions}{l9237}
For probability $p\In[0,1]$, and hyper-distributions $\Delta_{S_1}, \Delta_{S_2},\Delta_{I_1}, \Delta_{I_2}$ we have that  $\Delta_{S_i} \Ref \Delta_{I_i}$ implies 
\begin{equation*}
\Delta_{S_1} \PC{p} \Delta_{S_2} \Wide{\Ref} \Delta_{I_1} \PC{p} \Delta_{I_2} ~.
\end{equation*}
\begin{Proof}
From \Eqn{e0231} it is enough for each $v$ to find a refinement matrix $R$ such that 
\begin{equation*}\label{e93465}
 R \times \MatSem{\Delta_{S_1} \PC{p} \Delta_{S_2}} \times \IdM{v}
\Wide{\Similar}
\MatSem{\Delta_{I_1} \PC{p} \Delta_{I_2}} \times \IdM{v} ~.
\end{equation*}
We have:
\begin{Reason}
\Step{}{
 \MatSem{\,\Delta_{I_1} \PC{p} \Delta_{I_2}\,} \times \IdM{v}
}
\StepR{$\Similar$}{from \Eqn{e1503}}{
 (\,p{*}\MatSem{\Delta_{I_1}} \AboveRel 
 (1{-}p){*}\MatSem{\Delta_{I_2}}\,) 
\times \IdM{v}
}
%
%
\StepR{$=$}{distribute post-multiplication}{
                p{*}\MatSem{\Delta_{I_1}} \times \IdM{v} 
\AboveRel  (1{-}p){*}\MatSem{\Delta_{I_2}} \times \IdM{v}
}
\Space
\WideStepR{$\Similar$}{$\Delta_{S_i} {\Ref} \Delta_{I_i}$ implies 
$\MatSem{\Delta_{I_i}} {\times}  \IdM{v}
\Similar 
R_i {\times} \MatSem{\Delta_{S_i}} {\times} \IdM{v}$ for some refinement \\matrix $R_i$}{
                p * R_1 \times \MatSem{\Delta_{S_1}} \times \IdM{v} 
\AboveRel  (1{-}p)* R_2 \times \MatSem{\Delta_{S_2}} \times \IdM{v}
}
\Space
\WideStepR{$=$}{commute scalar multiplication; distribute post-multiplication}{
 (\,        R_1 \times      p{*}\MatSem{\Delta_{S_1}} 
 \AboveRel  R_2 \times (1{-}p){*}\MatSem{\Delta_{S_2}} \,)
 \times \IdM{v} 
}
\Space
\WideStepR{$=$}{\Lem{l35679} for some refinement matrix $R$}{
 R \times
 (\,              p * \MatSem{\Delta_{S_1}} 
 \AboveRel   (1{-}p)* \MatSem{\Delta_{S_2}} \,)
 \times \IdM{v} 
}
\StepR{$\Similar$}{from \Eqn{e1503}}{
  R \times \MatSem{\,\Delta_{S_1} \PC{p} \Delta_{S_2}\,} \times \IdM{v} ~.
}
\end{Reason}
\end{Proof}
\end{Lemma}

\begin{Lemma}{Pointwise monotonicity}{l02834}
For all program texts $P$ and hyper-distribut\-ions $\Delta_S$ and $\Delta_I$ such that $\Delta_S \Ref \Delta_I$, we have
\begin{equation*}
\Exp{(v,\delta)\In \Delta_S}{\Sem{P}.(v,\delta)} \Wide{\Ref} 
\Exp{(v,\delta)\In \Delta_I}{\Sem{P}.(v,\delta)} ~. 
\end{equation*}
\begin{Proof}
Let $\Set{i\In I}{}{M_i}$ be a set of $N{\times}N$ matrices giving the normal form for $\Sem{P}$ as at \Eqn{e1612} above, so that for any $(v,\delta)$ we have
\begin{equation*}
\MatSem{\,\Sem{P}.(v,\delta)\,} \Wide{\Similar}
\General{\Above}{i\In I}{}{\MatSem{v,\delta} \times M_i} ~.
\end{equation*}
From \Eqn{e0231} and \Eqn{l9745}, it is enough to show that for each $v'{\in}\VV$ there exists a refinement matrix $R$ such that
%
$
R {\times} \General{\Above}{i\In I}{}{\MatSem{\Delta_S} {\times} M_i} {\times} \IdM{v'}
~\Similar~
\General{\Above}{i\In I}{}{M_i {\times} \MatSem{\Delta_I}} {\times} \IdM{v'}
$. 
%
We have 
\begin{Reason}
\Step{}{
 \General{\Above}{i\In I}{}{\MatSem{\Delta_I}\times M_i}
 \times \IdM{v'} 
}
\StepR{$\Similar$}{$\MatSem{\Delta_I}$ is $\VV$-unique}{
  \General{\Above}{i\In I}{}{
    \General{\Above}{v\In \VV}{}{\MatSem{\Delta_I} \times \IdM{v}}
    \times M_i 
  }
  \times \IdM{v'} 
}
\StepR{$=$}{distribute post-multiplication}{
  \General{\Above}{i\In I; v\In \VV}{}{
    \MatSem{\Delta_I} \times \IdM{v} \times M_i
  }
  \times \IdM{v'} 
}
\Space
\WideStepR{$\Similar$}{$\Delta_{S} \Ref \Delta_{I}$ implies 
$\MatSem{\Delta_{I}} \times \IdM{v}
\Similar 
R_v \times \MatSem{\Delta_{S}} \times \IdM{v}$ for some\\ refinement matrix $R_v$}{
  \General{\Above}{i\In I; v\In \VV}{}{
    R_v \times \MatSem{\Delta_S} \times \IdM{v} \times M_i
  }
  \times \IdM{v'} 
}
\Space
\StepR{$=$}{\Lem{l35679} \\ for some refinement matrix $R$}{
  R \times 
  \General{\Above}{i\In I; v\In \VV}{}{
    \MatSem{\Delta_S} \times \IdM{v} \times M_i
  }
  \times \IdM{v'} 
}
\Space
\StepR{$\Similar$}{distribute post-multiplication; \\
$\MatSem{\Delta_S}$ is $\VV$-unique}{
  R\times \General{\Above}{i\In I}{}{\MatSem{\Delta_S}\times M_i} \times \IdM{v'} ~.
}
\end{Reason}
\end{Proof}
\end{Lemma}

\subsubsection{Monotonicity of secure programs w.r.t.\ $(\Ref)$} 
Using \Lem{l9237} and \Lem{l02834} we now prove \Thm{t1120} from \Sec{s9345}. We must show that if $S{\Ref}I$ then $\CC(S){\Ref}\CC(I)$ for all contexts $\CC$ built from programs as defined in \Fig{f1228}.

We use structural induction. For the base case, context $\CC(S)\Defs S$ is trivially monotonic. 

General probabilistic choice (and hence probabilistic and conditional choice) is trivially monotonic in either argument from monotonicity of addition over hyper-distributions (\Lem{l9237}). For example, for monotonicity in the first argument we have
\begin{Reason}
\Step{}{
  \Sem{S \PC{q.\Vv.\Vh} R}.(v,\delta)
}
%
\WideStepR{$=$}{Let $q_{\delta} \Defs \Exp{h\In\delta}{q.v.h}$; General choice from \Fig{f1228}}{
(\Sem{S}.(v, \PSet{h\In \delta}{q.v.h}{}) 
 \PC{q_{\delta}} 
 \Sem{R}.(v, \PSet{h\In \delta}{1{-}q.v.h}{}) )
}
\Space
\StepR{$\Ref$}{$S \Ref I$; \Lem{l9237}}{
(\Sem{I}.(v, \PSet{h\In \delta}{q.v.h}{}) 
 \PC{q_{\delta}} 
 \Sem{R}.(v, \PSet{h\In \delta}{1{-}q.v.h}{}) )
}
\StepR{$=$}{General choice from \Fig{f1228}}{
  \Sem{I \PC{q.\Vv.\Vh} R}.(v,\delta) ~.
}
\end{Reason}

To show monotonicity of sequential composition in its right-hand argument we have for any programs $R$ and $S \Ref I$ and initial state $(v,\delta)$ that
\begin{Reason}
\Step{}{
  \Sem{R;S}.(v,\delta)
}
%
%
\StepR{$=$}{Composition from \Fig{f1228}}{
  \Exp{(v',\delta')\In \Sem{R}.(v,\delta)}{\Sem{S}.(v',\delta')}
}
%
%
\StepR{$\Ref$}{$S \Ref I$; \Lem{l9237}}{
  \Exp{(v',\delta')\In \Sem{R}.(v,\delta)}{\Sem{I}.(v',\delta')}
}
%
%
\StepR{$=$}{Composition from \Fig{f1228}}{
  \Sem{R;I}.(v,\delta) ~.
}
\end{Reason}
For monotonicity in the first argument we have
\begin{Reason}
\Step{}{
  \Sem{S;R}.(v,\delta)
}
\StepR{$=$}{Composition from \Fig{f1228}}{
  \Exp{(v',\delta')\In \Sem{S}.(v,\delta)}{\Sem{R}.(v',\delta')}
}
\StepR{$\Ref$}{$S \Ref I$ and \Lem{l02834}}{
  \Exp{(v',\delta')\In \Sem{I}.(v,\delta)}{\Sem{R}.(v',\delta')}
}
\StepR{$=$}{Composition from \Fig{f1228}}{
    \Sem{I;R}.(v,\delta)~.
}
\end{Reason}
%

\section{Example of completeness construction}\label{s9347}

Here we illustrate the completeness proof set out in \Sec{s1634} by applying it to the example of \Sec{s1457}, where we claimed that $P_4{\NotRef}P_2$. We use \Sec{s1634} to find a $C$ such that indeed $P_4;C\NotERef P_2;C$.

\medskip
Our $v'$ is 1, since that is where we find the difference between $P_2$ and $P_4$ in the residual uncertainties of $\Vh$; with that, we extract the fractions
\[
 \textrm{when $v'{=}1$}\quad\left\{\quad\quad
 \begin{array}{rcl}
  \Pi_{P_4} &\Wide{\Defs}& \List{\PSet{}{}{1\Att{1}{4},3\Att{1}{12}}, \PSet{}{}{1\Att{1}{12},3\Att{1}{4}}} \\
  \Pi_{P_2} &\Wide{\Defs}& \List{\PSet{}{}{1\Att{1}{6}}, \PSet{}{}{1\Att{1}{6},3\Att{1}{6}}, \PSet{}{}{3\Att{1}{6}}}
 \end{array}
 \right.
\]
and find that there are two values of $\Vh$, two fractions in $\Pi_{P_4}$ and three fractions in $\Pi_{P_2}$. Accordingly we set $N$ to 3 and include an extra column for $\Vh{=}2$ and an extra, zero fraction in $\Pi_{P_4}$. Note that $\sum\Pi_{P_2}=\sum\Pi_{P_4}$ and that the total weight (of each) is $2/3$.

\parshape=5  0pt 1\linewidth 0pt .75\linewidth 0pt .75\linewidth 0pt .75\linewidth 0pt 1\linewidth
The $N{\times}N$ matrix corresponding to $\Pi_{P_2}$ is then as at right
\hfill
\raisebox{-1.5em}[0pt][0pt]{
\(
 \left(
 \begin{array}{c@{~}c@{~}c}
  1/6 & 0 & 0 \\
  1/6 & 0 & 1/6 \\
  0   & 0 & 1/6
 \end{array}
 \right)
\)
}\\
with its columns corresponding to values $1,2,3$ of $\Vh$ and the rows to $\Pi_{P_2}$'s three fractions. The point obtained by concatenating the rows is $(1/6,0,0,1/6,0,1/6,0,0,1/6)$, and is in 9-dimensional space; but to avoid a proliferation of fractions, we scale everything up from now on by a factor of 12, and so take $x_{P_2}$ to be the point $(2,0,0,2,0,2,0,0,2)$.

Now the scaled-up (and extended) matrix corresponding to $\Pi_{P_4}$, with a selection of refinement-forming matrices $M$ in $\FinerMatrices_N$, is given by
\[
 \left(
 \begin{array}{c@{~}c@{~}c}
  3 & 0 & 1 \\
  1 & 0 & 3 \\
  0 & 0 & 0
 \end{array}
 \right)
 \Wide{\textrm{and}}
 \left(
 \begin{array}{c@{~}c@{~}c}
  1 & 1 & 1 \\
  0 & 0 & 0 \\
  0 & 0 & 0
 \end{array}
 \right) ,
 \left(
 \begin{array}{c@{~}c@{~}c}
  1 & 0 & 1 \\
  0 & 1 & 0 \\
  0 & 0 & 0
 \end{array}
 \right) ,
 \left(
 \begin{array}{c@{~}c@{~}c}
  1 & 0 & 1 \\
  0 & 0 & 0 \\
  0 & 1 & 0
 \end{array}
 \right) ,
 \left(
 \begin{array}{c@{~}c@{~}c}
  0 & 1 & 1 \\
  1 & 0 & 0 \\
  0 & 0 & 0
 \end{array}
 \right) \cdots
\]
Carrying out the matrix multiplications gives us these four possible refinements of $\Pi_{P_4}$:
\[
  \left(
 \begin{array}{c@{~}c@{~}c}
  4 & 0 & 4 \\
  0 & 0 & 0 \\
  0 & 0 & 0
 \end{array}
 \right) ,
 \left(
 \begin{array}{c@{~}c@{~}c}
  3 & 0 & 1 \\
  1 & 0 & 3 \\
  0 & 0 & 0
 \end{array}
 \right) ,
 \left(
 \begin{array}{c@{~}c@{~}c}
  3 & 0 & 1 \\
  0 & 0 & 0 \\
  1 & 0 & 3
 \end{array}
 \right) ,
 \left(
 \begin{array}{c@{~}c@{~}c}
  1 & 0 & 3 \\
  3 & 0 & 1 \\
  0 & 0 & 0
 \end{array}
 \right) \cdots~.
\]
Doing all of them for $M$ in $\FinerMatrices_N$, and concatenating their rows to make points in 9-dimensional space, gives us this collection of refinements altogether:
\begin{equation} \label{e93475}
 \begin{array}[t]{clclclclr}
  \{ & (4,0,4,0,0,0,0,0,0) & ,
     & (3,0,1,1,0,3,0,0,0) & ,
     & (3,0,1,0,0,0,1,0,3) & , \\
     & (1,0,3,3,0,1,0,0,0) & ,
     & (0,0,0,4,0,4,0,0,0) & ,
     & (0,0,0,3,0,1,1,0,3) & , \\
     & (1,0,3,0,0,0,3,0,1) & ,
     & (0,0,0,1,0,3,3,0,1) & ,
     & (0,0,0,0,0,0,4,0,4) && \}~.
 \end{array}
\end{equation}
Our claim that $P_4{\NotRef}P_2$ suggests that the point $x_{P_2}$ (corresponding to the matrix derived from $\Pi_{P_2}$) should not lie in the convex closure of the points \Eqn{e93475} above.

We can see this easily by concentrating on the first and third dimensions only: for $\Pi_{_2}$ we get $(2,0)$; and for $\Pi_{P_4}$, that is \Eqn{e93475}, after removing duplicates we get $(4,4)$, $(3,1)$, $(1,3)$ and $(0,0).$ The $\Pi_{P_2}$-point $x_{P_2}$ is not in the convex closure of the other four because all of them have their two coordinates both positive or both zero, a property preserved by any convex combination but not shared by $(2,0)$.

\begin{wrapfigure}{l}{.5\textwidth}
\setlength{\unitlength}{.9cm}
\begin{picture}(5.5,4.8)(-0.5,-1)
\put(0,-0.5){\line(0,1){5}} 
\put(-0.5,0){\line(1,0){5}} 
\multiput(1,-.25)(1,0){4}{\line(0,1){0.5}} 
\multiput(-.25,1)(0,1){4}{\line(1,0){0.5}} 
\put(2,0){\circle{.2}} 
\put(0,0){\circle*{.2}} 
\put(1,3){\circle*{.2}} 
\put(3,1){\circle*{.2}} 
\put(4,4){\circle*{.2}} 
\put(0,0){\line(3,1){3}} 
\put(0,0){\line(1,3){1}} 
\put(3,1){\line(1,3){1}} 
\put(1,3){\line(3,1){3}} 
\thicklines
\put(-0.5,-0.5){\vector(3,1){5}} 
\put(4.5,1.25){\makebox(0.25,.5){\small$y=(x{-}1)/3$}}
\end{picture}

\small We insert a hyperplane (just a line, in 2-space) midway between the separated point and the convex shape, parallel to the boundary of the latter.
\caption{Finding a separating hyperplane $y=(x{-}1)/3$ in 2-space.}\label{f1446}
\end{wrapfigure}

Now that we can concentrate on just two dimensions, it's easy to find a separating hyperplane with a picture. \Fig{f1446} shows the $\Pi_{P_2}$-point as an open circle at $(2,0)$, while the filled circles give the vertices of the diamond-shaped convex closure corresponding to refinements of $\Pi_{P_4}$. Clearly the (degenerate) hyperplane $y=(x{-}1)/3$ separates the point from the diamond. The normal of that hyperplane (up and left, perpendicular to the line) has direction $(-1,3)$, and when we fill in the other seven dimensions as zero --since they're not needed for separation-- that gives us a candidate normal of $X \Defs (-1,0,3,0,0,0,0,0,0)$ in 9-space. By translating $X$ to matrix form and transposing it, we can then give a tentatitive definition of $D$ as shown at right. However this is not quite our final value for it. 
\hspace{\fill}
\raisebox{-1.8em}[0pt][0pt]{
\(
 \left(
 \begin{array}{r@{~}c@{~}c}
  -1 & 0 & 0 \\
   0 & 0 & 0 \\
   3 & 0 & 0
 \end{array}
 \right)
\)
}

\parshape=4  0pt .75\linewidth 0pt .75\linewidth 0pt .75\linewidth 0pt 1\linewidth
The dot-product of $X$ with $\Pi_{P_2}$, that is $\Tr.(\Pi_{P_2}{\times} D)$, turns out to be $-2$; and with the refinements of $\Pi_{P_4}$ shown at \Eqn{e93475} we get the dot-products $8$ and $0$ (multiple times), showing indeed the separation we expect, but in the wrong direction: the values 0 and 8 for $P_4$ are both strictly greater than the value $-2$ for $P_2$, and we want them to be strictly less. Accordingly we multiply the tentative $D$ above by $-1$
\footnote{The fact we can simply multiply by $-1$ to reverse the sense of the comparison does not mean we can just as easily construct a context to show $P_2{\NotRef}P_4$ --- which would indeed be a worry. In fact for $P_2{\NotRef}P_4$ we'd need a shape $X_{P_2}$ and a point $x_{P_4}$; and then we would find $x_{P_4}$ \emph{inside} the shape $X_{P_2}$, thus unable to be separated from it.}
and then add 3 to all its elements to make them non-negative; finally we divide everything by 10 to make its rows sum to no more than 1. To get the final $D$ from this we must then add a new ``zero-th column'' to make each row one-summing exactly. That gives
\vspace{1em}
\begin{displaymath}
 D ~\Defs~
 \left(
 \begin{array}{c@{~~}c@{~~}c@{~~}c}
  \raisebox{1.7em}[0pt][0pt]{\makebox[0pt][l]{\raisebox{-.3em}{$\downarrow$}~\tiny Zero'th column, for one-summing}}
  0 & 0.4 & 0.3 & 0.3 \\
  0.1 & 0.3 & 0.3 & 0.3 \\
  0.4& 0 & 0.3 & 0.3
 \end{array}
 \right) ~.
\end{displaymath}%
The distinguishing context $\SeqCC{C}$, say, must then overwrite $\Vh$ according to the distribution given by a row of $D$, the one selected by the value $\hat{h}$ of $\Vh$ incoming to $C$;
thus we construct $C$ to be
\[
 \begin{array}{ll@{~~}l}
  \If~\Vv{=}1 &\Then&
   \Vh\From
   (\PSet{}{}{1\At{\,0.4},2\At{\,0.3},3\At{\,0.3}}
   ~\If~\Vh{=}1~\Else~
   \PSet{}{}{0\At{\,0.4},2\At{\,0.3},3\At{\,0.3}}) \\
  &\Else&\Vh\Gets0~\Fi ~,
 \end{array}
\]
with the outer \If\ effectively restricting our attack to occur only when $v'{=}1$. (That allows us to ignore the $\Vh{=}2$ case in $D$, as well.) Thus we have our context $\SeqCC{C}$. Let us now check that it actually works.

We begin with $P_4;C$. Its output hyper-distribution is (after some calculation) given by
\[
 \begin{array}{llcr}
  \LeftPS & (0,\PSet{}{}{0})\Att{1}{3}~, \\
          & (1,\PSet{}{}{0\At{\,0.1}, 1\At{\,0.3},2\At{\,0.3},3\At{\,0.3}})\Att{1}{3} &, \\
          & (1,\PSet{}{}{0\At{\,0.3}, 1\At{\,0.1},2\At{\,0.3},3\At{\,0.3}})\Att{1}{3} && \RightPS~,
 \end{array}
\]
whose Bayes Vulnerability is $1/3{*}1 + 1/3{*}0.3 + 1/3{*}0.3 \approx 0.53$.
On the other hand, for $P_2;C$ the output hyper-distribution is
\[
 \begin{array}{llcr}
  \LeftPS & (0,\PSet{}{}{0})\Att{1}{3}~, \\
          & (1,\PSet{}{}{1\At{\,0.4},2\At{\,0.3},3\At{\,0.3}})\Att{1}{6} &, \\
          & (1,\PSet{}{}{0\At{\,0.2}, 1\At{\,0.2},2\At{\,0.3},3\At{\,0.3}})\Att{1}{3} &, \\
          & (1,\PSet{}{}{0\At{\,0.4},2\At{\,0.3},3\At{\,0.3}})\Att{1}{6} && \RightPS~,
 \end{array}
\]
and here the vulnerability is $1/3{*}1 + 1/6{*}0.4 + 1/3{*}0.3 + 1/6{*}0.3 \approx 0.55$. Note that in the third summand we took $0.3$ rather than the larger $0.4$ associated with 0, since as part of our construction we exclude guesses that $\Vh$ is 0.\,%
\footnote{\label{fn1223}Dealing with this detail would split the 0-case in half, uniformly distributed over $-1,-2$, since the resulting probability $0.4/2$ for each would then be small enough that a Bayes-Vulnerability attack would never choose it. The adjusted context $C'$ would contain
\[
 \Vh\From
 (\PSet{}{}{1\At{\,0.4},2\At{\,0.3},3\At{\,0.3}}
 ~\If~\Vh{=}1~\Else~
 \PSet{}{}{-2\At{\,0.2},-1\At{\,0.2},2\At{\,0.3},3\At{\,0.3}}) \\
\]
and the resulting output for $P_2;C'$ would be
\[
 \begin{array}{llr}
  \LeftPS & (0,\PSet{}{}{0})\Att{1}{3}~, \\
          & (1,\PSet{}{}{1\At{\,0.4},2\At{\,0.3},3\At{\,0.3}})\Att{1}{6}~, \\
          & (1,\PSet{}{}{-2\At{\,0.1},-1\At{\,0.1},1\At{\,0.2},2\At{\,0.3},3\At{\,0.3}})\Att{1}{3}~, \\
          & (1,\PSet{}{}{-2\At{\,0.2},-1\At{\,0.2},2\At{\,0.3},3\At{\,0.3}})\Att{1}{6} & \RightPS~,
 \end{array}
\]
in which neither $-2$ nor $-1$ would ever be chosen for a Bayes-Vulnerability attack.
}

\begin{wrapfigure}[18]{R}{.58\textwidth}
\includegraphics[scale=.32]{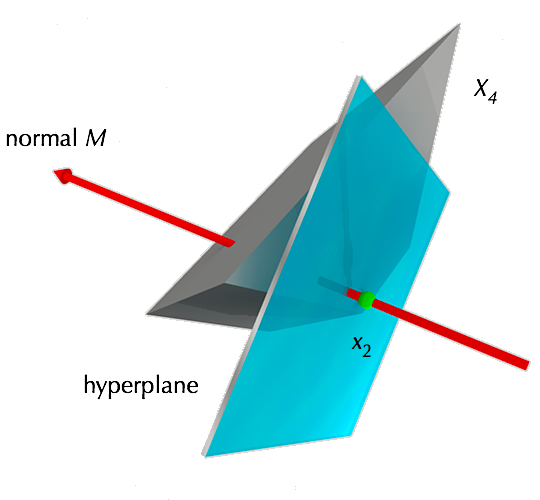}
\caption{Finding a separating hyperplane, with (2,1,1) as normal, in 3 of the 9 dimensions.}\label{f1854}
\end{wrapfigure}
Thus we have established that $P_4;C \NotERef P_2;C$ (for the adjusted $C$ --- see Footnote \ref{fn1223}), because the vulnerability of the former is $0.53$ but for the latter the vulnerability is the greater $0.55$. Hence when our refinement relation insists that $P_4{\NotRef}P_2$ --as we argued earlier above-- in fact it is not being too severe, but rather it is acting just as a compositional closure should. It protects us not only against the context $C$ we just made, but all other contexts too --- in spite of the fact that in isolation $P_4$ and $P_2$ are not distinguished by elementary testing.

Finally, in this example there are many hyperplanes with distinct normals that achieve the separation we need, and each of these may be used to construct different distinguishing contexts. For example, since there exists a separating hyperplane with normal
$(0,0,2,1,0,0,1,0,0)$
we can use it to define another distribution matrix
\[
 D ~\Defs~
 \left(
 \begin{array}{c@{~~}c@{~~}c}
  0.5 & 0.25 & 0.25 \\
  0 & 0 & 0 \\
  0 & 0.5 & 0.5
 \end{array}
 \right)
\]
from which we can specify the distinguishing context $\SeqCC{C}$, where $C$ is
\begin{equation}\label{e1947}
 \begin{array}{l}
 \If~\Vv{=}1~\Then \\
 \quad \Vh\From~
           (\,\PSet{}{}{1\Att{1}{2}, 2\Att{1}{4}, 3\Att{1}{4}}
           ~~\If~
           \Vh{=}1
           ~\Else~~
           \PSet{}{}{2\Att{1}{2}, 3\Att{1}{2}}\,) \\
 \Else~\Vh\Gets 1~\Fi~,
 \end{array}
\end{equation}
which requires no  $\Vh\Gets0$ case for $D$ since the rows of its defining normal just happen to have the same sum. (That's not true for the middle row; but as before we can ignore it since, in the $v'{=}1$ case we are considering, that row is never used.)
\footnote{
It can be shown that this is a legitimate counter-example by using  \Eqn{e1948} and \Eqn{e1947} to calculate the partitions
\begin{displaymath}
 \textrm{$v'{=}1$}~\left\{\quad\quad
 \begin{array}{r@{\hspace{1em}}l}
  \Pi_{P_2;C}\In
  & \List{\PSet{}{}{1\Att{1}{12},2\Att{1}{24},3\Att{1}{24}}, \PSet{}{}{1\Att{1}{12},2\Att{1}{8},3\Att{1}{8}}, \PSet{}{}{2\Att{1}{12},3\Att{1}{12}}} \\
  \Pi_{P_4;C}\In
  & \List{\PSet{}{}{1\Att{1}{8},2\Att{5}{48},3\Att{5}{48}}, \PSet{}{}{1\Att{1}{24},2\Att{7}{48},1\Att{7}{48}}} ~.
 \end{array}
 \right.
\end{displaymath}
and observing that the vulnerability of 
$\Pi_{P_4;C}$ is $1/8{+}7/48=13/48$, which is just smaller than the vulnerability of $\Pi_{P_2;C}$ at $1/12{+}1/8{+}1/12=7/24$. 
} 

Finding the normal that generates \Eqn{e1947}, however, is harder if done geometrically: it turns out that we would have had to specialise to three coordinate indices 3, 4 and 7 rather than just 1 and 3. 
The resulting inspection --to see just where to slip the hyperplane in between-- would then have had to be done in three- rather than two dimensions, as \Fig{f1854} illustrates (in a side view). In general such hyperplanes can of course be found, without drawing pictures, by using constraint solvers to deal with the linear inequalities symbolically.\par

\section{Proof of the Atomicity Lemma} \label{a29375}

To prove the atomicity distribution lemma (\Lem{l1222} from \Sec{s9374}) we use the matrix algebra from \App{a394759}.

Suppose we have matrix representations $\MatSemC{P_{\{1,2\}}}$ for the classical program texts $P_1$ and $P_2$ and a row-matrix representation $\MatSem{v,\delta}$ of an incoming split-state.

If from every initial and final $\Vv$-state of $P_1;P_2$ it is possible to determine the intermediate value of $\Vv$ (after $P_1$ and before $P_2$) then there must exist a total function $f\In \VV{\Fun}\VV{\Fun}\VV$ such that for all $v,v'$ we have
\begin{equation}\label{e9735}
\IdM{v} \times \MatSemC{P_1} \times \MatSemC{P_2} \times \IdM{v'}
 \Wide{=} 
\IdM{v} \times \MatSemC{P_1} \times \IdM{f.v.v'} \times \MatSemC{P_2} \times \IdM{v'}
\end{equation}
from which we have for all $\hat{v} \neq f.v.v'$ that
\begin{equation}\label{e57647}
\IdM{v} \times \MatSemC{P_1} \times \IdM{\hat{v}} \times \MatSemC{P_2} \times \IdM{v'}
\Wide{=}
\ZeroMatrix_N ~,
\end{equation}
where $\ZeroMatrix_N$ is the $N{\times}N$ matrix of zeros.

Assuming such an $f$ with properties \Eqn{e9735} and \Eqn{e57647}, we can calculate
\begin{Reason}
\Step{}{
 \MatSem[.9em]{\,\Sem{\Atomic{P_1;P_2}}.(v,\delta)\,}
}
\StepR{$\Similar$}{embedding}{
  \General{\Above}{v'\In\VV}{}{
   \MatSem{v,\delta}\times\MatSemC{P_1;P_2}\times\IdM{v'}}
}
\StepR{$=$}{classical composition}{
  \General{\Above}{v'\In\VV}{}{
   \MatSem{v,\delta}\times\MatSemC{P_1}\times\MatSemC{P_2}\times\IdM{v'}}
}
\StepR{$=$}{$\MatSem{v,\delta}{\times}\IdM{v}=\MatSem{v,\delta}$}{
  \General{\Above}{v'\In\VV}{}{
   \MatSem{v,\delta}\times\IdM{v}\times\MatSemC{P_1}\times\MatSemC{P_2}\times\IdM{v'}}
}
\StepR{$=$}{\Eqn{e9735}}{
  \General{\Above}{v'\In\VV}{}{
   \MatSem{v,\delta}\times\IdM{v}\times\MatSemC{P_1}\times\IdM{f.v.v'}\times\MatSemC{P_2}\times\IdM{v'}}
}
\Space
\StepR{$=$}{one-point rule\\ for $(\Above)$}{
  \General{\Above}{\Vpk,\hat{v}\In\VV}{\hat{v}{=}f.v.v'}{
   \MatSem{v,\delta}\times\IdM{v}\times\MatSemC{P_1}\times\IdM{\hat{v}}\times\MatSemC{P_2}\times\IdM{v'}}
}
\Space
\WideStepR{$\Similar$}{$\ZeroMatrix_{1{\times}N}$ is unit of concatenation, up to similarity}{
  &\General{\Above}{\Vpk,\hat{v}\In\VV}{\hat{v}{=}f.v.v'}{
   \MatSem{v,\delta}\times\IdM{v}\times\MatSemC{P_1}\times\IdM{\hat{v}}\times\MatSemC{P_2}\times\IdM{v'}} \\
  \Above& \General{\Above}{\Vpk,\hat{v}\In\VV}{\hat{v}{\neq}f.v.v'}{\ZeroMatrix_{1{\times}N}}
}
\Space
\WideStepR{$=$}{$\ZeroMatrix_{1{\times}N} = \MatSem{v,\delta}{\times}\ZeroMatrix_N$, \Eqn{e57647}}{
  &\General{\Above}{\Vpk,\hat{v}\In\VV}{\hat{v}{=}f.v.v'}{
   \MatSem{v,\delta}\times\IdM{v}\times\MatSemC{P_1}\times\IdM{\hat{v}}\times\MatSemC{P_2}\times\IdM{v'}} \\
  \Above& \General{\Above}{\Vpk,\hat{v}\In\VV}{\hat{v}{\neq}f.v.v'}{\MatSem{v,\delta}\times\IdM{v}\times\MatSemC{P_1}\times\IdM{\hat{v}}\times\MatSemC{P_2}\times\IdM{v'}}
}
\Space
\WideStepR{$\Similar$}{$\hat{v}{=}f.v.v'$ and $\hat{v}{\neq}f.v.v'$ are disjoint and exhaustive;\\
$(\Above)$ is commutative and associative up to similarity}{~\\
  \General{\Above}{\Vpk,\hat{v}\In\VV}{}{
   \MatSem{v,\delta}\times\IdM{v}\times\MatSemC{P_1}\times\IdM{\hat{v}}\times\MatSemC{P_2}\times\IdM{v'}} \\
}
\Space
\StepR{$=$}{$\MatSem{v,\delta}{\times}\IdM{v}=\MatSem{v,\delta}$}{
  \General{\Above}{\Vpk,\hat{v}\In\VV}{}{
   \MatSem{v,\delta}\times\MatSemC{P_1}\times\IdM{\hat{v}}\times\MatSemC{P_2}\times\IdM{v'}} \\
}
\StepR{$=$}{distribute $\Above$}{
  \General{\Above}{v'\In\VV}{}{
   \General{\Above}{\hat{v}}{}{\MatSem{v,\delta}\times\MatSemC{P_1}\times\IdM{\hat{v}}}\times\MatSemC{P_2}\times\IdM{v'}} \\
}
\StepR{$\Similar$}{composition, embedding}{
 \MatSem[.9em]{\Sem{\Atomic{P_1};\Atomic{P_2}}.(v,\delta)}~,
}
\end{Reason}
whence our result follows because $\MatSem{\cdot}$ is injective up to similarity and $(v,\delta)$ was arbitrary.

\section{Informal description of the Oblivious Transfer implementation\,\protect\footnotemark}\label{a1446}
\footnotetext{An even more informal description is this fairy tale. An Apprentice magician is about to graduate, and he must now choose between black- or white magic. His Sorcerer will allow him to read either the \emph{Black Tome} or the \emph{White Tome}, not both; and his choice must be his own, uncoerced, thus never revealed to the Academy.

The Sorcerer summons a \emph{trusted third party} Djinn who gives him two locks, one black and one white; and the Djinn gives a single, golden key to the Apprentice. On the key is a small dot that only the Apprentice can see: it is the colour of the matching lock. The Djinn then returns to his own dimension.

The Apprentice tells the Sorcerer to match the lock colours to the Tomes, or to reverse them: it depends on whether his choice matches the colour of the dot. The Sorcerer then leaves; the Apprentice can unlock only the Tome of his choice; and --provided he locks it again-- no one afterwards will know which one he read.} 

Given are two agents $B,C$; Agent $B$ has two messsages $\Vm_{\{0,1\}}$, bit-strings of the same length; Agent $C$ has a message variable $\Vm$ and a choice $\Vc\In\Set{}{}{0,1}$ of which of $\Vm_{\{0,1\}}$ is to be assigned to $\Vm$. The specification is thus
\[
 \begin{array}{l}
  \Vis_{B}~\Vm_{0},\Vm_{1};~\Vis_{C}~\Vm,\Vc; \\
  \hspace{1em}\Vm\Gets \Vm_{\Vc} ~.
 \end{array}
\]
Note that $B$ does not discover $\Vc$ and that $C$ does not discover $\Vm_{\neg \Vc}$.

\medskip
The implementation is, informally, as follows:
\begin{enumerate}
\item[]~
\item[] \hrulefill\textit{This is the prelude of the protocol}\hrulefill
\item\label{i1620} Agent $B$ chooses privately two random bit-strings $\Vm'_{\{0,1\}}$ to be used for $\Xor$-encrypting $\Vm_{\{0,1\}}$ respectively.
\item\label{i1621} Agent $C$ chooses privately in $\Vc'\In\Set{}{}{0,1}$ which of the encrypting strings $\Vm'_{\{0,1\}}$ will be revealed to her.
\item\label{i1619} A trusted third party collects both $\Vm'_{\{0,1\}}$ from $B$, collects $\Vc'$ from $C$ and then reveals (only) $\Vm'_{c'}$ to $C$. She throws $\Vm'_{\neg \Vc'}$ away, and then leaves.
\item[]~
\item[] \hrulefill\textit{From here is the main part of the protocol}\hrulefill
\item\label{i1622} Agent $C$ then tells $B$ to encrypt and send messages in the following way:
\begin{enumerate}
\item[]~
\item\label{i1610} If Agent $C$ wants $\Vm_{0}$ and has $\Vm'_{0}$, then she instructs $B$ to send her both $\Vm_{0}{\Xor}\Vm'_{0}$ and $\Vm_{1}{\Xor}\Vm'_{1}$. Because she has $\Vm'_{0}$ she can recover $\Vm_{0}$ via $(\Vm_{0}{\Xor}\Vm'_{0})\Xor \Vm'_{0}$; but she cannot recover $\Vm_{1}$.
\item\label{i1609} Similarly, if Agent $C$ wants $\Vm_{0}$ but has $\Vm'_{1}$ instead (of $\Vm'_{0}$), then she simply instructs $B$ to send her both $\Vm_{0}{\Xor}\Vm'_{1}$ and $\Vm_{1}{\Xor}\Vm'_{0}$, i.e.\ with the encryption the other way around.
\item If Agent $C$ wants $\Vm_{1}$ and has $\Vm'_{0}$ --- as for \Itm{i1609}.
\item\label{i1630} If Agent $C$ wants $\Vm_{1}$ and has $\Vm'_{1}$ --- as for \Itm{i1610}.
\end{enumerate}

The four cases (\ref{i1610}--\ref{i1630}) can be described succinctly --if cryptically-- simply by instructing $B$ to send $\Vm_{i}{\Xor}\Vm'_{i{\Xor}\Vc{\Xor}\Vc'}$ for $i=0,1$.
\end{enumerate}
Note that only Step \Itm{i1619} involves private messages (first between $B$ and the third party, and then between the third party and $C$), and that is only in the prelude, before any of the actual data $\Vm_{\{0,1\}},\Vc$ has appeared. Steps \Itm{i1620} and \Itm{i1621} involve no messages at all; and the messages occurring in Step \Itm{i1622} are $\Xor$-encrypted already. In effect the prelude has created a one-time pad.

A formal derivation of this implementation is given elsewhere \cite{Morgan:07}.

\section{Alternative uncertainty measures}\label{s3448}

\subsection{Shannon Entropy} \label{s3628}
The \emph{Shannon Entropy} of a (full) distribution $\delta \In \TDist{\cal X}$ is $\Ht.\delta \Defs \Exp{d\In\delta}{-\lg (\delta.d)}$, that is the weighted average of the negated base-2 logarithms of its constituent probabilities~\cite{Shannon:48}. By extension, for any hyper-distribution $\Delta$ we define the \emph{conditional} Shannon Entropy $\Ht.\Delta$ to be $\Exp{(v,\delta)\In \Delta}{\Ht.\delta}$, the expected value of the entropies of its support~\cite{Cachin:97}.

Going further, if we split up our hyper-distribution by $v$ into its partitions, we have an equivalent presentation of entropy as the sum of individual partition-entropies $\Ht.\Delta = \Sum{v\In\VV}{}{\Ht.(\Fracs.\Delta.v)}$, provided we define the entropy of a single partition, and of a single fraction, as follows:
\begin{equation}\label{e2058}
 \begin{array}{rc@{~~~}l}
  \Ht.\Pi &\Defs& \Sum{\pi\In\Pi}{}{\Ht.\pi} \\
  \Ht.\pi &\Defs& \Exp{d\In\pi}{\NLg(\Norm{\pi}.d)} ~,
 \end{array}
\end{equation}
where we write $\NLg$ for ${-}\kern-.2em\lg$ to avoid a proliferation of minus signs, and $\Norm{\pi}$ is normalisation of the fraction $\pi$, scaling it up (if necessary) to give a distribution again.

The ordering $(\HRef)$ based on hyper-distributions
\begin{equation*}
\Delta_S \HRef \Delta_I \Wide{\Defs}  
(\Ft.\Delta_S = \Ft.\Delta_I) \land 
(\Ht.\Delta_S \leq \Ht.\Delta_I)
\end{equation*}
is then specified, as for the Bayes order $(\ERef)$, so that $\Delta_S{\HRef}\Delta_I$ if they are functionally equivalent and the uncertainty (the Shannon Entropy in this case) of $\Delta_I$ is no less than that of $\Delta_S$. It extends pointwise to secure programs. Furthermore we write that $S \StrictHRef I$ when $S \HRef I$ but $I \NotHRef S$.

\subsubsection{Non-compositionality}
Consider again two functionally-equivalent programs from our three-box puzzle example from \Sec{s1121} and \Sec{s1201}:
\[
 \begin{array}{rc@{~~~}l@{~}l@{~}l}
  S &\Defs& 
   \Vh\Gets 0{\PC{}}1{\PC{}}2;&
   \Vv\From\PSet{}{}{w\Att{\Vh}{2},b\At{1{-}\frac{\Vh}{2}}};&
   \Vv\Gets\bot
  \\
  I_2 &\Defs& 
   \Vh\Gets0{\PC{}}1{\PC{}}2;&
   \Vv\From\PSet{}{}{w\At{(\Vh\div2)},b\At{1{-}(\Vh\div2)}};&
   \Vv\Gets\bot
 \end{array}  
\]
with final hyper-distributions
\medskip\par\noindent
\hfill\(
 \PSet{}{}{~(\bot,\PSet{}{}{1\Att{1}{3},2\Att{2}{3}}),~(\bot,\PSet{}{}{0\Att{2}{3},1\Att{1}{3}})~}
\)\hfill\makebox[0pt][r]{($\Delta'_S$)}
\par\noindent
\hfill\(
 \makebox[0pt][r]{and\hspace{7em}}
\PSet{}{}{~(\bot,\PSet{}{}{2})\Att{1}{3}~, ~(\bot,\PSet{}{}{0,1})\Att{2}{3}~}\makebox[0pt][l]{~.}
\makebox[0pt][l]{~.}
\)\hfill\makebox[0pt][r]{($\Delta'_{I_2}$)}%
\medskip\par\noindent

The Shannon entropy of $\Delta'_S$, calculated $2*\frac{1}{2}(\frac{1}{3}\NLg \frac{1}{3} + \frac{2}{3}\NLg \frac{2}{3})$, is slightly more than $0.918$, exceeding the entropy $\Delta'_{I_2}$ that, by the simpler calculation given by $\frac{1}{3}(\NLg\,1) + \frac{2}{3}(2{*}\frac{1}{2}\NLg \frac{1}{2})$, turns out to be exactly $\frac{2}{3}$; and so  $I_2{\HRef}S$.

However if we define context $\CC$ to be 
$\SeqCC{\Vh \Gets (1~\If~\Vh{=}2~\Else~\Vh)}$ 
then the entropy of $\CC(I_2)$ is the same as before at $\frac{2}{3}$; but the entropy of $\CC(S)$ is now only a half of what it was, at ${\approx}0.459$. Hence $\CC(I_2) \NotHRef \CC(S)$.

\subsubsection{Soundness}

We follow initially the structure of the soundness proof for Bayes Risk. Fix an initial split-state and construct the output hyper-distributions $\Delta'_{\{S,I\}}$ that result from $S,I$ respectively. Then since we assume $S{\Ref}I$ we must have $\Delta'_S{\Ref}\Delta'_I$. We now show that this implies $\Delta'_S{\HRef}\Delta'_I$.

Since $S{\Ref}I$ trivially guarantees that $\Ft.\Delta'_S {=} \Ft.\Delta'_I$, we need to show that the Shannon Entropy condition in $(\HRef)$ is satisfied. Since we have that $\Ht.\Delta'$ is $\Sum{v'\In \VV}{}{\Ht.(\Fracs.\Delta.v')}$, it is enough to show that for each $v'\In \VV$ the entropy of $\Pi'_S \Defs \Fracs.\Delta'_{S}.v'$ is no less than the entropy of $\Pi'_I \Defs \Fracs.\Delta'_{I}.v'$, provided that $\Pi'_S \Similar \Pi' \Finer \Pi'_I$ for some partition $\Pi'$ depending on $v'$.

For $\Pi'_S{\Similar}\Pi'$ we consider the unique $\Pi$ that is the reduction of both: it is formed in each case by adding together groups of similar fractions. From \Eqn{e2058} and arithmetic, we obtain immediately that $\Ht.\Pi'_S = \Ht.\Pi = \Ht.\Pi'$.\,%
\footnote{If $\pi_1{\Similar}\pi_2$ then $\Norm{\pi_1}{=}\Norm{\pi_2}{=}\Norm{\pi_1{+}\pi_2}$ and so the line marked $\dagger$ below becomes an equality.}

For $\Pi'{\Finer}\Pi'_I$ we know that the fractions of $\Pi'_I$ are sums of groups of not-necessarily-similar fractions in $\Pi'$. We consider the special case of just two fractions $\pi_{\{1,2\}}$ in $\Pi'$ summing to a single fraction $\pi\Defs\pi_1{+}\pi_2$ in $\Pi'_I$, and look at their relative contributions to the sum \Eqn{e2058}; we have
\begin{Reason}
\Step{}{
 \Ht.\pi
}
\Step{$=$}{
 \Ht.(\pi_1{+}\pi_2)
}
\Step{$=$}{
 \Exp{d\In(\pi_1{+}\pi_2)}{\NLg(\Norm{\pi_1{+}\pi_2}.d)}
}
\Step{$=$}{
 \Exp{d\In\pi_1}{\NLg(\Norm{\pi_1{+}\pi_2}.d)} + \Exp{d\In\pi_2}{\NLg(\Norm{\pi_1{+}\pi_2}.d)}
}
\StepR{$\HangLeft{$\dagger$~~~}{\geq}$}{see below}{
 \Exp{d\In\pi_1}{\NLg(\Norm{\pi_1}.d)} + \Exp{d\In\pi_2}{\NLg(\Norm{\pi_2}.d)}
}
\Step{$=$}{
 \Ht.\pi_1 + \Ht.\pi_2 ~,
}
\end{Reason}
that is that the contribution to the conditional entropy of $\pi$ on its own is at least as great as it was when was separated into $\pi_{\{1,2\}}$.

For ``see below'' we refer to the \emph{Key Lemma} \cite[p5]{Welsh:88} which states that for two total distributions $\delta,\delta'$ of equal support, the weighted sum $\Exp{d\In\delta}{\NLg(\delta'.d)}$ attains its minimum over $\delta'$ when $\delta{=}\delta'$.

Extending the argument similarly to multiple additions gives $\Ht.\Pi'\leq\Ht.\Pi'_I$ as required and thus we have $\Ht.\Pi'_S\leq\Ht.\Pi'_I$ overall. We note that the inequality at $\dagger$ is strict when $\pi_1{\NotSimilar}\pi_2$, because then e.g.\ $\Norm{\pi_1}{\neq}\Norm{\pi_1{+}\pi_2}$.\,%
\footnote{This follows from a strengthening of the Key Lemma to ``\ldots\emph{only} when $\delta{=}\delta'$,'' which is implied by the proof of Thm.~1 [\textit{op.\ cit.}]\ immediately before.}
We have established
\begin{Theorem}{Soundness of $(\Ref)$ w.r.t. $(\HRef)$}{t2112}
For all secure programs $S$ and $I$ and contexts $\CC$, we have that $S{\Ref}I$ implies $\CC(S){\HRef}\CC(I)$.
\end{Theorem}

Finally, when $S{\Ref}I$ but $S \neq I$ so that $\Delta'_S{\neq}\Delta'_I$ for some initial split-state we must have $\Pi'_S{\not\Similar}\Pi'_I$ for some final $v'$, since both those partitions are in reduced form: that is, reduced partitions cannot be similar without actually being equal. Thus also $\Pi'{\not\Similar}\Pi'_I$, and so we can find particular $\pi_1{\NotSimilar}\pi_2$ to realise the strict inequality at $\dagger$. That gives us
\begin{Lemma}{Strict soundness}{l2121}
For all hyper-distributions $\Delta_{\{1,2\}}$ we have that $\Delta_1{\Ref}\Delta_2$ but $\Delta_1 \neq \Delta_2$ implies $\Delta_1{\StrictHRef}\Delta_2$.
\end{Lemma}

\subsection{Marginal guesswork}

The \emph{Marginal guesswork}~\cite{Pliam:00} of a distribution $\delta\In \TDist{\cal X}$ is the least number of guesses an attacker requires to be sure that her chance of guessing some $\Vh$ chosen according to $\delta$ is at least a given probability $\alpha$. We define it
\begin{equation*}
\Gt{\alpha}.\delta \Wide{\Defs} \General{\MIN}{i\In1..N}{\UMax{i}\delta \geq \alpha}{}
\end{equation*}
where we write $\UMax{i}\delta$, or more generally $\UMax{i}\pi$ for fraction $\pi$ to mean the sum of the $i$ greatest probabilities in $\pi$, and $N$ is the cardinality of ${\cal X}$. Note that by super-distribution of maximum over addition we have $\UMax{i}(\pi_1{+}\pi_2)\leq\UMax{i}\pi_1+\UMax{i}\pi_2$ for any $i$ in the proper range. To avoid clutter, we will omit the range $1..N$ for $i$ from here on.

For a hyper-distribution $\Delta$ we define
\begin{equation}\label{e9874}
\begin{array}{ll}
& \Gt{\alpha}.\Delta ~\Defs~
\General{\MIN}{i}{\Exp{(v,\delta)\In\Delta}{\UMax{i}\delta} \geq \alpha}{} \\
\textrm{or equivalently~} & 
\Gt{\alpha}.\Delta ~\Defs~
\General{\MIN}{i}{\Sum{\Vv\In\VV; \pi \In \Fracs.\Delta.v}{}{\UMax{i}\pi} \geq \alpha}{}
\end{array}
\end{equation}
which is the least value $i$ such that if an attacker is allowed to make that many guesses then she can discover the value of $\Vh$ with probability at least $\alpha$.

Observe that our definition of $\Gt{\alpha}.\Delta$ is \emph{not} the same as the \emph{conditional marginal guesswork} $\Exp{\delta\In\Delta}{\Gt{\alpha}.\delta}$ as conventionally defined~\cite{Kopf:07}. We argue that conditional marginal guesswork \emph{is not} a reasonable measure of the number of guesses required by an attacker to ensure that the probability of guessing $\Vh$ in $\Delta$ is greater than $\alpha$. Consider for example the hyper-distributions
\begin{equation*}
\Delta_S \Defs \Uniform{(v,\Uniform{0}), (v,\Uniform{1..4})} \Wide{\textrm{and}}\\
\Delta_I \Defs \Uniform{(v, \Uniform{0} \PC{} \Uniform{1..4})}~.\footnotemark
\end{equation*}
\footnotetext{Alternatively we could write out $\Uniform{0} \PC{} \Uniform{1..4}$ with its explicit probabilities as $\PSet{}{}{0\Att{1}{2},1\Att{1}{8},2\Att{1}{8},3\Att{1}{8},4\Att{1}{8}}$, but we prefer to avoid the superscripts.}%
Note that $\Delta_S{\Ref}\Delta_I$ since the latter is obtained by merging the two split-states of the former.

Now an attacker has more information about how $\Vh$ was chosen in $\Delta_S$ than in $\Delta_I$: for $\Delta_S$ she knows not only that $\Vh$ is distributed according to the distribution $\Uniform{0} \PC{} \Uniform{1..4}$ overall (as for $\Delta_I$), but as well she knows when $\Vh$ was chosen from $\Uniform{0}$ and when $\Vh$ was chosen from $\Uniform{1..4}$. However, when we set $\alpha\Defs1/2$ the conditional marginal guesswork of $\Delta_S$ is $\frac{1}{2}(1) + \frac{1}{2}(4\alpha)$, that is $3/2$ --- which is higher than for $\Delta_I$, which gives only $1$. This suggests that it is \emph{harder} for an attacker to guess $\Vh$ in $\Delta_S$ than in $\Delta_I$, in spite of the fact that the attacker knows more about the final $\Vh$-distribution in $\Delta_S$ when launching an attack.

Using our $\Gt{\alpha}$ we have $\Gt{1/2}.\Delta_S = \Gt{1/2}.\Delta_I$, that is 1 in both cases: with just one guess at her disposal an attacker is guaranteed to guess $\Vh$ at least half the time. Applying her one guess to $\Delta_S$, half the time she can guess 0 and is sure to be right; in $\Delta_1$ she guesses 0 and will be right half the time.

The ordering between hyper-distributions based on marginal guesswork is 
\begin{equation*}
\Delta_S \GRef{\alpha} \Delta_I \Wide{\Defs} 
(\Ft.\Delta_S = \Ft.\Delta_I) \land 
(\Gt{\alpha}.\Delta_S \leq \Gt{\alpha}.\Delta_I)
\end{equation*}
which extends pointwise to programs.

\subsubsection{Non-compositionality}

When $\alpha$ is not zero, marginal guesswork --like the other measures-- is non-compositional for our subset of programs. For such an $\alpha{\neq}0$ take, for example, functionally equivalent programs
\begin{eqnarray*}
S   & \Defs & 
\Vh\PGets \Uniform{0,1,2} \PC{\alpha} \Uniform{{-}N..{-}1};\\
&& 
\Vv\From
 (\PSet{}{}{w\At{\frac{\Vh}{2}}, b\At{(1{-}\frac{\Vh}{2})}}
 ~~\If~\Vh{\geq}0~\Else~~
 \PSet{}{}{w\At{\frac{1}{2}}, b\At{\frac{1}{2}}}); \\
&&
\Vv\Gets\bot
\\\\
I & \Defs & 
\Vh\PGets \Uniform{0,1,2} \PC{\alpha} \Uniform{{-}N..{-}1};\\
&&
\Vv\From
 (\PSet{}{}{w\At{\Vh\div2}, b\At{1-\Vh\div2}}
 ~~\If~\Vh{\geq}0~\Else~~
 \PSet{}{}{w\At{\frac{1}{3}}, b\At{\frac{2}{3}}}); \\
&&
\Vv\Gets\bot
\end{eqnarray*}
such that if $\alpha {=} 1$ then $N {=} 0$ else $N \geq 3{\times}\frac{1{-}\alpha}{\alpha}$. These programs have the final output distributions
\medskip\par\noindent
\hfill\(
\Uniform{
(\bot,\PSet{}{}{1\Att{1}{3},2\Att{2}{3}} \PC{\alpha} \Uniform{{-}N...{-}1}),~
(\bot,\PSet{}{}{0\Att{2}{3},1\Att{1}{3}} \PC{\alpha} \Uniform{{-}N...{-}1})
}
\)\hfill\makebox[10pt][r]{($\Delta'_S$)}
\par\noindent
and 
\hfill\(
\PSet{}{}{
(\bot,\PSet{}{}{2}   \PC{\alpha} \Uniform{{-}N...{-}1})\Att{1}{3},~
(\bot,\PSet{}{}{0,1} \PC{\alpha} \Uniform{{-}N...{-}1})\Att{2}{3}
}
\makebox[0pt][l]{~.}
\)\hfill\makebox[10pt][r]{($\Delta'_{I}$)}%
\medskip\par\noindent
We can calculate that both $\Gt{\alpha}.\Delta'_S$ and $\Gt{\alpha}.\Delta'_I$ are $2$, and so $S \GRef{\alpha} I$, but that for context $\CC$ defined as $\SeqCC{\Vh \Gets (\Vh \div 2 ~\If~\Vh{\geq}0 ~\Else~ \Vh)}$ we have $\Gt{\alpha}.\Delta'_I$ is only $1$, while $\Gt{\alpha} \Delta'_S$ remains at $2$ --- and so $\CC(S) \NotGRef{\alpha} \CC(I)$. 

\subsubsection{Soundness}

\begin{Lemma}{$(\Ref)$ implies $(\GRef{\alpha})$}{l9734}
For all hyper-distributions $\Delta_S$ and $\Delta_I$ and probabilities $\alpha$, if $\Delta_S{\Ref}\Delta_I$ then also $\Delta_S{\GRef{\alpha}}\Delta_I$; consequently $S{\Ref}I$ implies$S{\GRef{\alpha}}I$.
\begin{Proof}
From \Eqn{e9874} and the definition of refinement (\Def{d1424}, \Sec{s1213}) it is enough to show that for any partition $\Pi$ and $i$ in range that (i) if the fractions in $\Pi$ are similar then $\Sum{\pi\In \Pi}{}{\UMax{i}\pi} = \UMax{i}\Sum{\pi\In\Pi}{}{}$ else (ii) $\Sum{\pi\In \Pi}{}{\UMax{i}\pi} \geq \UMax{i}\Sum{\pi\In\Pi}{}{}$. To show (ii) we have by generalising $\UMax{i}(\pi_1{+}\pi_2)\leq\UMax{i}\pi_1+\UMax{i}\pi_2$ that indeed
\[
 \Sum{\pi\In\Pi}{}{\UMax{i}\pi}
 \Wide{\geq}
 \UMax{i}\Sum{\pi\In\Pi}{}{} ~,
\]
and for (i) we can replace inequality by equality since $(\UMax{i})$ distributes over summation in that case. 
\end{Proof}
\end{Lemma}

\begin{Theorem}{Soundness of $(\Ref)$ w.r.t. $(\GRef{\alpha})$}{l9543}
For all probabilities $\alpha$, secure programs $S$, $I$ and contexts $\CC$ we have that $S \Ref I$ implies $\CC(S) \GRef{\alpha} \CC(I)$.
\begin{Proof}
\Lem{l9734} and monotonicity of $(\Ref)$ (\Thm{t1120} from \Sec{s9345}).
\end{Proof}
\end{Theorem}

\subsection{Guessing entropy}

The \emph{guessing entropy}~\cite{Massey:94} of a distribution $\delta$ is the (least) average number of guesses required to guess $\Vh$ in $\delta$. It is equivalent to the average $\alpha$-marginal guesswork over all values of $\alpha$ \cite{Pliam:00}, and we define it 
\begin{equation*}
\Gt{}.\delta \Wide{\Defs} \Sum{i\In1..N}{}{\UMin{i}\delta}~,
\end{equation*}
where $\UMin{i}\delta$ is the sum of the $i$ smallest probabilities in $\delta$.\,%
\footnote{If \Wlog\ the four probabilities $a,b,c,d$ are ordered greatest to least, then the best strategy is to guess (the value associated with) $a$ first, and then to go on to guess $b,c,d$ in order as necessary. The average number of guesses needed overall is then $a+2b+3c+4d$, that is $d + (d{+}c) + (d{+}c{+}b) + (d{+}c{+}b{+}a)$.}
Note that by subdistribution of minimum over addition we have $\UMin{i}(\pi_1{+}\pi_2)\geq\UMin{i}\pi_1+\UMin{i}\pi_2$ for any $i$ in range.
For hyper-distribution $\Delta$ we define the conditional guessing entropy, thus
\begin{equation*}
\begin{array}{ll}
& \Gt{}.\Delta \Wide{\Defs} \Exp{(v,\delta)\In \Delta}{\Gt{}.\delta} \\
\textrm{or equivalently}\quad 
& \Gt{}.\Delta \Wide{\Defs} \Sum{v\In \VV;\pi\In \Fracs.\Delta.v}{}{\Gt{}.\pi}~,
\end{array}
\end{equation*}
where $\Gt{}.\pi$ is defined in the same way as $\Gt{}.\delta$. We define the ordering by
\begin{equation*}
\Delta_S \GRef{} \Delta_I \Wide{\Defs} 
(\Ft.\Delta_S = \Ft.\Delta_I) \land
(\Gt{}.\Delta_S \leq \Gt{}.\Delta_I)~,
\end{equation*}
which extends pointwise to secure programs.

\subsubsection{Non-compositionality}

To show non-compositionality of ordering $(\GRef{})$ we refer again (as we did for Shannon entropy in \App{s3628}) to the functionally equivalent programs $S$ and $I_2$. First we calculate that 
\begin{equation*}
\begin{array}{ll@{~~}c@{~~}l@{~~}c@{~~}l}
& \Gt.\Delta'_S &=& 
2{\times}\frac{1}{2}(\frac{1}{3} + (\frac{1}{3}{+}\frac{2}{3})) &=& \frac{4}{3}  \\[.5ex]
\textrm{and} \quad  
& \Gt.\Delta'_{I_2} &=&
\frac{1}{3}(1) + \frac{2}{3}(\frac{1}{2} + (\frac{1}{2}{+}\frac{1}{2})) &=& \frac{4}{3} ~, 
\end{array}
\end{equation*}
so that we have $I_2 \GRef{} S$.
Again taking context $\CC$ to be $\SeqCC{\Vh \Gets (1~\If~\Vh{=}2~\Else~\Vh)}$ we get that the guessing entropy of $\CC(S)$ is reduced to 
$\frac{7}{6}$ while that of $\CC(I_2)$ is still $\frac{4}{3}$, and hence $\CC(I_2) \NotGRef{} \CC(S)$.

\subsubsection{Soundness}

\begin{Lemma}{$(\Ref)$ implies $(\GRef{})$}{l5433}
For all hyper-distributions $\Delta_S$ and $\Delta_I$, we have that $\Delta_S{\Ref}\Delta_I$ implies $\Delta_S {\GRef{}} \Delta_I$; and consequently $S{\Ref}I$ implies $S{\GRef{}}I$.
\begin{Proof}
As in the proof of soundness for marginal guesswork, it is enough to show
that for any partition $\Pi$ (i) if the fractions in $\Pi$ are similar then $\Sum{\pi\In\Pi}{}{\Gt{}.\pi} = \Gt{}.\Sum{\pi\In\Pi}{}{}$ else (ii) $\Sum{\pi\In\Pi}{}{\Gt{}.\pi} \leq \Gt{}.\Sum{\pi\In\Pi}{}{}$. For (ii) we reason:
\begin{Reason}
\Step{}{
 \Sum{\pi\In\Pi}{}{\Gt{}.\pi}
}
\StepR{$=$}{definition $\Gt{}$ for a partition}{
 \Sum{\pi\In\Pi}{}{\Sum{i}{}{\UMin{i}\pi}}
}
\StepR{$=$}{swap summations}{
 \Sum{i}{}{\Sum{\pi\In\Pi}{}{\UMin{i}\pi}}
}
\StepR{$\leq$}{subdistribute minimisation}{
 \Sum{i}{}{\UMin{i}\Sum{\pi\In\Pi}{}{\pi}}
}
\StepR{$=$}{definition $\Gt{}$ for partition}{
 \Gt{}.\Sum{\pi\In\Pi}{}{}~.
}
\end{Reason}
When all the fractions $\pi$ in $\Pi$ are similar, we can replace the inequality in the second-last step with equality, establishing (i).
\end{Proof}
\end{Lemma}

\begin{Theorem}{Soundness of $(\Ref)$ w.r.t. $(\GRef{})$}{l6855}
For all programs $S$ and $I$ and contexts $\CC$ we have that $S{\Ref}I$ implies $\CC(S){\GRef{}} \CC(I)$.
\begin{Proof}
Immediate from \Lem{l5433} and monotonicity of $(\Ref)$ (\Thm{t1120} in \Sec{s9345}).
\end{Proof}
\end{Theorem}

\end{document}